\documentclass[a4paper,12pt,english]{article}
\usepackage{amssymb,bbm,babel,graphicx,multicol}
\usepackage{psfrag}
\usepackage{color}
\usepackage{hyperref}

\makeatletter
\@addtoreset{equation}{section}
\makeatother

\def\dalemb#1#2{{\vbox{\hrule height .#2pt
        \hbox{\vrule width.#2pt height#1pt \kern#1pt
                \vrule width.#2pt}
        \hrule height.#2pt}}}

\def\cN{{\cal N}}
\def\cA{{\cal A}}
\def\cB{{\cal B}}
\def\cC{{\cal C}}
\def\cD{{\cal D}}
\def\ula{{\underline{a}}}
\def\ulb{{\underline{b}}}

\def\hp{ \frac{1}{2}}
\def\qu{\fr{1}{4}}

\let\a=\alpha \let\b=\beta \let\g=\gamma \let\d=\delta \let\e=\epsilon
 \let\h=\eta \let\q=\theta  
\let\l=\lambda \let\m=\mu \let\n=\nu \let\x=\xi \let\r=\rho
\let\s=\sigma \let\t=\tau  \let\f=\phi \let\c=\chi \let\y=\psi
\let\w=\omega  \let\D=\Delta  
    
 \let\W=\Omega   \let\G=\Gamma
  \let\re=\ref
  
\def\nn{\nonumber} \def\bd{\begin{document}} \def\ed{\end{document}}
\def\ds{\documentstyle} \let\fr=\frac \let\bl=\bigl \let\br=\bigr
\let\Br=\Bigr \let\Bl=\Bigl
\let\bm=\bibitem
\let\na=\nabla
\let\pa=\partial \let\ov=\overline
\let\ul=\underline
\newcommand{\be}{\begin{equation}}
\newcommand{\ee}{\end{equation}}
\def\ba{\begin{array}}
\def\ea{\end{array}}
\def\ft#1#2{{\textstyle{{\scriptstyle #1}\over {\scriptstyle #2}}}}
\def\fft#1#2{{#1 \over #2}}
\def\del{\partial}
\def\sst#1{{\scriptscriptstyle #1}}
 \def\oneone{\rlap 1\mkern4mu{\rm l}}
\def\ie{{\it i.e.\ }}
\def\via{{\it via}}
\def\semi{{\ltimes}}
\def\str{{\rm str}}
\def\Dm{{{D_{\sst{max}}}}}
\def\vac{ \left | 0 \right \rangle }
\def\kvac{ \left | k \right \rangle }

\def\sp{\; \; \;}

\def\bol{ \left | B (p^+) \right \rangle}
\def\bo1{ \left | B^0 (p^+) \right \rangle}

\def\bolt{ \left | B (p^+) \right \rangle_{\t}}

\def\boxl{ \left | B (x^-) \right \rangle}

\def\<{ \langle }
\def\>{ \rangle }

\def\vf{\varphi}

\def\ls{{(l,0)}}
\def\lv{{(l,\pm1)}}
\def\lt{{(l,\pm2)}}

\def\lse#1{{(l_{#1},0)}}
\def\lve#1{{(l_{#1},\pm1)}}
\def\lte#1{{(l_{#1},\pm2)}}

\def\lsg#1{{5(l_{#1},0)}}
\def\lvg#1{{5(l_{#1},\pm1)}}
\def\ltg#1{{5(l_{#1},\pm2)}}

\def\lsi#1{{5{(#1,0)}}}
\def\lvi#1{{5{(#1,\pm1)}}}
\def\lti#1{{5{(#1,\pm2)}}}

\def\lsr#1{{1{(#1,0)}}}
\def\lvr#1{{1{(#1,\pm1)}}}
\def\ltr#1{{1{(#1,\pm2)}}}

\def\cn{{\cal N}}
\def\cao{{\cal O}}
\def\cD{{\cal D}}
\def\cE{{\cal E}}
\def\cF{{\cal F}}
\def\cG{{\cal G}}
\def\cH{{\cal H}}
\def\cK{{\cal K}}
\def\cO{{\cal O}}
\def\cP{{\cal P}}
\def\cQ{{\cal Q}}
\def\cR{{\cal R}}
\def\cS{{\cal S}}
\def\cT{{\cal T}}
\def\cU{{\cal U}}
\def\cV{{\cal V}}
\def\cW{{\cal W}}

\newcommand{\nono}{\nonumber}
\newcommand{\eqref}[1]{(\ref{#1})}
\newcommand{\dtilde}[1]{\tilde{\tilde{#1}}}
\newcommand{\hatb}[1]{\hat{\ov{#1}}}
\newcommand{\hatt}[1]{\hat{\tilde{#1}}}
\newcommand{\emnr}{{e_\m}^{\n\r}}
\newcommand{\psm}{\int \frac{d^{11}\l}{{\l^+}^3}}

\newcommand{\hsp}{\hspace{0.5cm}}

\newcommand{\ho}[1]{$\, ^{#1}$}
\newcommand{\hoch}[1]{$\, ^{#1}$}
\newcommand{\bea}{\begin{eqnarray}}
\newcommand{\eea}{\end{eqnarray}}
\newcommand{\ra}{\rightarrow}
\newcommand{\lra}{\longrightarrow}
\newcommand{\Lra}{\Leftrightarrow}
\newcommand{\lera}{\leftrightarrow}
\newcommand{\ap}{\alpha^\prime}
\newcommand{\bp}{\tilde \beta^\prime}
\newcommand{\tr}{{\rm tr} }
\newcommand{\Tr}{{\rm Tr} }
\newcommand{\NP}{Nucl. Phys. }

\newcommand{\ams}{ ${}^1$ {\it Institute for Theoretical Physics,
University of Amsterdam, \\
Valckenierstraat 65, 1018XE Amsterdam, The Netherlands
 \\
{\tt J.Hoogeveen, K.Skenderis@uva.nl}}  \vspace{0.5 cm} \\
{${}^2$ {\it Kavli Institute for Theoretical Physics\\
University of California at Santa Barbara, Santa Barbara, CA 93106-4030, USA}}} 
\newcommand{\auth}{Joost Hoogeveen${}^1$ and Kostas Skenderis${}^{1,2}$}

\def\red{\color{red}}

\vspace{10pt}

\begin{document}

\begin{flushright}
\hfill{ITF-2009-15}
\\
\hfill{NSF-KITP-09-102}
\end{flushright}

\vspace{15pt}

\begin{center}

{\Large \bf Decoupling of unphysical states 
in the minimal pure spinor formalism I}

\vspace{20pt}

\auth

\vspace{15pt}

\vspace{8pt}

{\ams}

\vspace{20pt}

\underline{ABSTRACT}
\end{center}
This is the first of a series of two papers where decoupling of unphysical states in the minimal pure spinor formalism is investigated. The multi-loop amplitude prescription for the minimal pure spinor superstring formulated in hep-th/0406055 involves the insertion of picture changing operators in the path integral. These operators are BRST closed in a distributional sense and depend on a number of constant tensors. One can trace the origin of these insertions to gauge fixing, so the amplitudes are formally independent of the constant tensors. We show however by explicit tree-level and one-loop computations that the picture changing operators are not BRST closed inside correlators and the amplitudes do depend on these constant tensors. This is due to the fact that the gauge fixing condition implicit in the existing minimal amplitude prescription is singular and this can lead to Lorentz violation and non-decoupling of BRST exact states. 
As discussed in hep-th/0406055, a manifestly Lorentz invariant prescription can be obtained by integrating over the constant tensors and in the sequel to this paper, it is shown that when one includes these integrations unphysical states do decouple to all orders despite the fact that the PCO's are not BRST closed inside correlators.

\pagebreak

\tableofcontents
\addtocontents{toc}{\protect\setcounter{tocdepth}{2}}

\section{Introduction}

A new superstring formalism, the pure spinor formalism,
has been developed over the past ten years
\cite{Berkovits:2000fe,Berkovits:2000ph,Berkovits:2004px,Berkovits:2005bt,Berkovits:2006vi,Berkovits:2007wz}, see
\cite{Berkovits:2002zk, Mafra:2009wq} for reviews.
In this new formalism, the theory
exhibits manifest super Poincar\'{e} invariance,
as in the Green-Schwarz (GS) formalism, but in contrast
with the GS string the worldsheet theory in flat target space is
free, as in the Ramond-Neveu-Schwarz (RNS) formalism,
so the theory can be quantized straightforwardly (modulo the
issues with the pure spinor constraint that we discuss below). This
has opened
a new avenue for better understanding string perturbation theory.
Indeed, the new formalism has already produced a number of
interesting results in this direction, such as new non-renormalization theorems
and progress towards
proving finiteness of perturbative string theory
\cite{Berkovits:2004px,Berkovits:2006vc}, and one may anticipate more new results to
appear as the formalism is developed further.
On a different front, gauge/gravity dualities
and flux compactifications render an urgent need for a formalism
that can handle curved backgrounds with Ramond-Ramond fluxes
and the pure spinor formalism is
currently the best such candidate.

The new amplitude prescription has marked advantages over both
the RNS and GS formalisms. Compared to the RNS formalism,
the formalism does not involve
worldsheet fermions, so there is no need to sum over spin structures
and deal with supermoduli. 
Moreover, computations involving
external fermions and RR fields are markedly simpler than the
corresponding RNS ones and the manifest target space supersymmetry
automatically leads to expressions that incorporate the
entire supermultiplet. The GS formalism is also target space
supersymmetry but one must use the lightcone gauge
and contact term interactions \cite{Green:1987qu, Greensite:1987hm}
lead to complications in multi-loop computations.
The pure spinor superstring is free of these problems
and has already been used successfully in explicit computations 
\cite{Berkovits:2000ph,Berkovits:2002zk,Berkovits:2004px,Berkovits:2005df, Berkovits:2005bt, Berkovits:2005ng, Mafra:2005jh,Berkovits:2006vc,Policastro:2006vt,Mafra:2008ar,Mafra:2009wi}.
However, this formalism has not been derived by
gauge fixing a worldsheet diffeomorphism theory and as a
result not all aspects of the formalism are fully understood.
From the practical point of view, one would like to develop further the
computational tools relevant for the pure spinor sector.
This paper grew out of our
efforts to further develop and streamline the pure spinor formalism.
In this process
we encountered issues with decoupling of BRST exact states
which is the subject of this and of the companion paper \cite{new1}.

The pure spinor superstring has two versions, the
minimal \cite{Berkovits:2004px} and the non-minimal
formalism \cite{Berkovits:2005bt}. The two formalisms are formally
equivalent \cite{Hoogeveen:2007tu} with the former being
technically more intricate than the latter.
The non-minimal formalism however is known to have
a difficulty from genus three and higher:~one of the zero mode integrals in the path integral is
divergent due to poles in the composite $b$ field \cite{Berkovits:2005bt}.
Although there has been a proposal for dealing with these
divergences \cite{Berkovits:2006vi}, no explicit computation
in $g > 2$ has been completed with it to date, see however
\cite{Aisaka:2009yp,Grassi:2009fe} for recent work in this direction.
The minimal formalism on the other hand does not
appear to have such a problem: the corresponding composite
$b$ field does not have the poles that its non-minimal counterpart has.
This was one of the reasons that led us to revisit the minimal formalism.

The minimal formalism contains constant spinors
($C_{\a}$) and constant tensors ($B_{mn}$) in its amplitude
prescription. These constant tensors enter the theory via certain operators,
the picture changing operators (PCO's), which are needed to set up
the amplitude prescription. It was argued in \cite{Berkovits:2004px}
that amplitudes are independent of $C$ and $B$, because the Lorentz
variation of PCO's is BRST exact. 
In this paper we show by explicit computations that the amplitudes
do depend on the choice of the constant tensors and BRST exact
states do not decouple. This happens already at tree level,
but in this case one can show that there is a unique Lorentz invariant operator that can replace the PCO's in the tree-level
amplitude prescription. With this replacement BRST exact terms
do decouple and one can further show that this prescription
is equivalent to the
tree-level prescription obtained by integrating over $C$
\cite{Berkovits:2004px}, which correctly reproduces
known tree-level amplitudes.

Next we examine amplitudes at one loop. These should be independent of the constant tensors $B$ and $C$ but we find problems with Lorentz invariance and decoupling of $Q$ exact states just like at tree level. These
problems are not present when we integrate over $B$ and $C$, as will be discussed in the companion paper \cite{new1}.
Furthermore we prove a no-go theorem about finding new Lorentz covariant PCO's that are BRST closed inside correlators
that could be used to replace the original PCO's. Using such PCO's however one finds that all one-loop amplitudes are equal to zero.

The technical origin of the problem is that the PCO's
are BRST closed only in a distributional sense
and it turns out that the amplitudes are singular enough
so that distributional identities do not hold.
One should contrast this with the non-minimal
formalism where the corresponding object, the so-called
regularization factor, is BRST closed without subtleties.
Indeed, we show that the problems we found at tree and one-loop
level in the minimal formalism are not present in the non-minimal case.


To understand why the amplitudes are singular,
let us recall that the
PCO's originate from gauge fixing zero mode
invariances \cite{Hoogeveen:2007tu}. The
PCO's contain eleven delta functions of the form $\d(C^I_\a \lambda^\a)$,
where $C^I_\a$ are the constant spinors mentioned above.
It turns out that for any choice of $C^I$ that give an
irreducible set of eleven constraints, the solution
of  $C^I_\a \lambda^\a=0$ is given by $\lambda^{\a}=0$, which is the tip
of the cone that represents pure spinor space. As discussed in
\cite{Nekrasov:2005wg}, the  $\lambda^{\a}=0$ locus
should be removed from the pure spinor
space. Thus this prescription corresponds to a singular
gauge fixing condition and the problems we find reflect that
fact.

This paper is organized as follows. In the next section we review the
minimal pure spinor formalism with emphasis on the tree-level and the
one-loop amplitude prescription.  Then in section \ref{sec:c} we
demonstrate the dependence on the constant spinors, $C_{\a}$, by
performing a tree-levelcomputation with two different choices for
$C_{\a}$. In section \ref{sec:ratl} it is shown that after integrating
over $C$ BRST exact states do decouple and we show how to formulate the
prescription such that it does not contain constant spinors
anymore. Section \re{sec:oneloop} examines one-loop amplitudes with
unphysical states. We analyze these amplitudes both with and without
integrating over $B$. In the final part of this section the
computations are compared to their non-minimal
counterparts. 
Section \ref{sec:nogo} contains the no-go theorem, which states that a
Lorentz invariant picture changing operator leads to vanishing of all
one-loop amplitudes. In section \ref{sec:orig} we discuss the origin
of the problem as a singular fixing condition and we comment on
possible modifications such that the prescription would correspond to
a non-singular gauge. We conclude in section \ref{conclu}.  The paper
contains two appendices. In the first appendix we provide a
comprehensive and (in some cases) pedagogical review of many technical
aspects relevant for the pure spinor formalism and in appendix B we
compute several integrals needed for the one-loop discussion.

\section{Review of minimal pure spinor formalism}\label{sec:rev}
The worldsheet action in the minimal pure spinor formalism for the left movers in conformal gauge and flat target space is given by
\be \label{fl_action}
S= \int d^2z \left(\frac{1}{2}\del x^m \bar{\del}x_m +p_{\a} \bar{\del}\theta^{\a} - w_{\a}\bar{\del}\lambda^{\a}\right),
\ee
with $m=0,\ldots, 9$ and $\a=1,\ldots, 16$. The fields $p_{\a}$ and $w_{\a}$ have conformal weight one and are Weyl spinors, $\q^{\a}$ and $\lambda^{\a}$ have conformal weight zero and are Weyl spinor of opposite chirality. In addition $\lambda^{\a}$ is a pure spinor, i.e.~it satisfies
\be \label{eq:pscond}
\lambda^{\a}\g^m_{\a\b}\lambda^{\b}=0,
\ee
where $\g^m_{\a\b}$ are the ten dimensional Pauli matrices, which are
defined in appendix \ref{sec;clif}. The decomposition of a Weyl spinor
under the $SU(5)$ subgroup, ${\bf{16}} \rightarrow {\bf{1}} \oplus
{\bf{\bar{10}}} \oplus {\bf{5}}$, which is used extensively throughout
this work, is also discussed there. Since the worldsheet action
consists of two $\b\g$ systems quantization seems straightforward, but
$\lambda^{\a}$ is a pure spinor and therefore the $\lambda w$ part is actually a
curved $\b\g$ system \cite{Nekrasov:2005wg}. To deal with this, we work
on a patch in pure spinor space that is defined by $\lambda^+\neq 0$. On
this patch the pure spinor condition expresses $\lambda^a$ in terms of
$\lambda_{ab}$ and $\lambda^+$, with $a,b=1,\ldots,5$. The solution is (in
$SU(5)$ covariant components)
\be \label{solv_pur}
\lambda^a= \fr{1}{8}\fr{1}{\lambda^+}\e^{abcde}\lambda_{bc}\lambda_{de}.
\ee
A constraint on fields in the action induces a gauge invariance on the conjugate fields. In this case the gauge transformations are given by
\be \label{eq:delw}
\d w_{\a}= \Lambda_m \g^m_{\a\b}\lambda^{\b}.
\ee
In \cite{Berkovits:2004px} this gauge invariance is dealt with by
using gauge invariant quantities only. This means $w_{\a}$ can only appear
in the Lorentz current $N^{mn}$, the ghost number current $J$ and the
stress energy tensor $T_{(\lambda w)}$:
\be
N^{mn}=\hp w_{\a}(\g^{mn})^{\a}_{\ \b}\lambda^{\b},\quad J=w_{\a}\lambda^{\a},\quad T_{(\lambda w)}=w_{\a}\del \lambda^{\a}.
\ee
Since the $\lambda w$ part of the action is not free due to the pure spinor constraint it is not obvious what the OPE between $w$ and $\lambda$ will be. One way to proceed is by properly fixing the gauge invariance of \eqref{eq:delw}. In \cite{Hoogeveen:2007tu}, following \cite{Berkovits:2000fe},
 it was shown, by making the gauge choice $w_a=0$ and employing BRST methods, one can replace $\int d^2z w_{\a} \bar{\del} \lambda^{\a}$ by the free action,
\be \label{eq:gfaction}
\int d^2z (\w_+ \bar{\del} \lambda^+ +\hp \w^{ab}\bar{\del} \lambda_{ab}).
\ee
One might have expected BRST ghosts associated to the gauge fixing of $w_{\a}$. It turns out these can be integrated out. As a check of the validity of this procedure the OPE of the Lorentz currents ($N^{mn}|_{w_a=0}$) should give rise to the Lorentz algebra. Using \eqref{eq:gfaction} one finds
\be \label{eq:opewl}
N^{mn}(z)\lambda^{\a}(w) \sim \fr{1}{z-w}\hp (\g^{mn}\lambda)^{\a},\quad J(z)\lambda^{\a}(w)\sim \fr{1}{z-w}\lambda^{\a},
\ee
\[
N^{mn}(z)N^{pq}(w) \sim \fr{-3}{(z-w)^2}(\h^{n[p}\h^{q]m})+\fr{1}{z-w}(\h^{n[p}N^{q]m}-\h^{m[p}N^{q]n}),
\]
\[
J(z)J(w) \sim \fr{-4}{(z-w)^2},\quad J(z)N^{mn}(w) \sim {\rm regular},
\]
\[
N^{mn}(z) T(w) \sim \fr{1}{(z-w)^2}N^{mn}(w),\quad J(z)T(w) \sim \fr{-8}{(z-w)^3}+\fr{1}{(z-w)^2}J(w).
\]
The explicit computations can be found in appendix \ref{sec:pslg} and
it should be noted that there are subtleties regarding the double poles in
the OPE. Hence even though the gauge fixing condition is not Lorentz
covariant the OPE's of the gauge fixed currents are. The factor of
$-8$ of the triple pole in the $JT$ OPE implies at tree level only
correlators with total $J$ charge -8 will be non-zero
\cite{Friedan:1985ge}. The OPE's for the matter variables can be
straightforwardly derived from \eqref{fl_action}:
\be \label{eq:opemat}
x^m(z)x^n(w) \sim -\h^{mn}{\rm log}|z-w|^2,\quad p_{\a}(z)\q^{\b}(w) \sim \d^{\a}_{\b}\fr{1}{z-w}.
\ee

The action \eqref{fl_action} is invariant under a nilpotent fermionic symmetry generated by
\be
Q=\oint dz \lambda^{\a}d_{\a},
\ee
where
\be
d_\a = p_\a - \hp \g_{\a\b}^m  \q^\b \pa x_m - \frac{1}{8}\g_{\a \b}^m \g_{m \ \g \d} \q^\b \q^\g \pa \q^\d.
\ee
The transformations it generates are given by
\be \label{S-tr}
\d x^m= \lambda \g^m \q, \quad \d \q^{\a}= \lambda^{\a}, \quad \d \lambda^\a =0, \quad\d d_{\a}=-\Pi^m(\g_m \lambda)_{\a}, \quad \d w_{\a}= d_{\a},
\ee
where $\Pi^m = \pa x^m + \hp \q \g^m \pa \q$ is the supersymmetric momentum and again we restrict to the left movers (so in particular, the full transformation for $x^m$ contains a similar additive term with right moving fields). The cohomology of this operator (at ghost number one) indeed correctly reproduces the superstring spectrum \cite{Berkovits:2000nn}.

The gauge fixed action \eqref{eq:gfaction} is no longer invariant under $Q=\oint dz \lambda^{\a}d_{\a}$, but it is invariant under $\hat{Q}$ defined by
\be \label{eq:qhat}
\hat{Q} w_{\a}=d_{\a}-\fr{d_a}{\lambda^+}(\g^a \lambda)_{\a}.
\ee
On all other fields $\hat{Q}$ acts the same as $Q$. Note the second term in
\eqref{eq:qhat} is a gauge transformation with $\Lambda_a=\fr{d_a}{\lambda^+}, \Lambda^a=0$. This implies that when acting on gauge invariant quantities $Q=\hat{Q}$. Moreover $\hat{Q} w_a=0$. So that for instance
\be
\hat{Q} N^{mn}|_{w_a=0}=QN^{mn}=\hp \lambda\g^{mn}d.
\ee
$\hat{Q}$ also satisfies
\be
\hat{Q}^2=0,
\ee
on all fields including $w$, unlike $Q$.

It seems very natural to consider $Q$ as a BRST operator that appeared
after gauge fixing a local worldsheet symmetry that includes diffeomorphism
invariance. Despite considerable work, finding such a formulation
remains an open issue, see \cite{Matone:2002ft} for work
in this direction. There has also
been work in relating the pure spinor formalism to GS and RNS formalisms,
see \cite{Berkovits:2007wz} and references therein.

In \cite {Hoogeveen:2007tu} we presented a different perspective.
We considered the pure spinor action (\ref{fl_action})
as a $\sigma$-model action with a fermionic symmetry $Q$
and we coupled it to topological gravity in a way that preserves $Q$.
Gauge fixing worldsheet diffeomorphisms leads in a standard way to
a second nilpotent operator, the standard BRST operator. Then one
can proceed to derive the scattering amplitude prescription following
usual BRST methods.
From this perspective
the reason we start from
an action with $Q$ invariance is that the cohomology of $Q$ yields
the correct superstring spectrum.

\subsection{Tree-level prescription}

In this subsection we review the tree-level amplitude prescription of
\cite{Berkovits:2004px}.
The $N$ point open string tree-levelamplitude is given by
\[
\cA=\< V_1(z_1)V_2(z_2)V_3(z_3)\int dz_4 U_4(z_4) \cdots \int dz_N U_N(z_N)Y_{C_1}(y_1)\cdots Y_{C_{11}}(y_{11})\>=
\]
\[
\int [\cD^{10}x][\cD^{16}d][\cD^{16}\q][\cD^{11}\lambda][\cD^{11}w]V_1(z_1)V_2(z_2)V_3(z_3)\int dz_4 U_4(z_4) \cdots \int dz_N U(z_N)
\]
\be \label{eq:amplt}
Y_{C_1}(y_1)\cdots Y_{C_{11}}(y_{11})e^{-S},
\ee
where $[\mathcal{D}\f]$ denotes functional integration over the field $\f$. The functional integration over $x^m$ has been studied in detail and the same correlation functions appear in the RNS formalism. We will not include this factor in the computations in this paper because they are not relevant for us. $V$ and $U$ are the integrated and unintegrated vertex operators, i.e. they satisfy
\be
Q V(z)=0,\quad V(z) \sim V(z)+Q\Omega(z),
\ee
\be
Q \int dz U(z)=0, \quad \int dz U(z) \sim \int dz U(z) + Q \int dz \Omega'(z).
\ee
After using the gauge invariance to set a number of components to zero the solution to these equations is given by \cite{Berkovits:2000nn}
\bea
V&=&\lambda^{\a}A_{\a}(x,\q),
\\
U&=&\del \q^{\a}A_{\a}(x,\q)+\Pi^mA_m(x,\q)+d_{\a}W^{\a}(x,\q)+\hp N^{mn}\mathcal{F}_{mn}(x,\q),
\eea
with
\begin{eqnarray}
A_{\a}(x,\q)&=&e^{ik\cdot x}(\hp a_m(\g^m\q)_{\a}-\fr{1}{3}(\xi\g_m\q)(\g^m \q)_{\a}+\cdots), \\
A_m&=&\fr{1}{8}D_{\a}\g^{\a\b}_mA_{\b},\\
W^{\b}&=&\fr{1}{10}\g^{\a\b}_m(D_{\a}A^m-\del^m A_{\a}),\\
\mathcal{F}_{mn}&=&\fr{1}{8}D_{\a}(\g_{mn})^{\a}_{\ \b}W^{\b},
\end{eqnarray}
where $D_{\a}=\fr{\del}{\del \q^{\a}}+\hp \q^{\b}\g^m_{\a\b}\del_m$, $a_m$ and $\xi^{\a}$ are the polarizations and $k^m$ is the momentum. They satisfy $k^2=k^m a_m=k^m(\g_m\xi)_{\a}=0$, there is a residual gauge invariance $a_m \rightarrow a_m + k_m \omega$ and ... contains products of $k^m$ with $a_m$ or $\xi^{\a}$.

$Y_C$ are the picture changing operators (PCO):
\be
Y_C(y)=C_{\a}\q^{\a}(y)\d(C_{\b}\lambda^{\b}(y)),
\ee
where $C_{\a}$ is a constant spinor. We want to be absolutely explicit
about what we mean by a delta function, since we will see the problems with
decoupling of $Q$ exact states are intimately connected with these
delta functions. The definition we use in section \ref{sec:c} to
section \ref{sec:oneloop} is the usual one:
\be \label{eq:delfx}
\int dx \d(x) f(x)=f(0),\quad x\d'(x)=-\d(x).
\ee
The presence of the PCO's in the amplitude prescription is explained
from first principles in \cite{Hoogeveen:2007tu} and is
reviewed in section \ref{sec:orig}. In short, they come
from fixing a gauge invariance due to the zero modes of the weight
zero fields, $\lambda^{\a},\q^{\a}$. Note the weight one fields do not have
zero modes at tree level. At higher loops there will also be PCO's for
these fields. Since the PCO's are introduced as a gauge fixing term,
amplitudes should be independent of the constant tensors
$C_{\a}$. Moreover in all computations we will choose $y_i=\infty$ so
that the PCO's have no non-zero OPE with any other field.

The functional integral \eqref{eq:amplt} is evaluated by first using the OPE's of \eqref{eq:opewl} and \eqref{eq:opemat}. Note that this operation reduces the total conformal dimension of the worldsheet fields involved in the OPE. For example in the $p,\q$ OPE, the conformal weight of $p_{\b}(z)\q^{\a}(w)$ is one and the conformal weight of $\d^{\a}_{\b}$ is zero. Thus in the end the correlator only contains worldsheet fields of weight zero. This can be evaluated by replacing the fields by their zero modes and performing the zero mode integrations.

After integrating out the non-zero modes the amplitude reduces to
\be \label{eq:tlp}
\mathcal{A}=\int [d\lambda] d^{16}\q \lambda^{\a}\lambda^{\b}\lambda^{\g}f_{\a\b\g}(\q)(C^1\q)\d(C^1\lambda)\cdots (C^{11}\q)\d(C^{11}\lambda),
\ee
where $f_{\a\b\g}$ depends on all the polarizations and momenta. Note the functional integration of $x^m$ is omitted here as will be done in all computations in this paper. A priori $f_{\a\b\g}$ also depends on $z_1,z_2,z_3$. Of course we expect the final result to be independent of these coordinates. Also note all the fields are zero modes including those in the measure. $[d\lambda]$ is the unique Lorentz invariant measure of +8 ghost number on the space of pure spinors (cf. appendix \ref{sec:lim}). It is given by \cite{Berkovits:2004px}
\be \label{eq:measlam}
[d\lambda]\lambda^{\a}\lambda^{\b}\lambda^{\g}=d\lambda^{\a_1}\wedge \cdots \wedge d\lambda^{\a_{11}} ( \e T)^{\a\b\g}_{\a_1\cdots \a_{11}},
\ee
where
\be \label{eq:defet}
(\e T)^{\a\b\g}_{\a_{1}\cdots \a_{16}}= \e_{\a_1\cdots \a_{16}}\g_m^{\a\a_{12}}\g_n^{\b\a_{13}}\g_p^{\g\a_{14}}(\g^{mnp})^{\a_{15}\a_{16}}.
\ee
Note no gamma trace is subtracted. This tensor is already gamma matrix traceless as explained in appendix \ref{sec:lim}.

\subsection{One-loop prescription}
Compared to a tree-level amplitude, a one-loop one exhibits three new features, (1) PCO's for the weight one worldsheet fields $p,w$, (2) zero mode integrals over $p,w$ and (3) a composite $b$ ghost constructed out of the worldsheet fields from \eqref{fl_action}. The first two points are direct consequences of the presence of a zero mode of weight one fields on the torus. The new PCO's are given in terms of the gauge invariant quantities $N^{mn}$ and $J$:
\be
Z_B(z)=\hp B_{mn}\lambda(z)\g^{mn}d(z)\d(B_{mn}N^{mn}(z)),\quad Z_J(z)=\lambda^{\a}(z)d_{\a}(z) \d(J(z)).
\ee
All string theory amplitude prescriptions at one loop contain a $b$ ghost which satisfies
\be \label{eq:qbt}
\{Q,b(z)\}=T(z).
\ee
In the RNS formalism this field appears as reparametrization
antighost. In the pure spinor formalism the $b$ ghost is
composite \cite{Berkovits:2004px},
constructed out of the worldsheet fields from \eqref{fl_action},
as explained from first principles in \cite{Hoogeveen:2007tu}.
However, it is not
possible to solve equation \eqref{eq:qbt} in the minimal pure spinor
formalism \cite{Berkovits:2004px},
 because of ghost number ($J$ charge) conservation combined
with gauge invariance of objects containing $w_{\a}$. The former
implies $b$ must have ghost number minus one and since there are no
gauge invariant quantities with negative ghost number the latter rules
out any solution. A resolution to this problem is combining the
(composite) $b$ field with a PCO, $Z_B$, such that
\be \label{eq:qbbuz}
\{Q,\tilde{b}_B(u,z)\}=T(u)Z_B(z).
\ee
This equation ensures the $Q$ variation of the $b$ ghost vanishes after integrating over moduli space. The solution is given by  \cite{Berkovits:2004px}
\be
\tilde{b}_{B}(u,z)=b_{B}(u)+T(u)\int_u^z dv B_{pq}\del N^{pq}(v)\d(BN(v)).
\ee
The local $b$ ghost, $b_B(u)$, is a composite operator, constructed out of the worldsheet fields:
\be \label{eq:bghost}
b_B(z)={b_B}_0(z)\d(BN(z))+{b_B}_1(z)\d'(BN(z))+{b_B}_2(z)\d''(BN(z))+{b_B}_3(z)\d'''(BN(z)),
\ee
where the primes denote derivatives, $BN\equiv B_{mn}N^{mn}$ and
\bea \label{eq:bB0}
{b_B}_0&=&\hp G\g^{mn}dB_{mn}-\hp H^{\a\b}(\g^p\g^{mn})_{\a\b}\Pi_pB_{mn}+\\ \nn
&&\hp K^{\a\b\g}(\g^p\g^{mn})_{\b\g}(\g_p\del \q)_{\a}B_{mn}+\hp S^{\a\b\g}(\g^p\g^{mn})_{\b\g}(\g^p\del \lambda)_{\a}B_{mn},\\
{b_B}_1&=&\qu H^{\a\b} (Bd)_{\a}(Bd)_{\b}+\\ \nn
&&\qu K^{\a\b\g}(\g^p\g^{mn})_{\b\g}(Bd)_{\a}\Pi_pB_{mn}+\qu K^{\a\b\g}(\g^p\g^{mn})_{\a[\b}(Bd)_{\g]}\Pi_pB_{mn}+\\ \nn
&&\qu L^{\a\b\g\d}[((\g^p\g^{mn})_{\g\d}(Bd)_{[\a}(\g_p\del \q)_{\b]}-(\g^p\g^{mn})_{\b[\g}(Bd)_{\d]}(\g_p\del \q)_{\a})B_{mn}-\\ \nn
&&((\g^s\g^{rq})_{\a[\b}(\g^p\g^{mn})_{\g]\d}+(\g^s\g^{rq})_{\a\d}(\g^p\g^{mn})_{\b\g})\Pi_pB_{mn}\Pi_sB_{qr}],\\
{b_B}_2&=&-\fr{1}{8}K^{\a\b\g}(Bd)_{\a}(Bd)_{\b}(Bd)_{\g}-\fr{1}{8}L^{\a\b\g\d}((\g^p\g^{mn})_{\g\d}(Bd)_{\b}(Bd)_{\a}+\\ \nn
&&(\g^p\g^{mn})_{\b[\g}(Bd)_{\d]}(Bd)_{\a}+\hp (\g^p\g^{mn})_{\a[\d}(Bd)_{\g}(Bd)_{\b]})\Pi_pB_{mn},\\
{b_B}_3&=&-\fr{1}{16}L^{\a\b\g\d}(Bd)_{\a}(Bd)_{\b}(Bd)_{\g}(Bd)_{\d}, \label{eq:bB3}
\eea
where $(Bd)_{\a}\equiv B_{mn}(\g^{mn}d)_{\a}$ and $G,H,K,L$ are given in appendix \ref{sec:appco}.

The one-loop amplitude prescription in the minimal pure spinor formalism is given by
\be \label{eq:1looppre}
\cA^{(N)}=\int d^2 \t \< | \int d^2 u \m(u) \tilde{b}_{B^1}(u,z_1) \prod_{P=2}^{10}Z_{B^P}(z_P)Z_J(z_{11})\prod_{I=1}^{11}Y_{C_I}(y)|^2
\ee
\[
V_1(t_1)\prod_{T=2}^{N}\int d^2t_TU_T(t_T) \>.
\]
The Beltrami differential $\m(u)$ does not depend on the worldsheet coordinates on the torus. This implies the composite $b$ ghost only contributes through its zero mode:
\be
\int d^2u \m_{\t}(u) \tilde{b}(u,z)=\m_{\t}\int d^2u \tilde{b}(u,z).
\ee

A typical zero mode integral one encounters is given by \cite{Berkovits:2004px}:
\be \label{eq:cA}
\cA=\int [d\lambda][dB][dC]\prod_{R=1}^g[dN_R] f_B(\lambda,N_R,J_R,C,B)
\ee
where the zero mode measure for $[dN]$ is given by
\be
[dN]\lambda^{\a_1}\cdots \lambda^{\a_8}=dN^{m_1n_1}\wedge \cdots \wedge N^{m_{10}n_{10}}\wedge dJ R^{\a_1\cdots \a_{8}}_{m_1n_1\cdots m_{10}n_{10}},
\ee
with
\be \label{eq:defR}
R^{\a_1\cdots \a_8}_{m_1n_1\cdots m_{10}n_{10}}\equiv \g^{((\a_1\a_2}_{m_1n_1m_2m_3m_4} \g^{\a_3\a_4}_{m_5n_5n_2m_6m_7} \g^{\a_5\a_6}_{m_8n_8n_3n_6m_9} \g^{\a_7\a_8))}_{m_{10}n_{10}n_4n_7n_9}+ {\rm permutations}.
\ee
The permutations make $R$ antisymmetric under exchange in both $m_i \lera n_i$ and $m_in_i \lera m_jn_j$ and the double brackets denote subtraction of the gamma trace. The zero mode integral \eqref{eq:cA} is only non-zero if the function $f_B$ (called $f$ in \cite{Berkovits:2004px}) depends on $(\lambda,N,J,C,B)$ as
\be \label{eq:fb}
f_B(\lambda,N,J,C,B)=
\ee
\[
h_B(\lambda,N,J,C,B)\del^{M} \d(J)\prod_{P=1}^{10} \del^{L_{P}}\d(B^P N) \prod_{I=1}^{11} \del^{K_I}\d(C^I \lambda),
\]
where the polynomial $h_B$ assumes the form
\be \label{eq:hb}
(\lambda)^{\sum_{I=1}^{11}(K_I+1)}(J)^{M} (N)^{\sum_{P=1}^{10} L_{P}} \prod_{P=1}^{10} (B^P)^{L_{P}+1}\prod_{I=1}^{11} (C^I)^{K_I+1}.
\ee
The integration over the zero modes of the pure spinor variables and the constant tensors is defined in \cite{Berkovits:2004px} as
\be\label{eq:bint}
\cA=c \fr{\del}{\del \lambda^{\a_1}} \cdots \fr{\del}{\del \lambda^{\a_{3}}} (\e T)^{\a_1\cdots \a_{3}}_{\b_1\cdots \b_{11}} R^{\a_4\cdots \a_{11}}_{m_1n_1\cdots m_{10}n_{10}} \fr{\del}{\del \lambda^{\a_4}} \cdots \fr{\del}{\del \lambda^{\a_{11}}}  \fr{\del}{\del B^1_{m_1n_1}}\cdots \fr{\del}{\del B^{10}_{m_{10}n_{10}}}
\ee
\[
\fr{\del}{\del C^1_{\b_1}} \cdots \fr{\del}{\del C^{11}_{\b_{11}}}\prod_{I=1}^{11} (\fr{\del}{\del \lambda^\d}\fr{\del}{\del C^I_{\d}})^{K_I}\prod_{P=1}^{10} (\fr{\del}{\del B^P_{pq}} \fr{\del}{\del N^{pq}})^{L_{P}} (\fr{\del}{\del J})^{M} h_B(\lambda,N,J,C,B),
\]
for some proportionality constant $c$.

\section{Tree-level amplitudes}\label{sec:c}

In this section we will describe three problems with
\eqref{eq:amplt},  evaluated using the definitions
\eqref{eq:measlam} and \eqref{eq:delfx}. (1a) $\mathcal{A}$ is not
Lorentz invariant or equivalently (1b) $\mathcal{A}$ depends on the
choice of $C$'s and (2) $Q$ exact states do not decouple. The third
problem involves the position of the PCO's on the worldsheet.

\subsection{Lorentz invariance}

The prescription of \eqref{eq:tlp} appears to be Lorentz invariant and therefore independent of $C^I_{\a}$ because the Lorentz variation of the PCO's is BRST exact:
\be
M^{mn}Y_C=\hp(C\g^{mn}\q)\d(C\lambda)+\hp(C\q)(C\g^{mn}\lambda) \d'(C\lambda)=Q[\hp(C\g^{mn}\q)(C\q)\del \d(C\lambda)].
\ee
This argument requires vanishing of $\<QX\>$ for all $X$ and closedness of the PCO's. The first condition is satisfied because after integrating out the non-zero modes $\<QX\>$ reduces to
\be
\int [d\lambda] d^{16} \q  \lambda^{\a}\lambda^{\b}\lambda^{\g}D_{\a} f_{\b\g}(\q)C^1\q\d(C^1\lambda)\cdots C^{11}\q\d(C^{11}\lambda)=0,
\ee
because $\int d^{16}\q D_{\a} g(\q)=0$ for any function $g$. In order to see whether the PCO's are closed consider
\be \label{qvarcqcl}
QY_C=C_{\a}\lambda^{\a}\d(C_{\b}\lambda^{\b}).
\ee
This seems to be zero, but if we choose $C_{\a}=\d^+_{\a}$, we find $QY_C=\lambda^{+}\d(\lambda^{+})$. This is not zero because the measure contains $\fr{1}{(\lambda^+)^3}$. All we can use is ${\lambda^+}^4\d(\lambda^+)=0$. This problem is made even more explicit in the computation below. It will be shown that choosing particular $C$'s does not result into a Lorentz invariant answer.

Let us choose
\be \label{eq:Cgf}
C^1_{\a}=\d^+_{\a},\ \ (C^2)^{a_1a_2}=\d^{[a_1}_1\d^{a_2]}_2,\ldots ,(C^{11})^{a_1a_2}=\d^{[a_1}_4\d^{a_2]}_5,\quad {\rm all\ other}\ C^I_{\a}=0.
\ee
Note $C^I_{\a}$ has rank eleven for this choice, as it should.
As is discussed in section \ref{sec:orig}, within the present
formalism, the results below would be valid for any other choice, see
footnote \ref{singular}.
The three-point tree-level function is given by
\be
\cA=\< \lambda^{\a}{A_1}_{\a}(z_1)\lambda^{\b}{A_2}_{\b}(z_2)\lambda^{\g}{A_3}_{\g}(z_3) Y_{C_1}(\infty)\cdots Y_{C_{11}}(\infty) \>
\ee
\[
=\int[d\lambda]d^{16}\q \lambda^{\a}\lambda^{\b}\lambda^{\g}f_{\a\b\g}(\q)C^1_{\a_1}\q^{\a_1}\cdots C^{11}_{\a_{11}}\q^{\a_{11}}\d(C^1_{\a_1}\lambda^{\a_1})\cdots \d(C^{11}_{\a_{11}}\lambda^{\a_{11}})
\]
\[
=\int[d\lambda]d^{16}\q \lambda^{\a}\lambda^{\b}\lambda^{\g}f_{\a\b\g}(\q)\q^+\q_{12}\cdots\q_{45}\d(\lambda^{+})\d(\lambda_{12})\cdots \d(\lambda_{45})
\]
\[
=\int \fr{d\lambda^+\wedge d\lambda_{12}\wedge \cdots \wedge d\lambda_{45}}{{\lambda^+}^3}d^{16}\q \lambda^{\a}\lambda^{\b}\lambda^{\g}f_{\a\b\g}(\q)\q^+\q_{12}\cdots\q_{45}\d(\lambda^{+})\d(\lambda_{12})\cdots \d(\lambda_{45}).
\]
The only term that contributes is the one with $\a\b\g=+++$, in all other cases there is an integral of the form $\int d\lambda_{ab}\lambda_{ab}\d(\lambda_{ab})$ (no sum). There is a subtlety with these integrals, for instance
\[
\int [d\lambda](\lambda^{+})^2\lambda_{cd} \d(\lambda^+)\d(\lambda_{12})\cdots \d(\lambda_{45})=\int d\lambda^+d^{10}\lambda_{ab} \fr{\lambda_{cd}}{\lambda^+} \d(\lambda^+)\d(\lambda_{12})\cdots \d(\lambda_{45})=
\]
\be \label{eq:probl}
\int d\lambda^+ \fr{1}{\lambda^+}\d(\lambda^+)\int d\lambda_{cd} \lambda_{cd} \d(\lambda_{cd})=\infty 0.
\ee
Note however that \eqref{eq:probl} has $N$ charge one (cf. \eqref{eq:dec16}). Since the outcome of the integral (maybe after some regularization) must be a number, which does not transform under $N$, the integral has to vanish. In other words only integrals with zero $N$ charge, like $\int [d\lambda] (\lambda^+)^3\d(\lambda^+)\d(\lambda_{12})\cdots \d(\lambda_{45})$ can be non vanishing. After the integration over the $\lambda$ zero modes we are left with
\be
\cA=\int d^{16}\q f_{+++} \q^+\q_{12}\cdots \q_{45},
\ee
where $f_{+++}=A^1_+A^2_+A^3_+$ and this can be evaluated with the help of the explicit expressions for the gamma matrices from appendix \ref{sec:gama}. We choose as external states two gauginos and one gauge boson:
\be
\cA=\int d^{16}\q (\x_1^a \q_{ka}\q^k+\xi^1_{ka}\q^a\q^k)(\x_2^b\q_{lb}\q^l+\x^2_{lb}\q^b\q^l)\q^ca^3_c \q^+\q_{12}\cdots \q_{45}=\e^{abcde}\xi^1_{ab}\x^2_{cd}a^3_e.
\ee
This answer is not Lorentz invariant and different from the expected answer,
\be
\x^1\g^m\x^2a^3_m=2(\xi_1^+\xi_2^aa^3_a+\xi_1^a\xi_2^+a^3_a-\qu \e^{abcde}\xi^1_{ab}\xi^2_{cd}a^3_e+\x^1_{ab}\xi^a_2a_3^b+\xi_1^a\xi^2_{ab}a_3^b),
\ee
where $m$ is an $SO(10)$ index and all Latin letters that come before $m$ in the alphabet are $SU(5)$ indices. In conclusion this shows that using \eqref{eq:measlam} and \eqref{eq:delfx} does not lead to Lorentz invariant answers.

\subsection{Dependence on $C^I$}
We will now show that amplitudes are not invariant under $C^I_{\a} \rightarrow C^I_{\a}+\d C^I_{\a}$. In this computation it also becomes clear that not all BRST exact states decouple. Consider the same $C$'s as in \eqref{eq:Cgf} and $\d C^{11}_{\a}=\d_{\a}^1$, where the 1 is an $SU(5)$ index. The delta only has one non vanishing component. This changes $Y_{C_{11}}$ by
\be
\d Y_{C_{11}}= \d {C_{11}}_{\a}\q^{\a}\d(C_{11}\lambda)+{C_{11}}_{\a}\q^{\a}\d {C_{11}}_{\b}\lambda^{\b}\d'(C_{11}\lambda)
\ee
\[
=Q(\d {C_{11}}_{\a}\q^{\a}{C_{11}}_{\b}\q^{\b}\d'({C_{11}}_{\b}\lambda^{\b}))=Q(\q^1\q_{45}\d'(\lambda_{45})).
\]
Under this change in $C^I_{\a}$ the tree-level three-point function changes by
\be
\d\cA=\< V_1(z_1) V_2(z_2) V_3(z_3) Y_{C_1}(\infty)\cdots Y_{C_{10}}(\infty)\d Y_{C_{11}}(\infty) \>=
\ee
\[
\< V_1(z_1) V_2(z_2) V_3(z_3) Q(Y_{C_1}(\infty)\cdots Y_{C_{10}}(\infty))\q^1(\infty)\q_{45}(\infty)\d'(\lambda_{45}(\infty))\>
\]
\[
=\int d^{16}\q \fr{d^{11} \lambda}{(\lambda^+)^3} \lambda^{\a}\lambda^{\b}\lambda^{\g}A^1_{\a}A^2_{\b}A^3_{\g}Q(Y_{C_1}\cdots Y_{C_{10}})\q^1\q_{45}\d'(\lambda_{45}).
\]
There is a total of four $\lambda^{\a}$'s in the numerator (one hidden in $Q$) one of them has to be $\lambda_{45}$ and the other three have to be $\lambda^+$ to give a non vanishing answer. The term that contributes comes from $Q$ hitting $\q^+\d(\lambda^+)$, this $\lambda^+$ then cancels against a $\lambda^+$ in the denominator and the variation becomes
\be
\d\cA=\int d^{16}\q d^{11} \lambda A^{(1}_{+}A^2_+(A^{3)})^{45}\q^1\d(\lambda^+)\q_{12}\d(\lambda_{12})\cdots \q_{45}\d(\lambda_{45})=
\ee
\[
\int d^{16}\q A^{(1}_{+}A^2_+(A^{3)})^{45}\q^1\q_{12}\cdots \q_{45}.
\]
By choosing suitable polarizations it is not difficult to see this does not always vanish.

\subsection{Position of PCO's on the worldsheet} \label{sec:pospco}

In the prescription of \cite{Berkovits:2004px},
PCO's are inserted at arbitrary points on the worldsheet.
The derivative of the PCO's however is $Q$ exact:
\be \label{eq:dely}
\del Y_C(y)=Q[(C\del \q(y))(C\q(y))\d'(C\lambda(y))],
\ee
\be
\del  Z_B(z)=Q[-B_{pq}\del N^{pq}(z)\d(BN(z))],\quad
\del Z_J(z)=Q[-\del J(z)\d(J(z))].
\ee
and this suggests that the amplitudes do not depend on the
insertion points. As we have seen, however, BRST exact terms
do not decouple, so the amplitudes may depend on the insertion points.
In our computations of tree-levelamplitudes we will follow
\cite{Berkovits:2004px} and insert the PCO's at $y=\infty$. This is
equivalent to replacing the fields in the PCO's by their zero modes.

\section{Resolution at tree level}\label{sec:ratl}

Obtaining amplitudes which are not Lorentz invariant is a serious problem
and one might
ask why the tree-levelamplitude computations \cite{Berkovits:2004px,
  Policastro:2006vt} in the minimal pure spinor formalism gave Lorentz
invariant answers and why $Q$ exact states decoupled. Both these points are explained in the first part of this
section. In the second part we reformulate the tree-levelamplitude
prescription in a way that does not contain any constant spinors.

\subsection{Resolution in the literature}

Lorentz invariance is restored by integrating over all possible
choices of $C^I_{\a}$, and this also results in decoupling of $Q$ exact
states as will become apparent
in this section. The manifestly Lorentz invariant tree-level amplitude
in the minimal formalism is given by
\be
\cA=\int [dC] \< V_1(z_1)V_2(z_2)V_3(z_3)\int dz_4 U_4(z_4) \cdots \int dz_N U_N(z_N)Y_{C_1}(\infty)\cdots Y_{C_{11}}(\infty)\>.
\ee
After performing the OPE's and replacing the fields by their zero modes this becomes
\be
\cA=\int[dC]\int [d\lambda] d^{16}\q \lambda^{\a}\lambda^{\b}\lambda^{\g}f_{\a\b\g}(\q)(C^1\q)\d(C^1\lambda)\cdots (C^{11}\q)\d(C^{11}\lambda).
\ee
Now one uses
\be \label{eq:intcet}
\int[dC][d\lambda]\lambda^{\a}\lambda^{\b}\lambda^{\g} C^1_{\b_1} \cdots C^{11}_{\b_{11}} \d(C^1\lambda)\cdots \d(C^{11}\lambda)=(\e T)^{\a\b\g}_{\b_1\cdots \b_{11}}.
\ee
This is justified by Lorentz invariance, because the LHS is Lorentz invariant and the only invariant tensor with the appropriate symmetries is\footnote{
Incidentally, the following related integral can also be
computed using Lorentz invariance:
\be \label{corr}
\int [dC]d\lambda^{\a_1}\wedge \cdots \wedge d\lambda^{\a_{11}}C^1_{\b_1}\cdots C^{11}_{\b_{11}}\d(C^1\lambda)\cdots \d(C^{11}\lambda)=
\ee
\[
c_1 \d^{[\a_1}_{\b_1}\cdots\d^{\a_{11}]}_{\b_{11}}+c_2 \g_{mnp}^{[\a_1\a_2}\g^{mnp}_{[\b_1\b_2}\d^{\a_3}_{\b_3}\cdots\d^{\a_{11}]}_{\b_{11}]
},
\]
where $c_1$ and $c_2$ are non-zero numerical constants.
This structure follows from the fact
 ${\rm Asym}^{11}{\bf 16} \otimes {\rm Asym}^{11} {\bf 16'}$ contains
two scalars (see appendix \ref{app:ccp} for explanation about the notation and
the argument). The constants can be computed using
judicious choices of the indices. For example, the integral vanishes
for the choice $\a_1=\b_1, \cdots, \a_{11}=\b_{11} = +,12, \ldots, 35, 5$,
implying that
one needs a non-zero constant $c_2$.
Equation (\ref{corr}) corrects formula (3.25) of \cite{Berkovits:2004px}.
}
$(\e T)$, as can be verified with \cite{vanLeeuwen}.
Thus
\be \label{eq:maetcd}
\cA=(\e T)_{\a_1\cdots \a_{11}}^{\a\b\g}\int d^{16}\q f_{\a\b\g}(\q)\q^{\a_1}\cdots \q^{\a_{11}}.
\ee
The amplitude $\cA$ is manifestly Lorentz invariant.

This prescription also ensures the decoupling of unphysical states.
We will use $\cB$ to denote amplitudes with unphysical states throughout
this paper, while $\cA$ is used for any amplitude, so at tree level with
$V_1 = Q \Omega$,
\be
\cB=\int [dC]\< Q \W(z_1) V_2(z_2) V_3(z_3) \prod_{i=4}^N \int dz_i U(z_i) C^1_{\a_1}\q^{\a_1}\cdots C^{11}_{\a_{11}}\q^{\a_{11}}\d(C^1\lambda)\cdots\d(C^{11}\lambda) \>.
\ee
This can be written in the following form:
$$
\cB=\int [dC]\< \lambda^{\a}(z_2)\lambda^{\b}(z_3)g_{\a\b}(d,\q,N)Q(C^1_{\a_1}\q^{\a_1}\cdots C^{11}_{\a_{11}}\q^{\a_{11}})\d(C^1\lambda)\cdots\d(C^{11}\lambda)\> \sim
$$
\be
\int [dC]\< \lambda^{\a}(z_2)\lambda^{\b}(z_3)g_{\a\b}(d,\q,N)C^1_{\a_1}\lambda^{\a_1}\cdots C^{11}_{\a_{11}}\q^{\a_{11}}\d(C^1\lambda)\cdots\d(C^{11}\lambda)\>.
\ee
where in going from the first to the second line we omitted an overall
numerical factor of eleven. Such overall inconsequential
factors will be neglected throughout this work.
After using the OPE's to integrate out the non-zero modes one gets:
\bea
\cB&=&\int [dC] d^{16}\q [d\lambda] \lambda^{\a}\lambda^{\b}f_{\a\b}(\q)C^1_{\a_1}\lambda^{\a_1}C^2_{\a_2}\q^{\a_2}\cdots C^{11}_{\a_{11}}\q^{\a_{11}}\d(C^1\lambda)\cdots\d(C^{11}\lambda)=
\nonumber \\
&&
\int d^{16}\q f_{\a\b}(\q)(\e T)^{\a\b\a_1}_{\a_1\cdots \a_{11}}\q^{\a_2}\cdots \q^{\a_{11}}=0,
\label{eq:tret01}
\eea
where $f_{\a \b}(\q)$ is some function of $\q$ zero modes and we used \eqref{eq:intcet}. The integral vanishes because\footnote{Note ${\bf{126}}$ denotes a gamma matrix traceless symmetric rank two tensor
(recall that $\lambda^{\a}\lambda^{\b} \sim \lambda\g^{mnpqr}\lambda\g_{mnpqr}^{\a\b}$).} ${\bf{126}}\otimes {\rm{Asym}}^{10}{\bf{16}}$ does not contain a scalar (see appendix \ref{app:ccp} for explanation about the notation and the argument), in other words
\be \label{eq:trl}
(\e T)^{\b_1\b\g}_{\b_1\cdots \b_{11}}=0.
\ee
In this case one can also write out $(\e T)$ explicitly and check
that its trace contains a contraction of an antisymmetric tensor
($\e$) and a symmetric one ($\g_m^{\a\b}$).

\subsection{Lorentz invariant tree-levelprescription without constant spinors}\label{sec:tree}

We now present a new prescription for a tree-level amplitude, which does
not contain any constant spinors and is manifestly Lorentz
invariant.  This new prescription is equivalent to the one given in
\cite{Berkovits:2004px}, when the integral over $C$ in included.
The prescription is given by
\be
\cA=\< V_1(z_1)V_2(z_2)V_3(z_3)\int dz_4 U_4(z_4) \cdots
\int dz_N U_N(z_N)\Lambda_{\a\b\g}(\infty)
\ee
\[
(\e T)^{\a\b\g}_{\b_1\cdots \b_{11}}\q^{\b_1}(\infty)\cdots \q^{\b_{11}}(\infty)\>.
\]
In other words, we have replaced the eleven PCO's $Y_C$ by
$\Lambda_{\a\b\g}(\infty)$.
After integrating out the non-zero modes and replacing the fields by their zero modes $\cA$ reduces to
\be
\cA=\int d^{16}\q[d\lambda]\lambda^{\a}\lambda^{\b}\lambda^{\g}f_{\a\b\g}(\q)(\e T)^{\d_1\d_2\d_3}_{\b_1\cdots \b_{11}}\q^{\b_1}\cdots \q^{\b_{11}}\Lambda_{\d_1\d_2\d_3}.
\ee

The tensor $\Lambda_{\a\b\g}$ is defined by
\be \label{eq:Lam}
\int [d\lambda] \lambda^{\a}\lambda^{\b}\lambda^{\g}\Lambda_{\a'\b'\g'}=\d^{(\a}_{\a'}\d^{\b}_{\b'}\d^{\g)}_{\g'}-\fr{1}{40}\g_m^{(\a\b}\g^m_{(\a'\b'}\d^{\g)}_{\g')}\equiv \d^{((\a}_{\a'}\d^{\b}_{\b'}\d^{\g))}_{\g'},
\ee
and is a function of the $\lambda$'s only. More accurately, all components
contain eleven delta functions or derivatives thereof. The precise
form of \eqref{eq:Lam} follows from the fact that the integral must be
an invariant tensor combined with the pure spinor constraint.
Detailed arguments are provided in appendix
\ref{sec:detrel}. To see what conditions (\ref{eq:Lam}) imposes on
$\Lambda_{+++}$ note that choosing $\a\b\g=+++$ gives
\be
\int [d\lambda] {\lambda^+}^3 \Lambda_{+++}=6.
\ee
Moreover this is the only condition because for all other choices the
LHS of \eqref{eq:Lam} is not invariant under $M$, the generator of
a $U(1)$ subgroup of Lorentz group (see appendix \ref{sec:app_su} for
the definition of $M$).
Therefore the LHS is equal to
zero. In fact for all choices that lead to non-zero $M$ charge the RHS
vanishes by the charge conservation property of invariant tensors
(cf. appendix \ref{app:ccp}). The solution is given by
\be
\Lambda_{+++}=6\d(\lambda^+)\d(\lambda_{12})\cdots \d(\lambda_{45}).
\ee
To determine whether this object is indeed part of a representation of the Lorentz group one needs to check the Lorentz algebra holds when acting on $\Lambda_{+++}$. First note
\be
(N_S)^a_{\ b}\Lambda_{+++}=N_{ab}\Lambda_{+++}=0,\quad N\Lambda_{+++}=\fr{15}{4}\Lambda_{+++},
\ee
$N^{mn}$ denote the realization of Lorentz generators $M^{mn}$
in terms of pure spinors, see  appendix \ref{sec:pslg} for the precise
expressions. All Latin indices from the beginning of the alphabet are $SU(5)$ indices. The nontrivial commutation relations that remain to be checked are
\bea \label{eq:loralg1}
[N_{ab},N^{cd}]\Lambda_{+++}&=&-\hp \d^{[c}_{[a}{N}^{d]}_{\ \ b]}\Lambda_{+++}=-\fr{1}{5}\d^{c}_{[a}\d^d_{b]}N\Lambda_{+++}=-\fr{3}{4}\d^{c}_{[a}\d^d_{b]}\Lambda_{+++},\qquad
\\ \label{eq:loralg2}
[{N}^a_{\ b},N^{cd}]\Lambda_{+++}&=&\hp \d^{[c}_b N^{d]a}\Lambda_{+++}.
\eea
Because of the symmetric form of $\Lambda_{+++}$ it suffices to check
\bea \label{eq:n12n12}
[N_{12},N^{12}]\Lambda_{+++}&=&-\fr{3}{4}\Lambda_{+++},
\\
\label{eq:n12n13}
[N_{12},N^{13}]\Lambda_{+++} &=&0,
\\
\label{eq:nf12n23}
[{N}^1_{\ 2},N^{23}]\Lambda_{+++}&=&-\hp N^{13}\Lambda_{+++}.
\eea
Let us start with the LHS of \eqref{eq:n12n12}
\[
[N_{12},N^{12}]\Lambda_{+++}=N_{12}N^{12}\Lambda_{+++}=N_{12}\left[\hp6\lambda^+\d(\lambda^+)\d'(\lambda_{12})\d(\lambda_{13})\cdots \d(\lambda_{45})\right]=
\]
\[
(-\hp w_+\lambda_{12}-\qu\fr{1}{\lambda^+}w^{ab}\lambda_{ab}\lambda_{12}+\hp\fr{1}{\lambda^+}w^{ab}\lambda_{1a}\lambda_{2b})\left[\hp6\lambda^+\d(\lambda^+)\d'(\lambda_{12})\d(\lambda_{13})\cdots \d(\lambda_{45})\right]=
\]
\be
=(0-\fr{9}{4}+\fr{6}{4})\Lambda_{+++}=-\fr{3}{4}\Lambda_{+++},
\ee
Note that $N_{12}$ does not contain factors of $(\lambda_{12})^2$
(possible such factors cancel out). This is
useful when acting with $N_{12}$ in this second line.
In going from the second to the last line we used $x\d'(x)=-\d(x)$ twice. \eqref{eq:n12n13} and \eqref{eq:nf12n23} follow along the same lines.

It is instructive to compute the next two levels (distinguished by $N$ charge) of the components of $\Lambda_{\a\b\g}$. For the components on the second ($N=\fr{11}{4}$) level consider
\be
N^{a_1a_2}\Lambda_{+++}=-\hp\Lambda_{\ \ \ \ ++}^{a_1a_2}-\hp\Lambda_{+\ \ \ +}^{\ a_1a_2}-\hp\Lambda_{++}^{\ \ \ a_1a_2}=-\fr{3}{2}\Lambda_{\ \ \ \ ++}^{a_1a_2}\Rightarrow
\ee
\[
\Lambda_{\ \ \ \ ++}^{a_1a_2}=-\fr{2}{3}N^{a_1a_2}\Lambda_{+++}.
\]
The factor of $-\hp$ is consistent with $N^{ab}w_+=-\hp w^{ab}$. Going to the next level ($N=\fr{7}{4}$)
\be
N^{b_1b_2}\Lambda_{\ \ \ \ ++}^{a_1a_2}=-\hp\e^{a_1a_2b_1b_2e}\Lambda_{e++}-\hp \Lambda^{a_1a_2b_1b_2}_{\ \ \ \ \ \ \ \ +}-\hp \Lambda^{a_1a_2\ b_1b_2}_{\ \ \ \ +}=
\ee
\[
-\hp\e^{a_1a_2b_1b_2e}\Lambda_{e++}- \Lambda^{a_1a_2b_1b_2}_{\ \ \ \ \ \ \ \ +}.
\]
This seems to leave freedom to define one of the two components, which would indeed be true if $\Lambda_{\a\b\g}$ was just a symmetric rank three tensor and nothing more. However $\Lambda_{\a\b\g}$ is gamma matrix traceless,
\be
\g_m^{\a\b}\Lambda_{\a\b\g}=0.
\ee
This imposes one additional condition that relates components of equal $N$ charge to each other. Consequently all components of $\Lambda_{\a\b\g}$ are uniquely fixed in terms of $\Lambda_{+++}$. Note that this is consistent with the discussion under \eqref{eq:c=40}, where Lorentz invariance arguments were used to come to the same conclusion.

\subsubsection{Decoupling of $Q$ exact states}

The new insertion $\Lambda_{\a \b \g}$ was motivated by manifest Lorentz invariance,
but it also results in a prescription in which $Q$ exact states decouple.
Indeed, the tree-level amplitude with one BRST exact state,
\be
\cB=\< Q \W(z_1) V_2(z_2) V_3(z_3) \prod_{i=4}^N \int dz_i U(z_i) (\e T)^{\d_1\d_2\d_3}_{\b_1\cdots \b_{11}}\q^{\b_1}\cdots \q^{\b_{11}}(\infty)\Lambda_{\d_1\d_2\d_3}(\infty)\>,
\ee
can be written in the following form:
\be
\cB=\< \lambda^{\a}(z_2)\lambda^{\b}(z_3)f_{\a\b}(\q)Q((\e T)^{\d_1\d_2\d_3}_{\b_1\cdots \b_{11}}\q^{\b_1}\cdots \q^{\b_{11}}\Lambda_{\d_1\d_2\d_3}) \>=
\ee
\be
\< \lambda^{\a}(z_2)\lambda^{\b}(z_3)f_{\a\b}(\q)(\e T)^{\d_1\d_2\d_3}_{\b_1\cdots \b_{11}}\lambda^{\b_1}\q^{\b_2}\cdots \q^{\b_{11}}\Lambda_{\d_1\d_2\d_3} \>.
\ee
After using the OPE's to integrate out the non-zero modes one gets:
\be \label{eq:tret02}
\cB=\int d^{16}\q [d\lambda] \lambda^{\a}\lambda^{\b}f_{\a\b}(\q)(\e T)^{\d_1\d_2\d_3}_{\b_1\cdots \b_{11}}\lambda^{\b_1}\q^{\b_2}\cdots \q^{\b_{11}}\Lambda_{\d_1\d_2\d_3}=
\ee
\[
\int d^{16}\q f_{\a\b}(\q)(\e T)^{\a\b\b_1}_{\b_1\cdots \b_{11}}\q^{\b_2}\cdots \q^{\b_{11}}=0.
\]
The last line vanishes because all traces of $(\e T)$ vanish (cf. \eqref{eq:trl}).

\section{One-loop amplitudes} \label{sec:oneloop}

In this section we investigate one-loop amplitudes with one unphysical
state both in the prescription with an integral over $B$ and without. 
We first show that all such amplitudes are proportional to certain
zero mode integrals. Decoupling of BRST exact states would follow
if these zero mode integrals vanished. However, these integrals
do not vanish after the $\lambda$ and $N$ integrations have been performed
as one would expect based on $Q$ invariance of the PCO's, \eqref{qvarcqcl}.
We then focus on four-point functions,
these being the first two non-vanishing one-loop amplitudes.
We will find decoupling of unphysical states in this case.
In the prescription without an integral over $B$, however, these amplitudes vanish
because none of the remaining terms after the $\lambda$, $N$ integrals
contain precisely sixteen distinct components of the zero modes of
$d_{\a}$. Preliminary analysis suggests that
this mechanism is not operational in
higher-point functions. Furthermore, even the four-point functions
are not Lorentz invariant. The four-point function containing one unphysical state with an integral over $B$ is also analyzed and we prove it vanishes. In the companion to this paper \cite{new1}
we show using a different argument that unphysical states decouple to all orders,
when one integrates over $B$ and $C$.

Note that the picture raising
operators, $Z_B$, are $Q$-closed without subtleties:
\be \label{eq:qvarzb}
Q Z_B= \qu B_{mn}\lambda\g^{mn}dB_{m'n'}\lambda\g^{m'n'}d\d'(B_{pq}N^{pq})=\qu (B_{mn}\lambda\g^{mn}d)^2\d'(B_{pq}N^{pq})=0.
\ee
This vanishes because it contains the square of a fermionic quantity,
so one may anticipate that the problems are due to picture lowering
operators $Y$ not being $Q$-closed.
Let us also record the Lorentz variation of $Z_B$,
\be \label{eq:lorvarzb}
M^{mn} Z_B= Q[2\h^{p[m}\d^{n]}_rB_{pq}N^{qr}\d(BN)].
\ee



\subsection{Amplitudes with unphysical states without integrating over $B$.} \label{sec:cfwus}
A one-loop amplitude with one unphysical state is given by
\[
\cB^{(N)}=\< Q\W_1(z_1) \prod_{i=2}^N \int dz_i U_i (z_i) \int du \m (u) \tilde{b}_{B^1}(u,w) (\lambda B^2d)(y)\cdots (\lambda B^{10}d)(y) (\lambda d)(y)
\]
\be \label{eq:cala}
\d(B^1N(y))\cdots \d(B^{10}N(y))\d(J(y))\Lambda_{\d_1\d_2\d_3}(y)(\e T)^{\d_1\d_2\d_3}_{\b_1\cdots\b_{11}}\q^{\b_1}(y)\cdots \q^{\b_{11}}(y)\>,
\ee
where $\lambda Bd=B_{mn}\lambda\g^{mn}d$. Note that we have replaced the $Y_C$ insertions
by the Lorentz invariant insertion, $\Lambda_{\a \b \g}$, as in the tree-level
computation. This is equivalent with inserting  $Y_C$ and integrating
over $C$.
On the torus we cannot insert the PCO's such that all their OPE's would
vanish. We inserted them at some arbitrary point $y$. For later convenience we inserted $\tilde{b}$ at a different point, $w$.

We now integrate $Q$ by parts. When $Q$ acts on $\tilde{b}$ we
get a total derivative in moduli space, as usual. If this total derivative
is non-vanishing the theory has a BRST anomaly. These total derivative
terms will be suppressed below because they are not important for
our discussion.

The terms that are important for us are the ones one gets
by acting with $Q$ on the rest of the terms in \eqref{eq:cala}.
Formally, there should not be any such terms. The reason is that
both the vertex operators and the PCO's are BRST closed. More precisely,
the BRST variation of the PCO contains  delta functions of $\lambda$ and $N$
(cf. \eqref{qvarcqcl}, \eqref{eq:qvarzb}), so the
terms obtained by acting with $Q$ on the PCO should vanish after
integrating over $\lambda$ and $N$. The main result in this
section is that this does not happen. In contrast, as we will see
in section \ref{sec:nonmin}, these terms are indeed zero in
the non-minimal formulation.

More precisely, after integrating $Q$ by parts the amplitude
\eqref{eq:cala} becomes,
\be \label{eq:vanqex}
\cB^{(N)}=\< \W_1(z_1) \prod_{i=2}^N \int dz_i U_i (z_i) \int du \m (u) \tilde{b}_{B^1}(u,w) (\lambda B^2d)(y)\cdots (\lambda B^{10}d)(y) (\lambda d)(y)
\ee
\[
\d(B^2N(y))\cdots \d(B^{10}N(y))\d(J(y))\Lambda_{\d_1\d_2\d_3}(y)(\e T)^{\d_1\d_2\d_3}_{\b_1\cdots\b_{11}}(y)\lambda^{\b_1}(y)\q^{\b_2}(y)\cdots \q^{\b_{11}}(y)\>.
\]
where we emphasize again that we suppress the total derivative term
in moduli space originating from $Q$ acting on $\tilde{b}$.
In this subsection we will evaluate $\cB^{(N)}$ without integrating over $B$.
The choice we make is:
\be \label{eq:Bchoice}
(B^1)_{ab}=\d^{[1}_a\d^{2]}_b,\ldots, (B^{10})_{ab}=\d^{[4}_a\d^{5]}_b, \ \ \
(B^I)^{ab}=(B^I)^a_{\ b}=0
\ee

We demonstrate below that all such one-loop amplitudes can
be written as a sum of terms proportional to a certain zero mode integral
$I_{\b_2\cdots  \b_{11}}$. This is done by using the OPE's to remove all fields
of non-zero weight, in particular $N^{mn}$. This is a non trivial step
because of the complicated form of the $b$ ghost.

Had the zero mode integral $I_{\b_2\cdots \b_{11}}$ vanished,
this would have proven that
BRST exact states decouple at one loop (again modulo the total derivative
term from $Q$ acting on $\tilde{b}$). Non-vanishing of
$I_{\b_2\cdots \b_{11}}$ does not prove
that there exists a non-vanishing amplitude with a $Q$-exact state,
because there may be additional cancellations when one performs
the remaining integrals. It does show however that the PCO's
are not $Q$ closed.

\subsubsection*{Zero mode integral}

We will show below that all one-loop amplitudes (\ref{eq:vanqex}) can be written as a sum of terms that
are proportional to the following zero mode integral,
\be \label{eq:analog}
I^{\a_1}_{\b_1\b_2\cdots \b_{11}}\equiv\int [d\lambda][dN] \lambda^{\a_1}(\lambda\g^{13}d)\cdots (\lambda\g^{45}d)(\lambda d)\d(N^{12})\cdots \d(N^{45})\d(J)\Lambda_{\a\b\g}(\e T)^{\a\b\g}_{\b_1\cdots \b_{11}}.
\ee
Moreover we will show all one-loop amplitudes with an unphysical state can be written as a sum of terms proportional to the trace of $I^{\a_1}_{\b_1\cdots \b_{11}}$ which we call $I_{\b_2\cdots \b_{11}}$:
\be \label{eq:ib2b11}
I_{\b_2\cdots \b_{11}}\equiv I^{\a_1}_{\a_1\b_2\cdots \b_{11}}=
\ee
\[
\int [d\lambda][dN] \lambda^{\b_1}(\lambda\g^{13}d)\cdots (\lambda\g^{45}d)(\lambda d)\d(N^{12})\cdots \d(N^{45})\d(J)\Lambda_{\a\b\g}(\e T)^{\a\b\g}_{\b_1\cdots \b_{11}}.
\]
Thus \eqref{eq:ib2b11} is the one-loop analog of \eqref{eq:tret02} (or
\eqref{eq:tret01}).
Note that, in spite of the notation, $I^{\a_1}_{\b_1\b_2\cdots \b_{11}}$ is not
manifestly Lorentz invariant. Whether it is Lorentz invariant remains
to be seen. Our first task is to evaluate $I_{\b_2\cdots \b_{11}}$.

After using expression \eqref{eq:[dN]} for $[dN]$ to evaluate the $N$ integral in $I_{\b_2\cdots \b_{11}}$ we find
\be
I_{\b_2\cdots \b_{11}}=\int [d\lambda]\fr{1}{(\lambda^+)^8}\lambda^{\b_1}(\lambda\g^{13}d)\cdots (\lambda\g^{45}d)(\lambda d)\Lambda_{\a\b\g}(\e T)^{\a\b\g}_{\b_1\cdots \b_{11}}.
\ee
In this form it becomes apparent that the problems with factors of $\lambda^+$ in the denominator only become bigger at one loop. At this point we can only surmise this. To find a definitive answer we have to evaluate the $\lambda$ integral. This can be done by expanding the integrand by powers of $\lambda^+$:
\bea \label{eq:lamser}
&&\fr{1}{(\lambda^+)^8} (\lambda\g^{13}d)\cdots (\lambda\g^{45}d)(\lambda d)=(\lambda^+)^2D_{12}d_++
\\ \nn
&&\hp\lambda^+\lambda_{a_1a_2}( D_{12}d^{a_1a_2}+\hp\e^{aba_1a_2c}d_cD_{12ab}d^+)+\fr{1}{8}\lambda_{a_1a_2}\lambda_{a_3a_4}(D_{12}\e^{aa_1a_2a_3a_4}d_a+
\\ \nn
&&\e^{aba_1a_2c}d_cD_{12ab}d^{a_3a_4}+ \hp \e^{aba_1a_2c}\e^{dea_3a_4f}d_cd_fD_{12abde}d_+ )+
\\ \nn
&&\fr{1}{32}\fr{1}{\lambda^+}\lambda_{a_1a_2}\lambda_{a_3a_4}\lambda_{a_5a_6}(\e^{aba_1a_2c}d_cD_{12ab}\e^{da_3a_4a_5a_6}d_d+\e^{aba_1a_2c}\e^{dea_3a_4f}d_cd_fD_{12abde}d^{a_5a_6}+
\\ \nn
&&\hp \e^{aba_1a_2c}\e^{dea_3a_4f}\e^{gha_5a_6j}d_cd_fd_jD_{12abdegh}d_+ )+
\sum_{k=4}^6 \fr{1}{(\lambda^+)^{k-2}}\lambda_{a_1a_2}\cdots \lambda_{a_{2k-1}a_{2k}} Y^{a_1\cdots a_{2k}},
\eea
where
\be
D=d^{12}\cdots d^{45},\quad D_{a_1\cdots a_k}=\fr{\del}{\del d^{a_{k-1}a_k}}\cdots \fr{\del}{\del d^{a_1a_2}}D.
\ee
The $Y$'s can be expressed in terms of the $d$'s just like in the first four terms. Note that the minimal number of $d_a$'s in $Y^{a_1\cdots a_{2k}}$ is $k-1$. This is the reason the series stops at $k=6$. The maximum number of $d_a$'s in $Y^{a_1\cdots a_{2k}}$ is $k$. The $\lambda$ integration of \eqref{eq:lamser} can be evaluated term by term. $I_{\b_2\cdots \b_{11}}$ then becomes
\be
I_{\b_2\cdots \b_{11}}=\sum_{k=0}^6 (I_k)_{a_1\cdots a_{2k}\b_2\cdots \b_{11}}Y^{a_1\cdots a_{2k}}.
\ee
The integrals $I_k$ are investigated order by order in the sequel of this subsection.

For $k=0,1,2$ one can use \eqref{eq:Lam} and \eqref{eq:trl} to show the $\lambda$ integrals vanish:
\be \label{eq:I0}
(I_0)_{\b_2\cdots \b_{11}}=\int [d\lambda] \lambda^{\b_1}(\lambda^+)^2\Lambda_{\d_1\d_2\d_3} (\e T)^{\d_1\d_2\d_3}_{\b_1\cdots \b_{11}}= (\e T)^{++\b_1}_{\b_1\cdots \b_{11}}=0,
\ee
\be \label{eq:I1}
(I_1)_{a_1a_2\b_2\cdots \b_{11}}=\int [d\lambda] \lambda^{\b_1}\lambda^+\lambda_{a_1a_2} \Lambda_{\d_1\d_2\d_3}  (\e T)^{\d_1\d_2\d_3}_{\b_1\cdots \b_{11}}=(\e T)^{+\b_1}_{\ \ \ \ a_1a_2\ \b_1\cdots \b_{11}}=0,
\ee
\be \label{eq:I2}
(I_2)_{a_1\cdots a_4\b_2\cdots \b_{11}}=\int [d\lambda] \lambda^{\b_1}\lambda_{a_1a_2}\lambda_{a_3a_4} \Lambda_{\d_1\d_2\d_3} (\e T)^{\d_1\d_2\d_3}_{\b_1\cdots \b_{11}}=(\e T)^{\b_1}_{\ \ a_1a_2a_3a_4\ \b_1\cdots \b_{11}}=0.
\ee
If $k>2$, however, there are also factors of $\lambda^+$ in the denominator. As shown in appendix \ref{app:coeflint} the $\lambda$ integrals do not vanish anymore. $M$ charge conservation implies that $I_3$ can only be non vanishing if
\be
\b_2,\ldots ,\b_{11}= +,b_1b_2,\ldots ,b_9b_{10},c_1,c_2,c_3,c_4 \ {\rm or}\ \b_2,\ldots ,\b_{11}=b_1b_2,\ldots ,b_{13}b_{14},c_1,c_2,c_3.
\ee
This is explained in detail in the first part of appendix \ref{app:ik}. We explicitly compute $I_3$ for the first case. Since ${\rm Sym }^3 \bar{{\bf 10}} \otimes {\rm Asym}^5 \bf{10}\otimes {\rm Asym}^4 \bar{{\bf 5}}$ contains one scalar,
one finds
\[
(I_3)^{\ \ \ \ \ \ \ \ \ \ \ \ \ \ \ \ b_1b_2\cdots b_{9}b_{10}}_{a_1\cdots a_6 +c_1c_2c_3c_4}=\int [d\lambda] \fr{1}{\lambda^+}\lambda^{\b_1} \lambda_{a_1a_2}\cdots\lambda_{a_5a_6} \Lambda_{\a\b\g}(\e T)^{\a\b\g\ \ \ \ \ \ \ \ \ \ \ \ b_1\cdots b_{10}}_{\ \ \ \ \b_1+c_1c_2c_3c_4}=
\]
\be\label{eq:k=3}
c_1\e_{a_1a_2a_3a_4b_{14}}\e_{b_{16}a_5a_6b_{11}b_{12}}(\e_{10})^{b_1\cdots b_{20}}\e_{c_1c_2c_3c_4b_{17}}\e_{b_{13}b_{15}b_{18}b_{19}b_{20}}+{\rm 2\ perms},
\ee
where $(\e_{10})^{b_1\cdots b_{20}}$ is antisymmetric under both $b_{2i-1}
\lera b_{2i}$ and $b_{2i-1}b_{2i} \lera b_{2j-1}b_{2j}$ and $(\e_{10})^{12131415232425343545}=1$. The two permutations add terms to make the RHS symmetric under $a_{2i-1}a_{2i} \lera a_{2j-1}a_{2j}$. The constant $c_1$ is computed in
appendix \ref{sec:I3} and is given by
\be
c_1= \fr{129}{2} .
\ee
We will not compute any components of $I_4$ here.
Going to the next level, the only choice of $\b_2, \ldots, \b_{11}$
that leads to a non-zero answer for $I_5$ is
\[
(I_5)_{a_1\cdots a_{10}\ \ \ \ \ \ 12345}^{\ \ \ \ \ \ \ b_3\cdots b_{12}}=\int [d\lambda] \fr{1}{(\lambda^+)^3}\lambda^{\b_1}\lambda_{a_1a_2}\cdots \lambda_{a_9a_{10}}\Lambda_{\d_1\d_2\d_3}(\e T)^{\d_1\d_2\d_3\ \ b_3\cdots b_{12}}_{\ \ \ \ \ \ \b_1\ \ \ \ \ \ \ 12345}=
\]
\be
-\fr{2}{5}\e_{b_{13}a_1a_2a_3a_4}\e_{b_{15}a_5a_6a_7a_8}\e_{b_{17}a_9a_{10}b_1b_2} (\e_{10})^{b_1\cdots b_{20}}\e_{b_{14}b_{16}b_{18}b_{19}b_{20}}+{\rm 14\ perms}.
\ee
The details are given in appendix \ref{sec:I3}. Finally $I_6$ can be evaluated as:
\be
(I_6)_{a_1\cdots a_{12}\b_2\cdots \b_{11}}=\int[d\lambda] \fr{1}{(\lambda^+)^4}\lambda^{\b_1}\lambda_{a_1a_2}\cdots \lambda_{a_{11}a_{12}} \Lambda_{\a\b\g}(\e T)^{\a\b\g}_{\b_1\cdots \b_{11}}=
\ee
\[
\e_{b_1a_1a_2a_3a_4}\e_{b_2a_5a_6a_7a_8}\e_{b_3a_9a_{10}a_{11}a_{12}}(\e T)^{b_1b_2b_3}_{+\b_2\cdots \b_{11}} +{\rm permutations}=0.
\]
This vanished because $(\e T)^{b_1b_2b_3}_{+\b_2\cdots \b_{11}}=0$ and that follows from the $M$ charge conservation rule for invariant tensors. In other words it is not possible to choose $\b_2, \ldots, \b_{11}$ such that the total $M$ charge of the components is zero (cf. equation \eqref{eq:mchcon}). This concludes the computation of the pure spinor zero mode integrals that appear at one loop.

\subsubsection*{Non-zero mode integration}
We now demonstrate that all one-loop amplitudes with an unphysical state can be written as a sum of terms proportional to $I_{\b_2\cdots \b_{11}}$. After this proof we indicate how the argument can be modified to prove that $\cA^{(N)}$ can be written as a sum of terms proportional to $I^{\a}_{\b_1\cdots \b_{11}}$. In general the amplitude, $\cB^{(N)}$, becomes a sum of terms of the form
\be \label{eq:finex}
\cB^{(N)}_{i_1\cdots i_k}=\int [\cD \lambda][\cD N][\cD d][\cD \q](\prod_{i=2}^N\int dz_i)f_{m_1n_1\cdots m_kn_k}(z_1,\ldots,z_N)
\ee
\[
N^{m_1n_1}(z_{i_1})\cdots N^{m_kn_k}(z_{i_k})(\lambda\g^{13}d)(y)\cdots (\lambda\g^{45}d)(y)(\lambda d)(y)\lambda^{\b_1}(y)
\Lambda_{\a \b \g} (y)(\e T)^{\a\b\g}_{\b_1\cdots \b_{11}}
\]
\[
\q^{\b_2} (y)\cdots \q^{\b_{11}}(y)\int du \m(u)\tilde{b}_{B^1}(u,w)\d(N^{13}(y))\cdots \d(N^{45}(y))\d(J(y))e^{-S},
\]
where the indices in the PCO's are $SU(5)$ indices, $i_j \in\{2,\ldots,N\}$ and $f_{m_1\cdots n_k}$ does not contain any $\lambda$'s or $w$'s. The number $k$ indicates how many vertex operators provide an $N^{mn}$. The functional integrals over $\lambda$ and $N$ can be evaluated by performing the OPE's to remove all fields of non-zero weight. Then one replaces the fields by their zero modes and performs the integration over these modes. In order to perform the OPE between $N^{mn}$ and $\d(BN)$ we have to Taylor expand $\d(BN)$, as discussed in
\cite{Berkovits:2004px},
\be \label{eq:taylor}
\d((BN(y))=\d(BN_0\w(y)+B\hat{N}(y))=
\ee
\[
\d(BN_0\w(y))+(B\hat{N}(y))\d'(BN_0\w(y))+\hp (B\hat{N}(y))^2\d''(BN_0\w(y))+\cdots,
\]
where $\hat{N}$ denotes $N$ after omission of the zero mode. The holomorphic one form $\w(y)$ is constant on the torus:
\be
\w(y)=\fr{1}{4\pi^2 \t_2},
\ee
where $\t_2$ is the imaginary part of the modulus $\t$. The $b$ ghost also contains $N^{mn}$'s which have to be taken into account if one is removing all fields of non-zero weight. We first focus on the first term, the local $b$ ghost, $b_B(u)$. The second term of $\tilde{b}(u,y)$, with the integration in it, will be dealt with later. After replacing $\tilde{b}(u,y)$ by $b(u)$ in the amplitude, $\cB^{(N)}_{i_1\cdots i_k}$, becomes a sum over $n$, which counts the number of $N^{mn}$'s the local $b$ ghost provides, of the following objects:
\[
\cB^{(N)}_{i_1\cdots i_k,n}=\int [\cD \lambda][\cD N][\cD d][\cD \q](\prod_{i=2}^N\int dz_i)\int du \m(u)\sum_{j=0}^3 f_{jm_1n_1\cdots m_{k+n}n_{k+n}}(z,u,w)
\]
\[
N^{m_1n_1}(z_{i_1})\cdots N^{m_kn_k}(z_{i_k})N^{m_{k+1}n_{k+1}}(w)\cdots N^{m_{k+n}n_{k+n}}(w) (\lambda\g^{13}d)(y)\cdots (\lambda\g^{45}d)(y)
\]
\[
(\lambda d)(y)\lambda^{\b_1}(y)(\e T)^{\a\b\g}_{\b_1\cdots \b_{11}}\Lambda_{\a\b\g}(y)\q^{\b_2}(y)\cdots \q^{\b_{11}}(y)
\]
\be \label{eq:finex2}
\d^{(j)}(N^{12}(w))\d(N^{13}(y))\cdots \d(N^{45}(y))\d(J(y))e^{-S},
\ee
where $\d^{(j)}$ denotes the $j$th derivative of the delta function and the sum runs from zero to three because $b$ does not contain $\d^{(4)}(B^1N)$ or higher derivatives.

The product of the eleven delta functions, including the one from $b$, becomes a sum of products of eleven $\d^{(j)}(B^IN_0)$ after the Taylor expansion. We start with the first term in this sum, i.e. the one without $\hat{N}$'s and no derivatives on the delta functions. In this case the $N^{m_jn_j}(z_j)$'s from \eqref{eq:finex} have OPE's with themselves and with the $\lambda$'s from the PCO's. We first concentrate on the term in which all $N^{mn}$'s get contracted with an explicit $\lambda$. That term is given by\footnote{Since the distinction between worldsheet fields and their zero modes plays a central role in the argument, zero modes are denoted in an explicit way, unlike in other parts of this work.}
\[
\cC^{(N)}_{i_1\cdots i_k,n}=\int[d\lambda][dN][\cD^{16}d][\cD^{16}\q]\left[\prod_{i=2}^N\int dz_i\right ]\int du f_{m_1n_1\cdots m_{k+n}n_{k+n}}(z_1,\cdots,z_N,u)
\]
\[
\left[\prod_{l=1}^kF(z_{i_l},y)\right]F(w,y)^n\cN^{m_1n_1}\cdots \cN^{m_{k+n}n_{k+n}}\lambda_0^{\b_1}(\lambda_0\g^{13}d(y))\cdots (\lambda_0\g^{45}d(y))(\lambda_0 d(y))
\]
\be \label{eq:cC}
(\Lambda_0)_{\a\b\g}(\e T)^{\a\b\g}_{\b_1\cdots \b_{11}}\q^{\b_2}(y)\cdots \q^{\b_{11}}(y)\d(N_0^{12}) \cdots \d(N_0^{45})\d(J_0)e^{S_{p\q}},
\ee
where
\be
F(z,y)=\del_z {\rm{log}}E(z,y)
\ee
and $E(z,y)$ is the holomorphic prime form, which goes like $z-y$ when $z \rightarrow y$ \cite{D'Hoker:1988ta,Verlinde:1986kw}. $\cN^{mn}$ are abstract Lorentz generators for the $\lambda,w$ sector and they act to the right. They should not be thought of as containing (zero) modes of the $\lambda$ or $w$ worldsheet fields. The $\cN^{mn}$ merely multiply every index on a $\lambda$ or $w$ they hit by a two form gamma matrix. Up to now we only considered contractions between $N^{mn}$ and the explicit $\lambda$'s, but if two or more $N^{mn}$'s contract with each other in $\cB^{(N)}_{i_1\cdots i_k,n}$ we get a term of the form $\cC^{(N)}_{i_1\cdots i_l,m}$, with $l+m<k+n$, where the poles in $z_i-z_j$ are included in the unspecified function $f$.

The last step of our argument is showing all terms with derivatives on the delta functions can also be written as a sum of terms of the form $\cC^{(N)}_{i_1\cdots i_k,n}$. To see this note that if a derivative acts on $\d(N^{ab})$ one of the $N^{mn}$ must provide this zero mode, otherwise the integral vanishes. This step just reduces the number of $N^{mn}$'s in $\cB^{(N)}_{i_1\cdots i_k,n}$ that must be contracted, so in fact it becomes of the form $\cC^{(N)}_{i_1\cdots i_l,m}$ where $k+n-l-m$ is the number derivatives acting on the delta functions. Since the zero mode measures $[d\lambda]$ and $[dN]$ are Lorentz invariant we can pull the $\cN$ out of these integrals. This concludes the main part of the argument that a one-loop amplitude can be written as a sum of terms proportional to $I_{\b_2\cdots \b_{11}}$.

We still need to consider the second term in $\tilde{b}(u,w)$. This was not included in the above discussion because it contains $\del N^{mn}(v)$. This does not change the argument much, after the OPE's this part of the amplitude will also have the form of $\cC^{(N)}_{i_1\cdots i_k,n}$ where the effect of the $v$ derivative and the integral over $v$ are included in $f$.

To see $\cA^{(N)}$ can be written as a sum of terms proportional to $I^{\a_1}_{\b_1\cdots \b_{11}}$  one can use the above reasoning with a slight adjustment. This consists of replacing $\lambda^{\b_1}(y)$ by $\lambda^{\a_1}(z_1)$ in \eqref{eq:finex} and adding an $\a_1$ index to $f$. The only effect this has is the replacement of some $F(z_i,y)$ by $F(z_1,z_i)$ in \eqref{eq:cC}, apart from the fact $\a_1$ and $\b_1$ are not contracted anymore.

Thus we have shown that amplitudes with unphysical states do not
vanish by the $\lambda,N$ integration, opposite to expectations, but
nevertheless let us press on and explicitly compute a one-loop four-point
amplitude with an unphysical state. Perhaps we will find
some other mechanism that makes these amplitudes vanish.

\subsubsection{Four point function without integrating over $B$}
In this subsection we investigate two properties of the four-point one-loop function \eqref{eq:1looppre} in the minimal pure spinor formalism, namely decoupling of unphysical states and its Lorentz invariance in the formulation without integrating over $B$. We will find decoupling of $Q$ exact states, in spite of the results of the previous section. The vanishing is achieved after the integral over the $d$ zero modes. Lorentz invariance, however, does not follow in the same way.

\subsubsection*{Decoupling of unphysical states}
The one-loop four-point amplitude is an example of an amplitude in which only the zero modes contribute (cf. \cite{Berkovits:2004px}). It turns out only three terms have enough factors of $d_{\a}$ and $N^{mn}$ to give a non vanishing answer. This will become clear in equation \eqref{eq:bgtr} below. Thus we can immediately replace all the fields in \eqref{eq:cala} by their zero modes:
\be
\cB^{(4)}= \int [d\lambda][dN]d^{16}dd^{16}\q Q\W \prod_{i=2}^4 U_i \tilde{b}_{B^1} (\lambda B^2d)\cdots (\lambda B^{10}d) (\lambda d)
\ee
\[
\d(B^1N)\cdots \d(B^{10}N)\d(J)\Lambda_{\a\b\g}(\e T)^{\a\b\g}_{\b_1\cdots\b_{11}}\q^{\b_1}\cdots \q^{\b_{11}}.
\]
The only terms of $b_{B^1}$ that contributes are the ones with four $d$'s
and there are only three such terms:
\bea \label{eq:bgtr}
(b_B)|_{d^4}&=&-\fr{1}{1536} \g^{\a\b}_{mnp}(d\g^{mnp}d)(Bd)_{\a}(Bd)_{\b}\d'(BN)
\\
&&-\fr{1}{8}{c_1}^{\g\d\a\r}_{mn}N^{mn}d_{\r}(Bd)_{\a}(Bd)_{\b}(Bd)_{\g}\d''(BN) \nn
\\
&&-\fr{1}{16}{c_4}^{\d\g\b\a}_{mnpq}N^{mn}N^{pq}(Bd)_{\a}(Bd)_{\b}(Bd)_{\g}(Bd)_{\d}\d'''(BN), \nn
\eea
where the invariant tensors $c_1$ and $c_4$ can be read off from \eqref{eq:bB0}-\eqref{eq:bB3} and \eqref{eq:G}-\eqref{eq:L}. Note the $N$ integration will only be non vanishing if the fourth vertex operator provides an $N^{mn}$ zero mode. Moreover there are no terms in the $b$ ghost with three $d$'s and no derivatives on $\d(BN)$. Such terms could have contributed here. The three terms above turn out to all be proportional to (for $B_{ab}=\d^1_{[a}\d^2_{b]},B^a_{\ b}=B^{ab}=0$)
\be
d^{12}d_3d_4d_5 \d'(N^{12}).
\ee

For the first term this follows from direct computation using the
gamma matrices as listed in appendix \ref{sec:gama}. Actually,
one could have predicted the fact that three of
the four $d_{\a}$'s are $d_a$'s and one is a $d^{ab}$, by looking at
the $M$ charge of the full term. $\d'(N^{12})$ has $M$ charge two and
since $\g^{\a\b}_{mnp}(d\g^{mnp}d)(Bd)_{\a}(Bd)_{\b}\d'(BN)$ has $M$
charge zero, the $d$ part must have $M$ charge minus two. The only way
four $d$'s can give $M$ charge minus two is when three of them are a
$d_a$ ($M$ charge $-\fr{3}{4}$) and the fourth is a $d^{ab}$ ($M$
charge $\qu$).

The second term can be reduced as follows:
\be
(c_1)^{\g\b\a\r}_{mn}N^{mn}d_{\r}(Bd)_{\a}(Bd)_{\b}(Bd)_{\g} \d''(BN)=
\ee
\[
(c_1)_{12\ a_1\cdots a_8}d^{a_7a_8}\hp\e^{a_1a_212a}d_a\hp\e^{a_3a_412b}d_b\hp\e^{a_5a_612c}d_c \d'(N^{12}),
\]
where we used the $M$ charge conservation property of invariant tensors together with $(Bd)_a=0$. After observing that $(c_1)_{aba_1\cdots a_{8}}$ is an $SU(5)$ invariant tensor that is antisymmetric in the middle three pairs of indices $(a_1a_2,a_3a_4,a_5a_{6})$ and there is only one invariant tensor with these symmetries \cite{vanLeeuwen}, namely $(\e_{aba_1a_2[a_3}\e_{a_4]a_5a_{6}a_7a_8}+{\rm 5\ perms})$, we find that the second term in the $b$ ghost is proportional to
\be
(c_1)_{12343545a_7a_8}d^{a_7a_8}d_3d_4d_5 \d'(N^{12})=d^{12}d_3d_4d_5 \d'(N^{12}).
\ee
The same logic can be applied to the third term although this case is slightly simpler. $\a,\b,\g,\d$ has to be $+,ab,cd,ef$ and since $(Bd)_+=d^{12}$ we automatically get this factor.

The third integrated vertex operator must provide an $N^{12}$ zero
mode. It then follows that $\cB^{(4)}$ is proportional to
$I_{\b_2\cdots \b_{11}}$. This
integral can be written as a sum over $k$ just as in
\eqref{eq:lamser}. In this sum the $k=0,1,2,6$ terms vanish because of
the $\lambda$ integration and the $k=4,5$ terms vanish due to the $d$
integration (note that $b_{B^1}$ contains three $d_a$'s and
$Y_4,Y_5$ contain at least three $d_a$'s). The $k=3$ term is given
by
\be \label{eq:I3}
(b_{B^1})|_{d^4}(I_3)_{a_1\cdots a_6\b_2\cdots \b_{11}}(Y_3)^{a_1\cdots a_6}=d^{12}d_3d_4d_5(\fr{1}{32}\e^{aba_1a_2c}d_cD_{12ab}\e^{da_3a_4a_5a_6}d_d+
\ee
\[
\fr{1}{32}\e^{aba_1a_2c}\e^{dea_3a_4f}d_cd_fD_{12abde}d^{a_5a_6}+\fr{1}{64}\e^{aba_1a_2c}\e^{dea_3a_4f}\e^{gha_5a_6j}d_cd_fd_jD_{12abdegh}d_+ )
\]
\[
\int [d\lambda] \fr{1}{\lambda^+}\lambda^{\b_1}\lambda_{a_1a_2}\lambda_{a_3a_4}\lambda_{a_5a_6}\Lambda_{\d_1\d_2\d_3}(\e T)^{\d_1\d_2\d_3}_{\b_1\cdots \b_{11}}=
\]
\[
-\qu d^{12}d_3d_4d_5\e^{aba_1a_2c}d_cd_dD_{12ab} \int [d\lambda] \lambda^{\b_1}\lambda^d\lambda_{a_5a_6}\Lambda_{\d_1\d_2\d_3}(\e T)^{\d_1\d_2\d_3}_{\b_1\cdots \b_{11}}+
\]
\[
-d^{12}d_3d_4d_5\e^{aba_1a_2c}d_cd_fD_{12ab}\int [d\lambda] \lambda^{\b_1}\lambda_{a_1a_2}\lambda^f\Lambda_{\d_1\d_2\d_3}(\e T)^{\d_1\d_2\d_3}_{\b_1\cdots \b_{11}}=0,
\]
where we used
\be
D_{12abcd}d^{ef}=-\d^{[e}_{c}\d^{f]}_dD_{12ab}-\d^{[e}_{1}\d^{f]}_2D_{abcd}-\d^{[e}_{a}\d^{f]}_bD_{cd12}
\ee
and the integral vanishes because $\e T$ is traceless.

Thus, for the four-point one-loop amplitudes with a BRST exact state
the terms that do not vanish after the $\lambda, N$
integral now vanish because they contain a square of fermionic
quantity, namely  $d_{\a}d_{\a}\ ({\rm no\ sum})$. One may wonder
whether the same mechanism would work in higher point functions.
While we do not have a
definite answer to this, preliminary results suggest that this
is not the case. For example, the zero mode contribution to
the 5-point function with a $Q$-exact state does not vanish
in this way, but we should emphasize that our analysis
does not exclude possible cancellations between the contributions
of zero and non-zero modes.

\subsubsection*{Lorentz invariance}

In this subsection we study the Lorentz invariance of the amplitudes.
Recall that the Lorentz variation of the PCO's is $Q$ exact
(cf.~\eqref{eq:lorvarzb}). Thus one expects that the amplitude is Lorentz invariance. We have seen earlier however that $Q$ exact
states
may not decouple, so we will proceed to check explicitly whether the
$Q$ exact terms obtained from the Lorentz variation of the PCO's
evaluate to zero. We will focus on the term obtained by the Lorentz
variation of a single PCO. This term should be zero by itself
because it is $Q$ exact. Integrating $Q$ by parts one obtains
a total derivative in moduli space when $Q$ acts on $\tilde{b}$,
which will be suppressed as in our earlier discussion, and a
number of terms when $Q$ acts on the other PCO's. These terms
should evaluate to zero after the $\lambda$ and $N$ integrals are performed,
but we will see that they do not.

For the choice of the constant $B$ tensor in (\ref{eq:Bchoice}),
the PCO are invariant under $M^{ab}$ and under the $SU(5)$ generators
 $(M_S)^a_{\ b}$
transform into a $Q$-exact term (cf. (\ref{eq:lorvarzb})) (see appendix
\ref{sec:app_su} for the definition of the generators).
More specifically, the $SU(5)$ transformation of
$Z_{B^2}=(\lambda \g^{13}d) \d(N^{13})$ is given by
\be \label{mzb}
(M_S)^a_{\ b}((\lambda \g^{13}d) \d(N^{13}))=Q(\d_b^{[1} N^{3]a} \d(N^{13})).
\ee
As explained before the non-zero modes can be integrated out trivially
in the four point one-loop function:
\be
\cA^{(4)}=\int [d\lambda][dN]d^{16}dd^{16}\q \lambda^{\a}A_{\a} \prod_{i=2}^4 U_i \tilde{b}_{B^1} (\lambda B^2d)\cdots (\lambda B^{10}d) (\lambda d)
\ee
\[
\d(B^1N)\cdots \d(B^{10}N)\d(J)\Lambda_{\a\b\g}(\e T)^{\a\b\g}_{\b_1\cdots\b_{11}}\q^{\b_1}\cdots \q^{\b_{11}}.
\]
The Lorentz variation of the four point function can be written as a sum with one term for each $Z_B$:
\be
(M_S)^a_b \cA^{(4)}=\sum_I (\cA^{(4)}_{B^I})^a_b,
\ee
and using (\ref{mzb}) we obtain
\be \label{ab2}
(\cA^{(4)}_{B^2})^a_b=\int [d\lambda][dN]d^{16}dd^{16}\q \lambda^{\a}A_{\a} \prod_{i=2}^4 U_i \tilde{b}_{B^1} Q(\d_b^{[1} N^{3]a} \d(N^{13}))(\lambda B^3d)\cdots (\lambda B^{10}d)
\ee
\[
(\lambda d)\d(B^1N)\cdots \d(B^{10}N)\d(J)\Lambda_{\a\b\g}(\e T)^{\a\b\g}_{\b_1\cdots\b_{11}}\q^{\b_1}\cdots \q^{\b_{11}}.
\]
with similar formulas for the other terms.

Each of $(\cA^{(4)}_{B^I})^a_b$ should be zero separately, so we focus on
(\ref{ab2}).
After integrating $Q$ by parts and writing out the $b$ ghost one finds
\be
(\cA^{(4)}_{B^2})^a_b=\int d^{16}dd^{16}\q f_{\a}(\q,x)(\cA_{B^2,\lambda N}^{(4)})^{\a a}_{\ b},
\ee
for some $f_\a$ and
\be
(\cA_{B^2,\lambda N}^{(4)})^{\a a}_{\ b}=\d^3_b\d^a_2\int[d\lambda][dN] \lambda^{\a}d^{12}d_3d_4d_5 \d(N^{12})\cdots \d(N^{45})\d(J)
\ee
\[
 (\lambda\g^{14}d) \cdots (\lambda\g^{45}d) (\lambda d)\Lambda_{\d_1\d_2\d_3}(\e T)^{\d_1\d_2\d_3}_{\b_1\cdots\b_{11}}\lambda^{\b_1}\q^{\b_2}\cdots \q^{\b_{11}}.
\]
This integral can be evaluated in exactly the same fashion as the one
appearing in the four-point function with a $Q$ exact state. The first
step is to perform the $N$ integrations and then expand the
integrand in powers of $\lambda^+$:
\be \label{eq:lorvarw}
(\cA_{B^2,\lambda N}^{(4)})^{\a 2}_{\ 3}= d^{12}d_3d_4d_5\int [d\lambda]\Lambda_{\d_1\d_2\d_3}(\e T)^{\d_1\d_2\d_3}_{\b_1\cdots \b_{11}}[ \lambda^{\a}\lambda^{\b_1}\lambda^+ D_{1213}d_+ +
\ee
\[
\lambda^{\a}\lambda^{\b_1}\lambda_{a_1a_2}(\hp D_{1213}d^{a_1a_2}+\qu D_{1213ab}\e^{aba_1a_2c}d_c)+
\]
\[
\lambda^{\a}\lambda^{\b_1}\fr{1}{\lambda^+}\lambda_{a_1a_2}\lambda_{a_3a_4}(\fr{1}{8}D_{1213}\e^{a_1a_2a_3a_4a}d_a+\fr{1}{8}D_{1213ab}\e^{aba_1a_2c}d_cd^{a_3a_4}+
\]
\[
\fr{1}{16}D_{1213abcd}\e^{aba_1a_2e}\e^{cda_3a_4f}d_ed_fd_+)+
\]
\[
\lambda^{\a}\lambda^{\b_1}\fr{1}{(\lambda^+)^2}\lambda_{a_1a_2}\lambda_{a_3a_4}\lambda_{a_5a_6}(\fr{1}{32}D_{1213ab}\e^{aba_1a_2c}d_c\e^{a_3a_4a_5a_6d}d_d+
\]
\[
\fr{1}{32}D_{1213abcd}\e^{aba_1a_2e}\e^{cda_3a_4f}d_ed_fd^{a_5a_6})]\q^{\b_2}\cdots \q^{\b_{11}},
\]
all other terms vanish because they contain 6 or more $d_a$'s. The first two terms in the $\lambda^+$ expansion vanish by using \eqref{eq:Lam} and \eqref{eq:trl}. The next term reduces to
\bea
&& d^{12}d_3d_4d_5 \int [d\lambda] \lambda^{\a}\lambda^{\b_1}\fr{1}{\lambda^+}\lambda_{a_1a_2}\lambda_{a_3a_4}
(\fr{1}{8}D_{1213ab}\e^{aba_1a_2c}d_c\d^{a_3}_{[1}\d^{a_4}_{3]}+
\\
&&\fr{1}{16}D_{1213abcd}\e^{aba_1a_2e}\e^{cda_3a_4f}d_ed_fd_+)\Lambda_{\d_1\d_2\d_3}(\e T)^{\d_1\d_2\d_3}_{\b_1\cdots \b_{11}}\q^{\b_2}\cdots \q^{\b_{11}}.
\nonumber \\
&&\qquad \equiv (J_3)^{\a}_{\b_2\cdots \b_{11}}\q^{\b_2}\cdots \q^{\b_{11}}. \nonumber
\eea
To show that this contribution is non-zero, it suffices to prove that one
of its components $(J_3)^{\a}_{\b_2\cdots \b_{11}}$ is non-zero.
We will consider the case,
\be
\a=a_5a_6, \quad \b_2, \ldots, \b_{11}= +,
b_1b_2, \ldots, b_9b_{10},c_1, c_2, c_3, c_4.
\ee
To evaluate the $\lambda$ integral
we use \eqref{eq:k=3}. The first term vanishes after the $\lambda$
integration due to the $d$'s and the second term gives
\[
\fr{129}{2}\fr{1}{16}d^{12}d_3d_4d_5 D_{1213abcd} \e^{aba_1a_2e}\e^{cda_3a_4f}d_ed_fd_+(\e_{a_1a_2a_3a_4b_{14}} \e_{b_{16}a_5a_6b_{11}b_{12}}(\e_{10})^{b_1\cdots b_{20}}
\]
\be \label{Lor_vio}
\e_{c_1c_2c_3c_4b_{17}} \e_{b_{13}b_{15}b_{18}b_{19}b_{20}} +{\rm 2\ perms}).
\ee
Finally, we have the term containing
$\lambda^{\a}\lambda^{\b_1}\lambda_{a_1a_2}\lambda_{a_3a_4}\lambda_{a_5a_6}$. This term however
does not contain a factor of $d_+$, so it cannot interfere with
(\ref{Lor_vio}) (prior to the integration over $d$).

Thus we get a non-vanishing result after integrating over $\lambda,N$,
opposite to expectations. Note that (\ref{Lor_vio}) contains
13 $d$ zero modes. The remaining three $d$ zero modes can
be provided by the vertex operators, so the Lorentz variation does not
vanish in a similar fashion as in the discussion in the previous subsection,
although in principle there may still be a cancellation between
this term and terms originating from the term with
$\lambda^{\a}\lambda^{\b_1}\lambda_{a_1a_2}\lambda_{a_3a_4}\lambda_{a_5a_6}$ in (\ref{eq:lorvarw})
after integrating over $d$.

\subsection{Prescription including an integral over $B$}
At tree level decoupling of unphysical states was restored after integrating over the constant spinors $C$. In this section we analyze whether this is also the case at one loop, namely whether unphysical states decouple after integrating over $C$ and $B$. Similar to the tree-level case we show that all amplitudes are proportional to a certain invariant tensor (at tree level this was $(\e T)$) and amplitudes with $Q$ exact states are proportional to the trace of this invariant tensor. However, at one loop the trace of this tensor does not vanish.

Following the same steps as in the previous subsection (section \ref{sec:nogo} contains details of these steps), one can show that all amplitudes can be written as a sum of terms proportional to the following zero mode integral
\be \label{eq:anawib}
X^{\a_1\cdots \a_{11}}_{\b_1\cdots \b_{11}m_1n_1\cdots m_{10}n_{10}}\equiv \int [dB][dC][d\lambda][dN]\lambda^{\a_1}\cdots \lambda^{\a_{11}}
\ee
\[
B^1_{m_1n_1}\cdots B^{10}_{m_{10}n_{10}}C^1_{\b_1}\cdots  C^{11}_{\b_{11}}\d(C^1\lambda)\cdots \d(C^{11}\lambda)\d(B^1N)\cdots \d(B^{10}N)\d(J).
\]
Proportional here means in the sense of tensor multiplication: in the terms that appear after contractions, the tensor $X$ is multiplied by gamma matrices. Evaluating the integrals in \eqref{eq:anawib} is much easier than one might have anticipated, because we know that $X$ must be an invariant tensor, that is symmetric and gamma matrix traceless in the $\a$'s, antisymmetric in the $\b$'s and antisymmetric in both $m_i \lera n_i$ and $m_in_i\lera m_jn_j$. To find out how many independent invariant tensors with these properties exist, we compute the number of scalars in the relevant tensor product, which is one (see also section \ref{sec:dyl}). As a matter of fact we already know such a nonvanishing tensor:
\be
(\e TR)^{\a_1\cdots \a_{11}}_{\b_1\cdots \b_{11}m_1n_1\cdots m_{10}n_{10}}\equiv (\e T)^{((\a_1\a_2\a_3}_{\b_1\cdots \b_{11}}R^{\a_4\cdots \a_{11}))}_{m_1n_1\cdots m_{10}n_{10}},
\ee
where the double brackets denote gamma matrix traceless, see appendix \ref{sec:detrel}. We stress that Lorentz invariance has completely fixed $X$, there is no freedom remaining.

Starting from a correlator with an unphysical state and integrating $Q$ by parts, it will hit a $\q$ from a PCO (where again we suppress the total derivative in moduli space obtained when $Q$ acts on $\tilde{b}$, which does not play a role here). This means all amplitudes with an unphysical state can be written as a sum of terms proportional to the trace of $(\e TR)$:
\[
\int [dB][dC][d\lambda][dN]\lambda^{\a_2}\cdots \lambda^{\a_{11}}B^1_{m_1n_1}\cdots B^{10}_{m_{10}n_{10}}\lambda^{\b_1}C^1_{\b_1}C^2_{\b_2}\cdots  C^{11}_{\b_{11}}
\]
\be \label{eq:tretr}
\d(C^1\lambda)\cdots \d(C^{11}\lambda)\d(B^1N)\cdots \d(B^{10}N)\d(J)=(\e TR)^{\a_1\cdots \a_{11}}_{\a_1\b_2\cdots \b_{11}m_1n_1\cdots m_{10}n_{10}}.
\ee
There are two independent invariant tensors with indices and symmetries of the trace of $(\e TR)$, so one expects a non-vanishing trace. Indeed, 
it is proven  in section \ref{sec:tretr} that this trace does not vanish, which implies the PCO is not $Q$ closed. One might want to replace $(\e TR)$ by its traceless part to restore $Q$ invariance, but this is not possible since all invariant tensors with the symmetries and indices of $X$ are proportional to $(\e TR)$. In other words removing the trace of $(\e TR)$ would set the entire tensor to zero.

We conclude that
the proof of decoupling of unphysical states at tree level does not generalize to one loop
and one needs a new argument. Such a new argument is presented in \cite{new1},
where it is shown that unphysical states decouple to all loop order.

\subsection{Comparison to non-minimal formalism} \label{sec:nonmin}
In this subsection we briefly compare with the non-minimal formalism \cite{Berkovits:2005bt}. None of the problems that were found, when we examined the prescription without integration over $B$, are present in this case. 

In the non-minimal formalism one introduces a set of non-minimal variables,
the complex conjugate $\bar{\lambda}_\a$ of $\lambda^\a$,
a fermionic constrained spinor $r_\b$ satisfying
\be
\bar{\lambda}_\a \g_m^{\a \b} \bar{\lambda}_\b =0, \qquad \bar{\lambda}_\a \g_m^{\a \b} r_\b =0
\ee
and their conjugate momenta, $\bar{w}^\a$ and $s^a$.
Analogous to the minimal formalism these conditions induce a gauge invariance:
\be
\d\bar{w}^{\a}=\bar{\Lambda}^m(\g_m\bar{\lambda})^{\a}-\f^m(\g_mr)^{\a},\quad \d s^{\a}=\f^m(\g_m\bar{\lambda})^{\a}.
\ee
This implies $\bar{w}^{\a}$ and $s^{\a}$ can only appear in the gauge
invariant quantities
\be
\bar{N}^{mn}=\hp (\bar{\lambda}\g^{mn}\bar{w}-s\g_{mn}r),\quad \bar{J}=\bar{\lambda}\bar{w}-s r,\quad T_{\bar{\lambda}\bar{w}}=\bar{w}^{\a}\del \bar{\lambda}_{\a}-s^{\a}\del r_{\a},
\ee
\[
S_{mn}=\hp s\g_{mn}\bar{\lambda},\quad S=s \bar{\lambda}.
\]
The action (\ref{fl_action}) is modified by the addition of the term $S_{nm}$:
\be \label{nn_ac}
S \to S+ S_{nm}, \qquad S_{nm} = \int d^2 z \left(-\bar{w}^{\a}\bar{\del}\bar{\lambda}_{\a}+s^{\a}\bar{\del}r_{\a}\right)
\ee
and the generator $Q$ by
\be
Q \to Q + \oint dz \bar{w}^\a r_\a.
\ee
This acts on the non-minimal variables as follows
\be \label{qs_nm}
\d \bar{\lambda}_{\a}= r_{\alpha}, \qquad \d r_{\a} =0,\qquad \d s^{\a}= \bar{w}^{\a}, \qquad \d \bar{w}^{\a} = 0.
\ee
These transformation rules imply that the cohomology is independent of
the non-minimal variables. In other words the vertex operators can
always be chosen such that they do not include these variables.

The non-minimal variables can also be understood as originating
from the BRST treatment of the gauge freedom due to
shifts of the zero modes of the worldsheet fields
\cite{Hoogeveen:2007tu}. This also explains why vertex operators do
not depend on the non-minimal fields and why only the zero modes of
these fields appear in the path integral. Furthermore the OPE's given
in section \ref{sec:rev} still comprise a complete list, since the new
fields do not have non-zero modes. Note however that in more recent
work \cite{Aisaka:2009yp} that aims at dealing with divergences
as $\bar{\lambda} \lambda \to 0$, non-zero modes  of $\bar{\lambda}$ do play a role.
It would be interesting to understand how this fits with the discussion
in \cite{Hoogeveen:2007tu}.

In the non-minimal formalism the PCO's are replaced by
\be \label{reg}
\cN= e^{-(\lambda\bar{\lambda}+r\q+\hp N_{mn}\bar{N}^{mn}+\qu S_{mn}\lambda\g^{mn}d+J\bar{J}+\qu S \lambda d)}.
\ee
This is invariant under $Q$:
\be
Q\cN=(\lambda r-\lambda r +\bar{N}^{mn}\hp\lambda\g_{mn}d-\bar{N}^{mn}\hp\lambda\g_{mn}d+\bar{J}(\lambda d)-\bar{J}(\lambda d))\cN=0.
\ee
Thus, all problematic terms of the minimal formalism are manifestly absent here and  BRST exact states decouple. In other words, these amplitudes vanish because two equal terms are subtracted.


\section{A no-go theorem for Lorentz invariant $Q$-closed PCO's} \label{sec:nogo}
This section is a result of an investigation into possibilities of replacing the PCO's by ones that are Lorentz invariant and $Q$ closed. It turns out that any such PCO's would trivialize the entire formalism. More precisely {\it if all formal properties of the picture changing operators were to hold then all one-loop amplitudes would vanish}.

A Lorentz invariant $Q$ closed PCO is defined as an operator $Y$ that satisfies
\begin{itemize}
\item $Y=f_{\b_1\cdots \b_{11}}(\lambda)\q^{\b_1}\cdots \q^{\b_{11}}$,
\item $f_{\b_1\cdots \b_{11}}(\lambda)$ has ghost number $-11$,
\item $f_{\b_1\cdots\b_{11}}(\lambda)$ is a Lorentz tensor,
\item $Q Y =0$.
\end{itemize}
The original proposal in \cite{Berkovits:2004px} is the special case where the function $f$ is given by\footnote{The $C$ integral can be evaluated to give
\be
f_{\b_1\cdots \b_{11}}=(\e T)^{\a\b\g}_{\b_1\cdots \b_{11}}\Lambda_{\a\b\g}.
\ee
}
\be \label{eq:ex}
f_{\b_1\cdots \b_{11}}=\int [dC]
C^1_{[\b_1}\cdots C^{11}_{\b_{11}]}\d(C^1\lambda)\cdots \d(C^{11}\lambda).
\ee
This satisfies the first three conditions, but although 
$Q Y \sim \lambda \delta(\lambda)$ the fourth bullet does not hold for \eqref{eq:ex},
as we have seen.

Using the fact that $f$ is a Lorentz tensor one finds,
\bea
&&\int [dB][d\lambda][dN]\lambda^{\a_1}\cdots \lambda^{\a_{11}}B^1_{m_1n_1}\cdots B^{10}_{m_{10}n_{10}}f_{\b_1\cdots \b_{11}}(\lambda)\d(B^1N)\cdots \d(B^{10}N)\d(J)=
\nonumber \\
&& \qquad \qquad \label{qint}
c_1(\e TR)^{\a_1\cdots \a_{11}}_{\b_1\cdots \b_{11}m_1n_1\cdots m_{10}n_{10}},
\eea
for some $c_1$. This follows from the fact that $(\e TR)$ is the
unique Lorentz tensor with the indicated tensor structure. Now the
crucial observation is that for functions $f$ such that $QY=0$ the
integral (\ref{qint}) must be equal to zero. Indeed, using
\be
0=QY=f_{\b_1\cdots \b_{11}}(\lambda)\lambda^{\b_1}\q^{\b_2}\cdots \q^{\b_{11}}.
\ee
we compute
\bea
0&& =\int [dB][d\lambda][dN]\lambda^{\a_2}\cdots \lambda^{\a_{11}}
B^1_{m_1n_1}\cdots B^{10}_{m_{10}n_{10}}
\left(f_{\b_1\cdots \b_{11}} \lambda^{\b_1} \q^{\b_2}\cdots \q^{\b_{11}}\right)
\nn
\\ \label{eq:ght}
&& \d(B^1N)\cdots \d(B^{10}N)\d(J)=
c_1(\e TR)^{\a_1\cdots \a_{11}}_{\a_1\b_2\cdots \b_{11}m_1n_1\cdots m_{10}n_{10}} \q^{\b_2}\cdots \q^{\b_{11}}. \label{qy}
\eea
We will show shortly that the trace of $(\e TR)$ does not vanish,
so we conclude that
\be
c_1=0.
\ee

To prove that this implies vanishing of all one-loop amplitudes the
above result is not enough, because there are also zero mode integrals
with derivatives on the delta functions and $N$ insertions. After the
non-zero mode integration is performed,
an arbitrary amplitude is reduced to a sum
of zero mode integrals, all of which are of the form
\be \label{eq:intbzm}
\mathcal{E}^{\a_1\cdots\a_{11}p_1q_1\cdots p_{L}q_{L}}_{\b_1\cdots\b_{11}m_1n_1\cdots m_{10}n_{10}r_1s_1\cdots r_{L}s_{L}}=
\ee
\[
\int [dB][dN] [d\lambda] \prod_{j=1}^L N^{p_{j}q_{j}}\prod_{i_1=1}^{L_1}B^1_{r_{i_1}s_{i_1}}\prod_{i_2=L_1+1}^{L_1+L_2} B^2_{r_{i_2}s_{i_2}}\cdots \prod_{i_{10}=L_1+\cdots+L_9+ 1}^{L} B^{10}_{r_{i_{10}}s_{i_{10}}}
\]
\[
\lambda^{\a_1}\cdots \l^{\a_{11}}f_{\b_1\cdots \b_{11}}(\l)B^1_{m_1n_1}\cdots B^{10}_{m_{10}n_{10}}
\d^{(L_1)}(B^1N)\cdots \d^{(L_{10})}(B^{10}N)\d(J),
\]
where all the fields are zero modes and $L=\sum_{P=1}^{10} L_P$ and
$\d^{(m)}(x)$ denotes the $m$-th derivative of $\d(x)$.
In the previous section we saw all zero mode integrands had to be of the form \eqref{eq:fb}, \eqref{eq:hb} for a non vanishing answer. In writing down the above zero mode integrand we started from $f_B,h_B$ and used the following four arguments.
\begin{itemize}
\item For each $P$ the total number of $B^P$'s outside the delta functions
  is equal to the number of derivatives on $\d(B^PN)$ plus one. This
  can be inferred from the explicit form of the $b$ ghost,
  \eqref{eq:bghost}, and the Taylor expansion of the delta functions.
This is reflected in \eqref{eq:intbzm} because
  $L_P$ appears in two places.

\item For a non-zero answer the total number of $N$ zero modes must equal the total number of derivatives on the delta functions. This gives the restriction $L=\sum L_P$.
\item One might have expected derivatives on $\d(J)$ as well, but for
  a non vanishing answer there must also be enough $J$ zero modes,
so one can always reduce the amplitude to contain only $\d(J)$.
\item Compared to \eqref{eq:fb} the $\lambda$ dependence is less general. It is possible to restrict to this class of integrands because
$f_{\b_1\cdots \b_{11}}(\lambda)$ is a Lorentz tensor. To see this note the OPE's of $N$ and $J$ with $f$ do not introduce derivatives:
\bea
N^{mn}(z) f_{\b_1\cdots \b_{11}}(\lambda(w))&\sim& \sum_{i=1}^{11} (\g^{mn})_{\b_i}^{\a}f_{\b_1\cdots \a\cdots \b_{11}}(\lambda(w))\fr{1}{z-w},
\\
J(z)f_{\b_1\cdots \b_{11}}(\lambda(w))&\sim& -11 f_{\b_1\cdots \b_{11}}(\lambda(w))\fr{1}{z-w},
\eea
where the $\a$ index is in the $i^{\rm th}$ position.
\end{itemize}
Note that the free indices on $\cE$ can be either contracted among each other or with $d$ or $\q$ zero modes. The integral in \eqref{eq:intbzm} can be evaluated by using the definition of $B$ integration in \eqref{eq:bint}. Let us call the integrand of \eqref{eq:intbzm} $g$ and write it as
\be
g(\lambda,N,J,B^P)=\lambda^{\a_1}\cdots \lambda^{\a_{11}}h_{\a_1\cdots \a_{11}}^{\b_1\cdots \b_{11}}(N,J,B^P)\prod_{P=1}^{10}\d^{(L_P)}(B^PN)f_{\b_1\cdots \b_{11}}(\lambda),
\ee
where $h$ is a polynomial depending on $(N,J,B)$ as
\be
(N)^{L}\prod_{P=1}^{10}(B^P)^{L_P+1}.
\ee
It also contains other fields (e.g.~$\q,d$) but these are suppressed.

The integrations can be performed using (\ref{eq:bint}):
\be \label{eq:defbint}
\int [dB] [d\lambda][dN]g(\lambda,N,J,B^I)\equiv \fr{\del}{\del\lambda^{\a_1}}\cdots\fr{\del}{\del\lambda^{\a_{11}}} (\e TR)^{\a_1\cdots \a_{11}}_{\b_1\cdots \b_{11}m_1n_1\cdots m_{10}n_{10}}
\ee
\[
\fr{\del}{\del B^1_{m_1n_1}}\cdots \fr{\del}{\del B^{10}_{m_{10}n_{10}}}\prod_{P=1}^{10}(\fr{\del}{\del B^P_{pq}}\fr{\del}{\del N^{pq}})^{L_P}\lambda^{\g_1}\cdots \lambda^{\g_{11}}h^{\b_1\cdots \b_{11}}_{\g_1\cdots \g_{11}}(\lambda,N,J,B^P)=
\]
\[
(\e TR)^{\a_1\cdots \a_{11}}_{\b_1\cdots \b_{11}m_1n_1\cdots m_{10}n_{10}}\fr{\del}{\del B^1_{m_1n_1}}\cdots \fr{\del}{\del B^{10}_{m_{10}n_{10}}} \prod_{P=1}^{10}(\fr{\del}{\del B^P_{pq}}\fr{\del}{\del N^{pq}})^{L_P}h^{\b_1\cdots \b_{11}}_{\a_1\cdots \a_{11}}(\lambda,N,J,B^P)
\]
This reduces to \eqref{eq:bint} with $K_I=0$ if one chooses $f_{\b_1\cdots \b_{11}}(\lambda)$ as in \eqref{eq:ex} and uses
\be
h^{\b_1\cdots \b_{11}}_{\a_1\cdots \a_{11}}=\fr{\del}{\del C^1_{\b_1}} \cdots \fr{\del}{\del C^{11}_{\b_{11}}}(h_B)_{\a_1\cdots \a_{11}}.
\ee

Using the above definition the integral in \eqref{eq:intbzm} can be evaluated as
\[
\mathcal{E}^{\a_1\cdots\a_{11}p_1q_1\cdots p_{L}q_{L}}_{\b_1\cdots\b_{11}m_1n_1\cdots m_{10}n_{10}r_1s_1\cdots r_{L}s_{L}}=c_{L_1\cdots L_{10}}\d^{([p_1}_{r_1}\d^{q_1]}_{s_1}\cdots \d^{[p_{L}}_{r_{L}}\d^{q_{L}])}_{s_{L}}(\e TR)^{\a_1\cdots \a_{11}}_{\b_1\cdots \b_{11}m_1n_1\cdots m_{10}n_{10}}
\]
\be \label{eq:intbzmans}
 +{\rm symmetrization\ in}([r_{L_{P-1}+1},s_{L_{P-1}+1}],\ldots, [r_{L_P},s_{L_P}],[m_Pn_P]),
\ee
for some constant $c_{L_1\cdots L_{10}}$. Note the round brackets denote symmetrization in
\be
[p_1q_1],\ldots,[p_Lq_L].
\ee
Also note the second line above includes ten symmetrizations, one for
each $P$. $\mathcal{E}$ is symmetric in these indices because they all
appear on $B^I$. (Note that by definition $L_0=0$). To get some insight how to obtain \eqref{eq:intbzmans} consider the case $L_1=L=1$. In that case the rhs of \eqref{eq:intbzm} is given by
\be
(\e TR)_{m'_1n'_1\cdots m'_{10}n'_{10}} \fr{\del}{\del B^1_{p'q'}}\fr{\del}{\del N^{p'q'}}\fr{\del}{\del B^1_{m'_1n'_1}}\cdots \fr{\del}{\del B^{10}_{m'_{10}n'_{10}}}N^{pq} B^1_{r_1s_1}B^1_{m_1n_1}\cdots B^{10}_{m_{10}n_{10}},
\ee
where the spinor indices on $(\e TR)$ are suppressed. The last nine $B$ differentiations are trivial resulting in:
\be
(\e TR)_{m'_1n'_1m_2n_2\cdots m_{10}n_{10}} \fr{\del}{\del B^1_{p'q'}}\fr{\del}{\del N^{p'q'}}\fr{\del}{\del B^1_{m'_1n'_1}}N^{pq} B^1_{r_1s_1}B^1_{m_1n_1}.
\ee
Now we first perform the $N$ differentiation followed by the last two $B$ differentiations:
\[
(\e TR)_{m'_1n'_1m_2n_2\cdots m_{10}n_{10}} \fr{\del}{\del B^1_{pq}}\fr{\del}{\del B^1_{m'_1n'_1}} B^1_{r_1s_1}B^1_{m_1n_1}=(\e TR)_{m'_1n'_1m_2n_2\cdots m_{10}n_{10}} \d^{([p}_{r_1}\d^{q]}_{s_1}\d^{[m'_1}_{tm_1}\d^{n'_1])}_{n_1}=
\]
\be
\d^{[p}_{r_1}\d^{q]}_{s_1}(\e TR)_{m_1n_1\cdots m_{10}n_{10}}+(r_1s_1 \lera m_1n_1),
\ee
which agrees with \eqref{eq:intbzmans}. The above computation clarifies the appearance of the Kronecker delta's. It is a consequence of the fact $\fr{\del}{\del B^1_{pq}}$ and $\fr{\del}{\del N^{pq}}$ appear contracted. The symmetrizations in \eqref{eq:intbzmans} follow from the product rule of differentiation.

With these preliminaries we are ready to prove that if $Q Y =0$ then
all one-loop amplitudes vanish:
\paragraph{No go theorem}
\bea 
\label{eq:th2}
QY=0 \Longrightarrow c_{D_1\cdots D_{10}}=0,
\\ \label{eq:th3}
c_{D_1\cdots D_{10}}=0\Longrightarrow {\rm all\ one\ loop\ amplitudes\ vanish},
\eea
{\it Proof of \eqref{eq:th2}.} In terms of $f$ the condition on the LHS of \eqref{eq:th2} reads
\be
0=QY=f_{\b_1\cdots \b_{11}}(\lambda)\lambda^{\b_1}\q^{\b_2}\cdots \q^{\b_{11}}.
\ee
This implies
\be \label{eq:nztrddetr}
0=\mathcal{E}^{\a_1\cdots\a_{11}p_1q_1\cdots p_{L}q_{L}}_{\a_1\b_2\cdots\b_{11}m_1n_1\cdots m_{10}n_{10}r_1s_1\cdots r_{L_1}s_{L_1}}=
\ee
\[
c_{L_1\cdots L_{10}}\d^{([p_1}_{r_1}\d^{q_1]}_{s_1}\cdots \d^{[p_{L}}_{r_{L}}\d^{q_{L}])}_{s_{L}}(\e TR)^{\a_1\cdots \a_{11}}_{\a_1\b_2\cdots \b_{11}m_1n_1\cdots m_{10}n_{10}}
\]
\[
 +{\rm symmetrization\ in}([r_{L_{P-1}+1},s_{L_{P-1}+1}],\ldots, [r_{L_P},s_{L_P}],[m_Pn_P]),
\]
As we discuss below the trace $\tr (\e TR)$
of $(\e TR)$ does not vanish, so in particular $\tr (\e TR)$
has at least one non vanishing component.
Let us denote this index choice by hats. If one chooses
\bea
r_is_i=\hat{m}_P\hat{n}_P,\quad i=L_{P-1}+1,\cdots, L_P,
\\
p_iq_i=\hat{m}_P\hat{n}_P,\quad i=L_{P-1}+1,\cdots, L_P,
\eea
the tensor on the RHS of \eqref{eq:nztrddetr} is non vanishing. Therefore
\be
c_{L_1\cdots L_{10}}=0.
\ee
{\it Proof of \eqref{eq:th3}.} As explained around \eqref{eq:intbzm} all amplitudes can be written as a sum of terms, where all terms contain a $c_{L_1\cdots L_{10}}$.

\subsection{Non vanishing of the trace of $(\e TR)$} \label{sec:tretr}
In this subsection we compute the trace  $\tr (\e TR)$ of the tensor
$(\e TR)$.
To show that this trace does not vanish
we define a tensor $Y$ and an operator $\hat{X}$:
\bea
Y_{m_1\cdots n_{10}}&\equiv&
\bar{\lambda}_{\a_4}\cdots \bar{\lambda}_{\a_{11}}R^{\a_4\cdots \a_{11}}_{m_1\cdots n_{10}},
\\
\hat{X}&\equiv&\y_{\b_{12}}\cdots \y_{\b_{16}}\bar{\lambda}_{\a_1}\cdots \bar{\lambda}_{\a_{3}}T^{\b_{12}\cdots \b_{16},\a_1\a_2\a_3}\y_{\a}\fr{\del}{\del \bar{\lambda}_{\a}},
\eea
where $\y_{\a}$ is a fermionic Weyl spinor and $\bar{\lambda}_{\a}$ is a
pure spinor of opposite chirality to $\lambda^{\a}$. Note that, because
$\bar \lambda_\a$ is a contrained spinor,
$\fr{\del}{\del \bar{\lambda}_{\a}}$ is only defined
up to a gauge transformation 
\be
\d \fr{\del}{\del \bar{\lambda}_{\a}}=A^m(\g_m\bar{\lambda})^{\a}.
\ee
The operator $\hat{X}$, however, is well defined, since it is gauge invariant. This follows from
\be \label{eq:ljed}
\bar{\lambda}\g^q\y\y_{\b_{12}}\cdots \y_{\b_{16}}\bar{\lambda}_{\a_1}\cdots \bar{\lambda}_{\a_{3}}T^{\b_{12}\cdots \b_{16},\a_1\a_2\a_3}=0.
\ee
That can be shown be noting there are no scalars in ${\rm Asym}^{6} {\bf 16'} \otimes {\bf 10} \otimes {\rm Gam}^{4}{\bf 16'}$, where Gam means the symmetric and gamma matrix traceless tensor product. Note we can use
\be
\fr{\del}{\del \bar{\lambda}_{\a}}\bar{\lambda}_{\b}=\d^{\a}_{\b}
\ee
when $\fr{\del}{\del \bar{\lambda}_{\a}}$ is part of a gauge invariant quantity, $S_{\a}\fr{\del}{\del \bar{\lambda}_{\a}}$, because
\be
S_{\a}\fr{\del}{\del \bar{\lambda}_{\a}} \bar{\lambda}\g^m\bar{\lambda}=S\g^m\bar{\lambda}=0,
\ee
the last equality is a consequence of gauge invariance.

First we show that $\hat{X}Y \neq 0$. We finish the argument by proving this implies the trace of $(\e TR)$ does not vanish. Consider the following component of $\hat {X}Y$ in a Lorentz frame in which the only non-zero component of $\bar{\lambda}$ is $\bar{\lambda}_+$:
\bea
&&\hat{X} Y_{a_1b_1a_2b_2\cdots a_{10}b_{10}}=(\bar{\lambda}\g_m\y)(\bar{\lambda}\g_n\y)(\bar{\lambda}\g_p\y)(\y\g^{mnp}\y)
\\ \nn
&&[2(\y\g_{a_1b_1a_2a_3a_4}\bar{\lambda})(\bar{\lambda}\g_{a_5b_5b_2a_6a_7}\bar{\lambda})(\bar{\lambda}\g_{a_8b_8b_3b_6a_9}\bar{\lambda})(\bar{\lambda}\g_{a_{10}b_{10}b_4b_7b_9}\bar{\lambda})
\\ \nn
&&2(\bar{\lambda}\g_{a_1b_1a_2a_3a_4}\bar{\lambda})(\y\g_{a_5b_5b_2a_6a_7}\bar{\lambda})(\bar{\lambda}\g_{a_8b_8b_3b_6a_9}\bar{\lambda})(\bar{\lambda}\g_{a_{10}b_{10}b_4b_7b_9}\bar{\lambda})
\\ \nn
&&2(\bar{\lambda}\g_{a_1b_1a_2a_3a_4}\bar{\lambda})(\bar{\lambda}\g_{a_5b_5b_2a_6a_7}\bar{\lambda})(\y\g_{a_8b_8b_3b_6a_9}\bar{\lambda})(\bar{\lambda}\g_{a_{10}b_{10}b_4b_7b_9}\bar{\lambda})
\\ \nn
&&2(\bar{\lambda}\g_{a_1b_1a_2a_3a_4}\bar{\lambda})(\bar{\lambda}\g_{a_5b_5b_2a_6a_7}\bar{\lambda})(\bar{\lambda}\g_{a_8b_8b_3b_6a_9}\bar{\lambda})(\y\g_{a_{10}b_{10}b_4b_7b_9}\bar{\lambda}) + {\rm{permutations}}],
\eea
where the permutations make the RHS antisymmetric in $a_ib_i\lera a_jb_j$. This reduces, up to an overall constant which is not zero\footnote{We omitted constants in the following two relations:
\be
(\bar{\lambda}\g_m\y)(\bar{\lambda}\g_n\y)(\bar{\lambda}\g_p\y)(\y\g^{mnp}\y) \propto \e^{c_1\cdots c_5}\y_{c_1}\cdots \y_{c_5}(\bar{\lambda}_+)^{3},
\ee
\be
(\g^{++}_{a_1b_1a_2a_3a_4}\g^{++}_{a_5b_5b_2a_6a_7}\g^{++}_{a_8b_8b_3b_6a_9} \g^{++}_{a_{10}b_{10}b_4b_7b_9} +{\rm{permutations}})\propto (\e_{10})_{a_1b_1\cdots a_{10}b_{10}}.
\ee
}, to
\be
\hat{X} Y_{a_1b_1a_2b_2\cdots a_{10}b_{10}}=\e^{c_1\cdots c_5}\y_{c_1}\cdots \y_{c_5}(\bar{\lambda}_+)^{10}\y_+\g^{++}_{a_1b_1a_2a_3a_4}\g^{++}_{a_5b_5b_2a_6a_7}
\ee
\[
\g^{++}_{a_8b_8b_3b_6a_9}\g^{++}_{a_{10}b_{10}b_4b_7b_9}+{\rm{permutations}}=
\e^{c_1\cdots c_5}\y_{c_1}\cdots \y_{c_5}(\bar{\lambda}_+)^{10}\y_+(\e_{10})_{a_1\cdots b_{10}}\neq 0.
\]
What remains is to show the non vanishing of this tensor implies the non vanishing of the trace of $(\e TR)$.
\be
\hat{X}Y_{m_1n_1\cdots m_{10}n_{10}}=\e^{\b_1\cdots \b_{16}}[(\e T)^{((\a_1\a_2\a_3}_{\b_1\cdots \b_{11}}\y_{\a_{11}}\y_{\b_{12}}\cdots \y_{\b_{16}}]R^{\a_4\cdots \a_{11}))}_{m_1n_1\cdots m_{10}n_{10}}\bar{\lambda}_{\a_1}\cdots \bar{\lambda}_{\a_{10}}.
\ee
For the term in the square brackets we can move the $\a_{11}$ to $(\e T)$ by using
\be
0=(\e T)^{\a_1\a_2\a_3}_{[\b_1\cdots \b_{11}}\y_{\b_{12}}\cdots \y_{\b_{16}}\y_{\a_{11}]}=
\ee
\[
6(\e T)^{\a_1\a_2\a_3}_{[\b_1\cdots \b_{11}}\y_{\b_{12}}\cdots \y_{\b_{16}]}\y_{\a_{11}}+11(\e T)^{\a_1\a_2\a_3}_{\a_{11}[\b_1\cdots \b_{10}}\y_{\b_{11}}\cdots \y_{\b_{16}]}.
\]
The first line is zero because we are fully antisymmetrizing seventeen indices that only take sixteen values.
\be
\hat{X}Y_{m_1n_1\cdots m_{10}n_{10}}=\e^{\b_1\cdots \b_{16}}[(\e T)^{((\a_1\a_2\a_3}_{\a_{11}\b_1\cdots \b_{10}}\y_{\b_{11}}\cdots \y_{\b_{16}}]R^{\a_4\cdots \a_{11}))}_{m_1n_1\cdots m_{10}n_{10}}\bar{\lambda}_{\a_1}\cdots \bar{\lambda}_{\a_{10}}.
\ee
Since $(\e TR)^{\a_1\cdots \a_{11}}_{\a_{11}\b_2\cdots \b_{11}m_1n_1\cdots m_{10}n_{10}}$ is fully antisymmetric in $\b_2\cdots \b_{11}$ and symmetric and gamma matrix traceless in $\a_1\cdots \a_{10}$, we can conclude from the non vanishing of $\hat{X} Y$ that
\be
(\e T)^{((\a_1\a_2\a_3}_{\a_{11}\b_1\cdots \b_{10}}R^{\a_4\cdots \a_{11}))}_{m_1n_1\cdots m_{10}n_{10}} \neq 0.
\ee

\section{Origin of the problems and possible resolutions} \label{sec:orig}
To understand the origin of the problems encountered when one does not integrate over $C$ (and $B$), we go back to the first principles derivation of the amplitude prescription in \cite{Hoogeveen:2007tu}. We will see that there is a singular gauge choice implicit in the prescription of \cite{Berkovits:2004px}.


\subsection{Derivation of the amplitude prescriptions}
In \cite{Hoogeveen:2007tu} the minimal and non-minimal amplitude prescriptions were derived by coupling the pure spinor sigma model to topological gravity and then proceeding to BRST quantize this system. Following \cite{Craps:2005wk}, the BRST treatment included the gauge invariance due to zero modes. The singular gauge fixing refers to the gauge fixing of the invariance due to pure spinor zero modes.

The BRST quantization led to the following generating functional of scattering
amplitudes
\be \label{gen}
Z[\r^i] =
\int d \mu_\s d \mu \exp \left(- {\cal S} - L_1 -  L_2-L_3\right),
\ee
where $\r^i$ are sources that couple to vertex operators, $d \mu_\s$ is the path integral
measure for the original sigma model fields, $d \mu$ is the path integral measure for the fields introduced
in the BRST quantization procedure and ${\cal S}$ is the worldsheet action with two dimensional coordinate invariance. $L_1$
contains the gauge fixing
terms due to the diffeomorphism and Weyl symmetry and $L_2$ contains the gauge fixing terms for the
invariances due to the zero modes of the ghost fields. In the case of the bosonic string \cite{Craps:2005wk}, $L_1$ leads to the usual ghost action
and $L_2$ to the usual ghost and antighost insertions in the path integral. In our case these
contributions cancel out. To understand the cancellations recall that the pure spinor sigma model has a fermionic nilpotent symmetry generated by $Q$, the pure spinor BRST operator.
After coupling to topological gravity and gauge fixing all
symmetries, there is a second nilpotent operator $Q_V$, the standard BRST operator related with gauge
fixing local symmetries.
$Q_V$ in particular contains the standard terms related to diffeomorphisms and Weyl transformations
and it also has terms related to the invariance due to zero modes of the worldsheet fields.
Since we want to keep the $Q$ symmetry manifest, all fields are introduced in $Q$-pairs. In particular,
together with the $b,c$ ghosts we also introduce their $Q$ partners, $\b,\g$. These have opposite statistics and it turns out these fields can be integrated out and the $b,c$ part cancels against the $\b,\g$ part.
Even though all terms related to gauge fixing of worldsheet diffeomorphism cancel out, this procedure
explains why the pure spinor amplitude prescription is so similar to the bosonic string amplitude prescription.

The main object of interest here is $L_3$, which
is related to gauge fixing of the invariance due to the zero modes of $\lambda^\a$ and its conjugate momentum
$w_\a.$ The part relevant to the $\lambda$ zero modes is given by
\be \label{l3}
L_3 = Q_V Q (b_\a \theta^\a )= Q_V (-b_\a \lambda^\a + \tilde{b}_\a \theta^\a)= \pi_\a \lambda^\a + \tilde{\pi}_\a \theta^\a
+b_\a c^\a + \tilde{b}_\a \g^\a.
\ee
The field $\pi_\a$ is the BRST auxiliary field that enforces the gauge fixing condition
for the invariance due to zero modes of $\lambda$. Since there are eleven zero modes we need
eleven gauge fixing conditions and the BRST auxiliary field $\pi_\a$ must contain only eleven independent components.
The gauge condition
implicit in  (\ref{l3}) will be discussed shortly.
$(b_\a,c^\a)$ and their Q-partners $(\tilde{b}_{\a},\g^{a})$ are the corresponding BRST ghosts.
These fields can be integrated out and cancel each other. Then we are left with
\be \label{eq:L3'}
L'_3= \pi_\a \lambda^\a + \tilde{\pi}_\a \theta^\a.
\ee

\paragraph{Minimal formalism.}
To express the fact that $\pi_{\a}$ and $\tilde{\pi}_{\a}$ have eleven independent components
we parametrize them as follows
\be  \label{piI}
\pi_\a = p_I C^I_\a, \qquad \tilde{\pi}_\a = \tilde{p}_I C^I_\a, \quad I=1, \ldots, 11.
\ee
where $C^I_\a$ is a matrix that must have maximal rank. Thus the gauge fixing condition is given by
\be \label{gaugechoice}
C^I_{\a} \lambda^{a} =0
\ee
We will shortly show that this is a singular gauge condition.

The eleven constant spinors $C^I_\a$
are the ones that enter in the minimal pure spinor prescription. Indeed,
using (\ref{piI}) we find that the path integral contains
\be \label{PCO}
\int [d p_I] [d \tilde{p}_I] \exp \left(p_I C^I_{\a} \lambda^{a}+\tilde{p}_I C^I_{\a}\q^{\a}\right)
= \prod_{I=1}^{11} (C^I_{\a}\q^{\a}) \d (C^I_{\a}\lambda^{\a})
\ee
which are the eleven picture changing operators $Y_C$ we discussed earlier.

Implicit in \eqref{PCO} there is an analytic continuation in the field variables.
Recall that the solution to the pure spinor constraint \eqref{eq:pscond} requires that $\lambda$ is
complex and in the minimal formulation only the holomorphic part appears. In equation (\ref{PCO})
one analytically continues $\lambda$ to be real and considers $\pi_I$ to be purely imaginary.
This can be done if the explicit expressions appearing in the amplitude
computations are not singular. Typical integrals in the minimal formalism
at tree level are of the form
\be \label{minint}
\int_{-i\infty}^{i\infty}[dp]\int_{-\infty}^{\infty}[d\lambda] f(\lambda) e^{p_I C^I_{\a} \lambda^{a}}=\int_{-\infty}^{\infty}[d\lambda] f(\lambda)\d(C^1\lambda)\cdots \d(C^{11}\lambda).
\ee
where $f(\lambda)$ contains $\lambda$ but not its complex conjugate. For this expression to be well-defined
$f(\lambda)$ should not contain any $(C^I \lambda)$ poles and moreover there should not be any poles
that obstruct the analytic continuation of $\lambda$ to real values.

At higher loops the conjugate momentum has zero modes as well and gauge fixing this invariance
leads exactly to the insertion of PCO's $Z_B, Z_J$, where the tensors $B_{mn}$ enter through the gauge fixing
condition, see \cite{Hoogeveen:2007tu} for the details. In addition, one needs a composite $b$ field satisfying \eqref{eq:qbt}.
In the minimal formulation, a solution of (\ref{eq:qbt}) is given by \cite{Berkovits:2001us}
\be \label{bmin}
b = \frac{\lambda^\a G^\a}{C_\a \lambda^\a}
\ee
where $G^\a$ is given in \eqref{eq:G}. This is however too
singular to be acceptable. One can obtain a non-singular $\tilde{b}$ field
by combining the $b$ field with the PCO and solving instead \eqref{eq:qbbuz}. Note that
this $\tilde{b}$ field now depends on the $B_{mn}$ constant tensors but not on $C_\a$.

\paragraph{Non-minimal formalism.}
We now show that the same  expression \eqref{eq:L3'} leads to the so-called regularization factor
in  \eqref{reg}.
This time we choose $\pi_{\a}$ to be a pure spinor of opposite chirality to $\lambda^{\a}$, usually called $\bar{\lambda}_{\a}$. This indeed has 11 independent components, as required.
The field $\tilde{\pi}_{\a}$, usually called $r_\a$,
 automatically follows because it is the $Q$ variation of $\pi_{\a}$,
\be
r_{\a}=Q\bar{\lambda}_{\a}.
\ee
This leads to the non-minimal formalism. To see this explicitly note that the factor $e^{-L_3}$ in \eqref{gen}, which is given by
\be
e^{-\bar{\lambda}_{\a}\lambda^{\a}-r_{\a}\q^{\a}},
\ee
is precisely $\cN$. The additional factors $N_{mn}\bar{N}^{mn}+\qu S_{mn}\lambda\g^{mn}d+J\bar{J}+\qu S \lambda d$
originate from gauge fixing the zero modes of $w_\a$, see \cite{Hoogeveen:2007tu} for the details.

Note that $\lambda$ is now holomorphic and $\pi_\a\equiv \bar{\lambda}_\a$ is considered as its complex conjugate variable.
Typical integrals one encounters at tree level in the non-minimal formalism are therefore
\be \label{nonminint}
\int [d\lambda][d\bar{\lambda}]f(\lambda)e^{-\bar{\lambda}\lambda}.
\ee

At higher loop order we also need the $b$ field. In the non-minimal formalism, equation (\ref{eq:qbt})
has a solution that depends on both $\lambda$ and $\bar{\lambda}$. It is however singular as $\bar{\lambda} \lambda \to 0$
and this causes problems starting from three loops. Note that the $b$ field does not depend on how we treat the gauge
invariances due to the zero modes of $w_\a$. This is similar to the $b$ field in \eqref{bmin} but
different than $\tilde{b}$ which depends on the gauge fixing of the invariance due to zero modes of
the conjugate momentum through $B_{mn}$.

To summarize, the minimal and non-minimal are related by field redefinitions and an analytic continuation in field space. In particular, starting from the non-minimal formalism
one obtains the minimal formalism by taking $\bar{\lambda}_\a = C_\a^I \pi^I$ and analytically continuing
$\pi^I$ to be imaginary while at the same time analytically continuing $\lambda$ to be real.
There are similar redefinitions and
analytic continuations in the sector related with the conjugate momentum.
Furthermore, the non-minimal $b$ field combined with part of $\cN$ is related to $\tilde{b}$.
Clearly, the two formalisms would be equivalent if the analytic continuations had
not been obstructed by singularities in the amplitudes.
Finally, note that the underlying gauge choice for the invariance due to pure spinor
zero modes is the same: the gauge fixed action is the same, only the reality condition of the fields
is different.

\subsection{Toy example}

Given the formal equivalence between  the minimal and non-minimal formalisms one may wonder why we found problems at one loop in the
one formalism but not the other. We discuss this issue here by analyzing a toy example that has almost
all features of the actual case.
Consider the following integral
\be
I=\int dx dp e^{-xp}.
\ee
To compare with the expressions in the previous subsection $p$ corresponds to the BRST auxiliary field
and $x$ to the pure spinor.

If one wants to evaluate the above integral, contours have to be chosen for $x$ and $p$. If we choose $p=ip_1$ and $x=x_1$ with $p_1,x_1$ real, we get
\be \label{eq:conmin}
I=i \int dx_1dp_1 e^{ix_1p_1}=i \int_{-\infty}^{\infty}dx_1 2\pi \d(x_1)=2\pi i.
\ee
Another choice is to consider $x$ complex and take $p=x^*$. In this case $I$ becomes
\be \label{eq:connm}
I = \int dxdx^* e^{-xx^*}= 2 i \int_0^\infty r dr \int_0^{2 \pi} d\q e^{-r^2}=2\pi i.
\ee
This agrees nicely with the general property of contour integrals, that one is free to deform them as long as no poles are encountered. Note that \eqref{eq:conmin} resembles a zero mode integral in the minimal formalism and \eqref{eq:connm} a non-minimal one.

The difference between the two prescriptions is exposed by considering the integral $I$ with a function $f$ in the integrand.
\be \label{minimal}
I_{{\rm min}}[f]=i \int_{-\infty}^{\infty} dx_1 \int_{-\infty}^\infty dp_2  e^{i x_1 p_1} f(x_1) = i \int_{-\infty}^{\infty} dx_1 2\pi \d(x_1) f(x_1) = 2 \pi i f(0).
\ee
Now rotate the contour, $p =x^*$, so that the integral becomes
\be \label{nonm}
I_{{\rm non-min}}[f]=\int dx dx^* e^{-|x|^2} f(x) =  2 i \int_0^\infty r d r e^{-r^2}
\int_0^{2 \pi} d \theta  f(r e^{i \theta}),
\ee
$I_{{\rm min}}$ is the analogue of (\ref{minint}) and $I_{{\rm non-min}}$ the analogue of (\ref{nonminint}).
$I_{{\rm min}}$ and $I_{{\rm non-min}}$ give exactly the same answer if $f(x)$ is non singular but (\ref{minimal}) is ill defined for any choice of singular $f(x)$ whereas (\ref{nonm}) may be well defined. For example, for the function
\be
f(x) = \frac{1}{x},
\ee
(\ref{minimal}) yields $\infty$ but (\ref{nonm}) gives 0. More precisely, \eqref{nonm} is well defined for all functions $f(z)=\sum_n c_nz^n$, with $c_n=0$ for $n<-1$. For the $n<-1$ terms the $\q$ integral vanishes and the $r$ integral diverges, which makes $I_{\rm non-min}$ ambiguous for these kind of functions.

A third representation is obtained by noticing that the $\theta$ integral can be rewritten as a contour
integral
\be
\int_0^{2 \pi} d \theta = -i \oint_C \frac{dz}{z}
\ee
where $z=r e^{i \theta}$ and the contour $C$ is a circle of radius $r$. Thus for any meromorphic
function $f(z)$ the integral over theta is independent of $r$ and
\be \label{newmin}
I[f] = 2 i \left(\int_0^\infty r d r e^{-r^2}\right)\left( -i \oint_C \frac{dz}{z} f(z) \right)
= \oint_C \frac{dz}{z} f(z)
\ee
The expression (\ref{newmin}) are well-defined for all meromorphic functions $f(z)$
whereas (\ref{minimal}) and (\ref{nonm}) are not.

Going back to pure spinors and working on the patch with $\lambda^+ \neq 0$ we see that because of the factor
$(\lambda^+)^{-3}$ in the measure (cf. \eqref{eq:measlc}) the minimal formalism
is expected to have a singularity
unless the integrand provides a factor of $(\lambda^+)^3$,
but the expressions (\ref{nonm}) and
(\ref{newmin}) are not necessarily singular.

\subsection{Singular gauge and possible resolution}

We show in this subsection that the gauge (\ref{gaugechoice}) is singular for any
choice of the constant spinors $C^I_\a$. To see this, recall that the space of pure spinors
can be covered with 16 coordinate patches and on each patch at least one of the components of $\lambda^\a$
is non-zero. Let us call this component $\lambda^+$ and solve the
pure spinor condition as in (\ref{solv_pur}). Then,
\[
0=C^I_{\a}\lambda^{\a}=C^I_+\lambda^++C^{I,ab}\lambda_{ab}+C^I_a\lambda^a=C^I_+\lambda^++C^{I,ab}\lambda_{ab}+\frac{1}{8}C^I_a\e^{abcde}\lambda_{bc}\lambda_{de}\frac{1}{\lambda^+} \Rightarrow
\]
\be \label{eqnC}
C^I_+(\lambda^+)^2+C^{I,ab}\lambda^+\lambda_{ab}+\frac{1}{8}C^I_a\e^{abcde}\lambda_{bc}\lambda_{de}=0.
\ee
This system of equations however does not have a solution with $\lambda^+ \neq 0$
and the gauge is singular. To see this, we first solve ten of the above
equations to obtain $\lambda_{ab}$ as a function of $\lambda^+$. A scaling
argument implies that these functions
are linear in $\lambda^+$. Then we plug the relation
$\lambda_{ab}=b_{ab}\lambda^+$ in the eleventh equation to find that $\lambda^+$ vanishes.
Thus we find that for any choice  $C^I_\a$ of maximal rank , the path integral
localizes at the $\lambda^\a=0$ locus\footnote{This also shows that the
choice of $C$ in \eqref{eq:Cgf} that manifestly leads to a factor $\d(\lambda^+)$
is not special. Any other choice of $C$ will also contain this factor.
\label{singular}},
which is the point that should be excised from the
pure spinor space for the theory to be non-anomalous \cite{Nekrasov:2005wg}.

As discussed above, the minimal and non-minimal formalisms are related
by analytic continuation in field space. In the toy example in the
previous subsection, we saw that the analytic continuation from the
``minimal variables'' $x_1,p_1$ to the ``non-minimal variables'' $x,
x^*$ sets to zero certain singular contributions (functions $f(x) \sim
x^{-1}$) but the integral still localizes at $x=0$. One would
thus expect that the zero mode integrals in the non-minimal formalism
localize at the $\lambda^\a=0$ locus, as the minimal ones do, and the
problems with the $\bar{\lambda} \lambda$ poles at 3 loops and higher are a
manifestation of this fact.

To avoid these problems\footnote{We would like to thank Nathan
  Berkovits and Nikita Nekrasov for discussions and suggestions about
  this point.} one must find a way to gauge fix the zero mode
invariances such that the zero mode integrals do not localize at
$\lambda^\a=0$.  Let us discuss how to achieve this in the minimal
formulation.  First, in order to avoid the unnecessary analytic
continuation to real $\lambda$ one should work with the analogue of the
contour representation of the delta function in \eqref{newmin} which
is appropriate for holomorphic $\lambda$ (and is less singular
than (\ref{minimal}) and (\ref{nonm})) In this language the choice of
$C$'s translates into a choice of position of poles. Secondly, one
must take global issues into account.  In particular, as mentioned
above, the space of pure spinors can be covered with sixteen
coordinates patches.  In order to avoid landing in the singular gauge
discussed above, one should arrange such that the expression for the
path integral insertions valid in any given patch always contains at
least one pole that lies in another patch.  Work in this direction is
in progress \cite{progress}. We also refer to \cite{Berkovits:2004bw}
for related relevant work.

\section{Conclusion} \label{conclu}

In this paper we have studied tree-level and one-loop amplitudes in
the minimal pure spinor formalism, in particular those with a $Q$
exact state. The amplitude prescription includes constant spinors
$C^I_{\a}$ and constant tensors $B^J_{mn}$ that are used to define the
picture changing operators which are necessary to absorb the zero
modes of the worldsheet fields. Amplitudes should be independent of
these constant tensors because the Lorentz variation of the PCO's is
$Q$ exact. The first computation we performed demonstrated that this
argument does not hold, 
because a tree-level amplitude does depend on the choice of $C$'s and
a certain amplitude for a given choice of $C^I_{\a}$ is not Lorentz
invariant. In the subsequent section it was shown that integrating
over the $C$'s, which was originally done to make the formalism
manifestly Lorentz invariant, results in a prescription that decouples
$Q$ exact states. We also introduced a formulation of
the minimal formalism at tree level in which the insertions of the
picture changing operators are replaced by a (unique) Lorentz tensor,
so the formalism is manifestly Lorentz invariant. BRST exact states
are shown to decouple and it also turned out that this formulation is
equivalent to the formulation in which one integrates over $C^I_{\a}$.

At one loop we found similar problems in the case we did not include the integral over $B$. Although the Lorentz variation of the PCO's is $Q$ exact, the Lorentz variation of a one-loop amplitude does not vanish. At least not after the $\lambda,N$ integrations, as one would expect. One expects the Lorentz variation to vanish after the $\lambda, N$ integrals because the formal argument for decoupling of Q-exact states uses that picture changing operators are BRST closed. In the minimal formalism however the picture changing operators are BRST closed in a distributional sense, $Q Y \sim x \d(x)$ with $x$ that depends on $\lambda$ and $N$, so the amplitudes should vanish if distributional identities hold and this requires performing the integrations of $\lambda, N$ but none of the other integrations.
The case with an integral over $B$ is dealt with in the companion paper \cite{new1}.
That paper contains a proof of decoupling of unphysical states in the minimal pure spinor formalism including an integral over $B$.

In the tree-level case one could reformulate the prescription so that the picture changing operators
are replaced by a BRST closed Lorentz tensor, as mentioned above. So one may wonder whether something similar
can also be done at one loop. We showed in section \ref{sec:nogo} that this is not possible.
More precisely, we showed that if the picture changing operators are Lorentz invariant and BRST closed then all one-loop amplitudes vanish.

Note that the problems we find in the minimal formulation at tree level and one loop
are not present in the non-minimal formalism. In this formalism the PCO's are replaced by the regularization factor $\cN$. In contrast to the PCO's, $\cN$ is $Q$ closed without subtleties. Hence the non-minimal formalism does not suffer from such problems. 

In \cite{Hoogeveen:2007tu} we showed that both the PCO's and the regularization factor $\cN$ come from a proper BRST treatment of fixing the gauge invariance generated by shifting the zero modes of the worldsheet fields. The difference between the minimal and non-minimal formalism can be understood as choosing different contours for the zero modes integrations. As became apparent in this work the choice that leads to the minimal formalism gives rise to anomalies. Moreover we saw that the gauge condition implicit in the current formulation of the amplitude prescriptions is singular and localizes the pure spinor zero mode integrals at the $\lambda^\a=0$
locus, which should be excised from the pure spinor space for the theory to be non-anomalous. We suspect that
the three-loop problems in the non-minimal formalism are also due to this singular gauge choice.
To avoid these problems one should reformulate the theory in a non-singular gauge.
We hope to report on this in the future.

\section*{Acknowledgments}
We would like to thank Nathan Berkovits, Carlos Mafra and Jun-Bao Wu for discussions.
KS is supported in part by NWO. KS would like to thank KITP, the 
Galileo Galilei Institute for Theoretical Physics and the Aspen 
Center of Physics for
hospitality during the completion of this work. 
This work was supported in parts by the National Science Foundation under
Grant No. NSF PHY05-51164.

\appendix

\section{Definitions, conventions and technical results}\label{sec:app}
This appendix contains detailed derivations of arguments we used in the main text. The purpose of the first subsection is to explain the notion of an invariant tensor and the meaning of the position of indices, which are important for the main text. The second subsection introduces spinors and in particular pure spinors. Moreover it contains details about the decomposition of $SO(10)$ representations under its $SU(5)\times U(1)$ subgroup. The following two subsections deal with the Lorentz generators and measures for the pure spinor sector. Their main purpose is to set the conventions, however they
contain more than just that. The fifth subsection is about gamma matrix traceless invariant tensors. Finally there is a subsection on the chain of operators that is used in the construction of the composite $b$ ghost.

\subsection{Invariant tensors}
Before we give the definition of an invariant tensor it is useful to recall what a representation of $SO(N)$ is. A generic $d$ dimensional representation can be denoted as
\be
v^a\rightarrow (g(A))^a_{\ b}v^b,\quad a,b=1,\cdots,d
\ee
where $A\in SO(N)$ and $g(A)$ is a linear map from $\mathbb{C}^d \rightarrow \mathbb{C}^d$. The fundamental representation is given by $d=N$ and $g$ is the identity map ($g(A)=A$):
\be
v^a \rightarrow A^a_{\ b}v^b,
\ee
A second representation of $SO(N)$ is given by
\be
v_a \rightarrow v_b (A^{-1})^b_{\ a}\ {\rm or}\ v \rightarrow (A^{-1T})v.
\ee
In fact this can be generalized to construct a second representation from any given one. One just replaces $v \rightarrow g(A)v$ by
\be
v \rightarrow (g(A))^{-1T}v.
\ee
This is called the conjugate representation. Note the position of the indices on the conjugate representation is opposite to the original representation. This is very convenient because together with the rule that indices can only be summed over if one is up and one is down, tensors transform as indicated by their free indices. In particular combinations without free indices are invariant. For example for an arbitrary representation and its conjugate
\be
w_a v^a \rightarrow w_b ((g(A)^{-1})^b_{\ a}g(A)^a_{\ c}v^c=w_b \d^b_c v^c=w_av^a.
\ee

An invariant tensor is a tensor that transforms into itself under all elements of the group. For example $\d^a_b$ is an invariant tensor for any representation. Note the range of $a$ and $b$  depends on the (dimension of the) representation. Its transformation is given by
\be
\d^a_b \rightarrow g(A)^a_{\ c}\d^c_d((g(A)^{-1})^d_{\ b}=\d^a_b.
\ee
For $SO(N)$ $\d^{ab}$ is also an invariant tensor where $a,b$ denote the vector representation, hence they run from 1 to $N$. Invariant tensors can be used to construct invariants from tensors. Objects that consist of (covariant) tensors and invariant tensors transform according to their free indices. In particular combinations without free indices are invariant. For example,
\be
v_aw_b\d^{ab} \rightarrow v_cw_d (B^{-1})^c_{\ a}(B^{-1})^d_{\ b}\d^{ab}=v_cw_d\d^{cd},
\ee
where the definition of $SO(N)$ was used.

The complex conjugate of a representation, $g(A)$, is given by $g^*(A)$. One can check this always defines a representation if $g(A)$ did. If a representation is equivalent to its complex conjugate it is real. For $SU(N)$ the conjugate of the fundamental representation is equivalent to the complex conjugate because $A^{-1T}=A^*$.

\subsection{Clifford algebra and pure spinors}\label{sec;clif}
The Clifford algebra in ten dimensions with Euclidian signature is given by
\be
\{\G^m,\G^n\}^{\underline{a}}_{\ \underline{b}}=2\d^{mn} \d^{\underline{a}}_{\underline{b}},\quad m,n=0,...,9\ \ \ula,\ul{b}=1,..,32.
\ee
These $\G^m$'s can be used to construct a representation of the Lorentz algebra and by exponentiating also of the Lorentz group. $\Sigma^{mn}=\fr{1}{4}[\G^m,\G^n]$ satisfy the Lorentz algebra. This representation is the (Dirac) spinor representation. Furthermore $(\G^m)^{\ula}_{\ \ulb}$ is an invariant tensor. A proof for the four dimensional case can be found in \cite{Peskin:1995ev}. The Clifford algebra has a representation in which the 32 by 32 components $\G$ matrices are off diagonal:
\be
\G^m=\left(
\begin{array}{cc}
0  & {\g^m}^{\a\b} \\
 \g^m_{\a\b} & 0 \\
\end{array}\right),
\ee
where $\a,\b=1,\cdots,16$. The Lorentz generators $\Sigma$ become
\be
\Sigma^{mn}=\fr{1}{4}\left(
\begin{array}{cc}
(\g^{[m}\g^{n]})^{\a}_{\ \b}  & 0 \\
 0 & (\g^{[m}\g^{n]})_{\a}^{\ \b} \\
\end{array}\right).
\ee
This implies the representation of the Lorentz group is reducible:
\be
{\bf 32} = {\bf 16} \oplus {\bf 16'}
\ee
The sixteen dimensional representation is the Weyl representation and it is not equivalent to its conjugate, hence the prime. The $\g^m_{\a\b}$ are invariant tensors with respect to Weyl representation.

Note the Clifford algebra now reduces to
\be \label{eq:clifgam}
{\g^{(m}}^{\a\b}\g^{n)}_{\b\g}=2\d^{mn}\d^{\a}_{\g}.
\ee
In particular $(\g^m)^{\a\b}$ is the inverse of $(\g^m)_{\a\b}$. An explicit solution to \eqref{eq:clifgam} is given in the next section.

\subsection{The $SU(N)$ subgroup of $SO(2N)$} \label{sec:app_su}

In this section we show that $SO(2N)$ has an $SU(N)$
subgroup and discuss how several representations of $SO(10)$
decompose into representations of $SU(5)$. Part of this analysis is
based on \cite{Georgi:1982jb}. To start, let us define for any $SO(2N)$ vector
$v$:
\be \label{eq:u5ind}
v^{a}=\hp(v^{2a} - i v^{2a+1}),\quad v_{a}=\hp(v^{2a} + i v^{2a+1}),\quad a=1,\ldots,N.
\ee
We now express
the $SO(2N)$ algebra in terms of generators labelled by the indices defined in
\eqref{eq:u5ind},
\bea
[M^{ab},M_{cd}]&=&-\hp\d^{[a}_{[c}M^{b]}_{\ d]},\quad a,b,c,d=1,..,N,
\\ \label{eq:u5sg}
[M^a_{\ b},M^c_{\ d}]&=&\hp(\d^a_dM^c_{\ b}-\d^{c}_bM^a_{\ d}),
\\  \label{eq:u5qw}
 [M^a_{\ b},M^{cd}]&=&\hp\d^{[c}_bM^{d]a},\quad [M^a_{\ b},M_{cd}]=-\hp \d^{a}_{[c}M_{d]b},
\\ \label{eq:u5sd}
[M^{ab},M^{cd}]&=&[M_{ab},M_{cd}]=0.
\eea
From \eqref{eq:u5sg} we see that the $SO(2N)$ algebra has an $N^2$ dimensional subalgebra. This subalgebra contains a $U(1)$ generated by $M \equiv M^a_{\ a}$ and the other $N^2-1$ generators,
\be \label{eq:ms}
U(1): \quad M \equiv M^a_{\ a}, \qquad SU(N): \quad
(M_S)^a_{\ b}\equiv M^a_{\ b}-\fr{1}{5}\d^a_bM^c_{\ c},
\ee
The generators $(M_S)^a_{\ b}$ are traceless and generate an $SU(N)$,
\be
[(M_S)^a_{\ b},(M_S)^c_{\ d}]=
-\fr{1}{2}(\d^a_d(M_S)^c_{\ b}-\d^{c}_b(M_S)^a_{\ d}).
\ee
The $U(1)$ charges of the generators are given by
\be
[M,M^{ab}]=-M^{ab}\ \ [M,M^a_{\ b}]=0\ \ [M,M_{ab}]=M_{ab}.
\ee

Every representation of $SO(2N)$ can be decomposed into representations
of $SU(N)$. In our case we will be interested in decomposition of
representations
the (Wick rotated) Lorentz group $SO(10)$ under $SU(5)$.
For several cases of interest the decomposition reads,
\bea \label{eq:dec16}
{\bf{16}} \rightarrow {\bf{1}}_{-\fr{5}{4}} \oplus {\bf{\bar{10}}}_{-\fr{1}{4}} \oplus {\bf{5}}_{\fr{3}{4}}&\quad& \lambda^{\a} \rightarrow \lambda^+,\lambda_{a_1a_2},\lambda^a,
\\ \label{eq:dec16'}
{\bf{16'}} \rightarrow {\bf{1}}_{\fr{5}{4}} \oplus {\bf{10}}_{\fr{1}{4}} \oplus {\bf{\bar{5}}}_{-\fr{3}{4}}&\quad& w_{\a} \rightarrow w_+,w^{a_1a_2},w_a,
\eea
\bea \label{eq:dec10}
{\bf{10}}\rightarrow {\bf{5}}_{-\hp}\oplus \bar{{\bf{5}}}_{\hp}&\quad& v^m \rightarrow v^a,v_a,
\\ \label{eq:dec45}
{\bf{45}} \rightarrow {\bf{1}}_0\oplus {\bf{24}}_0\oplus {\bf{10}}_{-1}\oplus \bar{{\bf{10}}}_{1}&\quad& M^{mn} \rightarrow M^a_{\ a},(M_S)^a_{\ b},M^{ab},M_{ab}.
\eea
where the subscripts are the $U(1)$ charges.

\subsubsection{Charge conservation and tensor products} \label{app:ccp}
The $M$ charge conservation property of invariant tensors can be used to prove that a large number of components of invariant tensors is zero, which is very useful if one is doing computations by using the explicit expressions of the tensors. An invariant tensor $T^{\a\b}_{\g\d}$ satisfies
\be \label{eq:mchcon}
0=M T^{\a\b}_{\g\d}=(M^u(\a)+M^u(\b)+M^d(\g)+M^d(\d))T^{\a\b}_{\g\d},
\ee
where $M^u(+)=-\fr{5}{4},M^u(a_1a_2)=-\fr{1}{4},M^u(a)=\fr{3}{4}$, $M^d(+)=\fr{5}{4},M^d(a_1a_2)=\fr{1}{4},M^d(a)=-\fr{3}{4}$. The $u$ is for up and the $d$ for down. This refers to the position of the Weyl index not the $SU(5)$ indices. So if the $M$ charges of the indices of a components do not sum up to zero the component vanishes. In this case one can for instance conclude $T^{+}_{\ \ b_1b_2,c,d}=0$, because the $M$ charge of the components is $-\qu(5+1+3+3)\neq 0$.

In this paper we are often interested in questions like: how many independent invariant tensors $T^{m\d}_{(\a\b\g)}$ exist? The upper index $\d$ denotes the Weyl representation, the lower indices stand for the conjugate Weyl representation and $m$ is the ten dimensional vector. To answer this question first of all note that the space of all tensors with the index structure and symmetries of $T$ forms a representation of $SO(10)$. The question how many independent invariant tensors exist in that space now translates to what the dimension of the invariant subspace is. This number can be obtained by computing the number of scalars in the relevant tensor product. This is one of the features of the computer algebra program LiE \cite{vanLeeuwen}. For the case of $T$ we compute
\be \label{eq:dec}
{\bf 10} \otimes {\bf 16} \otimes {\rm Sym}^3{\bf 16'}= {\bf 1} \oplus{\bf 45}\oplus{\bf 45}\oplus{\bf 45}\oplus \cdots,
\ee
where the dots are higher dimensional irreducible representations. The above decomposition shows that the space of invariant tensors with the symmetries of $T$ is one dimensional. Based on this result we can for example conclude
\be \label{eq:r3td}
\g^m_{(\a\b}\d^{\d}_{\g)} \propto \g^n_{(\a\b}\g^m_{\g)\e}\g^{\e\d}_n.
\ee
In order to find the constant of proportionality, computing a single component on both sides suffices.

\subsubsection{Dynkin labels and gamma matrix traceless tensors} \label{sec:dyl}
Throughout this work we denote irreducible representations by their dimensions. This is slightly ambiguous, therefore we clarify what we mean exactly by specifying the Dynkin labels of the highest weight state of the representation.
\be
{\bf 10} \lera (1,0,0,0,0),\quad{\bf 16} \lera (0,0,0,1,0),\quad {\bf 16'} \lera (0,0,0,0,1),\quad {\bf 45} \lera (0,1,0,0,0).
\ee
There is one further irreducible representation of interest, which is given by symmetric and gamma matrix traceless tensors:
\be \label{eq:gamn}
T^{((\a_1\cdots \a_n))} \lera (0,0,0,n,0) \lera {\rm Gam}^n {\bf 16},
\ee
where the Dynkin labels are specified. These representations are discussed in more detail in \cite{Cederwall:2001bt}. There are three gamma matrix traceless tensors that interest us in particular:
\be
(T_1)^{((\a_1\a_2\a_3))}_{[\b_1\cdots \b_{11}]},\quad (T_2)^{((\a_1\cdots \a_8))}_{[[m_1n_1],\cdots ,[m_{10}n_{10}]]},\quad (T_3)^{((\a_1\cdots \a_{11}))}_{[\b_1\cdots \b_{11}][[m_1n_1],\cdots ,[m_{10}n_{10}]]}.
\ee
For the three tensors above the computer algebra program LiE can be used to conclude there is only one independent invariant tensor. Note this is consistent with the arguments in \cite{Berkovits:2006bk}, where it is argued that a tensor which is symmetric and gamma matrix traceless, let us say in some indices $\a_i$, is completely specified by the components where the $\a$'s are all +. In order to see that this implies there is only one independent invariant tensor of the form of $T_1$ note that for an invariant tensor the components
\be \label{eq:indcom}
(X_1)^{+++}_{\b_1\cdots \b_{11}}
\ee
are only nonvanishing if
\be
\b_1, \ldots, \b_{11}=+,12,13,\cdots,45.
\ee
This follows from the charge conservation property of invariant tensors. By antisymmetry of the $\b$'s there is only one independent component in \eqref{eq:indcom}. Thus the argument of \cite{Berkovits:2006bk}
implies that the entire invariant tensor is completely specified by a single component and therefore the space of invariant tensors of the form of $T_1$ is one dimensional. The above argument applies equally well to $T_2$ and $T_3$.

\subsubsection{Explicit expression for gamma matrices and pure spinors} \label{sec:gama}
A solution to \eqref{eq:clifgam} is given by
\be
(\g^k)_{\a\b}=\left(
\begin{array}{ccc}
0  &0& \d^k_b \\
0& -\e^{ka_1a_2b_1b_2} & 0 \\
\d^k_a &0&0\\
\end{array}\right),
\ \ \ \ \
(\g_k)_{\a\b}=\left(
\begin{array}{ccc}
0  &0& 0 \\
0& 0 & \d^{[a_1}_k\d^{a_2]}_b \\
0&\d^{[b_1}_k\d^{b_2]}_a&0\\
\end{array}\right),
\ee
\be
(\g^k)^{\a\b}=\left(
\begin{array}{ccc}
0  &0& 0 \\
0& 0 & \d_{[a_1}^k\d_{a_2]}^b \\
0&\d_{[b_1}^k\d_{b_2]}^a&0\\
\end{array}\right),
\ \ \ \ \
(\g_k)^{\a\b}=\left(
\begin{array}{ccc}
0  &0& \d_k^b \\
0& -\e_{ka_1a_2b_1b_2} & 0 \\
\d_k^a &0&0\\
\end{array}\right).
\ee
Note these matrices are skew diagonal, this is a consequence of the charge conservation property of invariant tensors.

A pure spinor is a Weyl spinor that satisfies
\be
\lambda^{\a}\g^m_{\a\b}\lambda^{\b}=0.
\ee
After plugging in the explicit expression for the gamma matrices this becomes
\begin{eqnarray}
2\lambda^+\lambda^a-\qu \e^{abcde}\lambda_{bc}\lambda_{de}=0,\\
2\lambda^b\lambda_{ab}=0.
\end{eqnarray}
These equations are solved by
\be
\lambda^a=\fr{1}{8}\fr{1}{\lambda^+}\e^{abcde}\lambda_{bc}\lambda_{de}
\ee
The explicit expression of the three form gamma's is:
\bea
(\g_{k_1k_2k_3})^{\a\b}&=&\fr{1}{6}(\g_{[k_1}\g_{k_2}\g_{k_3]})^{\a\b}=
\left(
\begin{array}{ccc}
0 & \e_{k_1k_2k_3b_1b_2} &0 \\
 -\e_{k_1k_2k_3a_1a_2} &  0&0\\
 0 & 0 & 0
\end{array}\right)
\\ \nn
(\g^{k_1}_{\ \ k_2k_3})^{\a\b}&=&\fr{1}{6}((\g^{k_1}\g_{[k_2}\g_{k_3]})^{\a\b}-(\g_{[k_2}\g^{k_1}\g_{k_3]})^{\a\b}+(\g_{[k_2}\g_{k_3]}\g^{k_1})^{\a\b})=
\\
&&\hp \left(
\begin{array}{ccc}
0 & 0 &- \d^{k_1}_{[k_2}\d^{b}_{k_3]} \\
 0 &  \d^{k_1}_{[a_1}\e_{a_2]k_2k_3b_1b_2}-\d^{k_1}_{[b_1}\e_{b_2]k_2k_3a_1a_2} &0\\
 \d^{[k_1}_{k_2}\d^{a]}_{k_3} & 0 & 0
\end{array}\right),
\\
(\g^{k_1k_2}_{\ \ \ \ k_3})^{\a\b}&=&\fr{1}{6}((\g^{[k_1}\g^{k_2]}\g_{k_3})^{\a\b}-(\g^{[k_1}\g_{k_3}\g^{k_2]})^{\a\b}+(\g_{k_3}\g^{[k_1}\g^{k_2]})^{\a\b})=
\\ \nn
&&\left( \begin{array}{ccc}
0 & 0 & 0 \\
 0 & 0 & \d^{k_1}_{[a_1}\d^{k_2}_{a_2]}\d^b_{k_3}+\hp\d^b_{[a_1}\d^{[k_1}_{a_2]}\d^{k_2]}_{k_3} \\
 0& -\d^{k_1}_{[b_1}\d^{k_2}_{b_2]}\d^a_{k_3}-\hp\d^a_{[b_1}\d^{[k_1}_{b_2]}\d^{k_2]}_{k_3} & 0
\end{array}\right),
\\
(\g^{k_1k_2k_3})^{\a\b}&=&\fr{1}{6}(\g^{[k_1}\g^{k_2}\g^{k_3]})^{\a\b}=
\left(
\begin{array}{ccc}
0 & 0&0 \\
 0&  0&0\\
 0& 0 & -\e_{k_1k_2k_3ab}
\end{array}\right).
\eea
\subsection{Pure spinor Lorentz generators} \label{sec:pslg}
In our computations we often need the explicit form of the pure spinor Lorentz generators, $N^{mn}$, which are given by
\be \label{eq:nexpl}
N^{mn}=\hp w_{\a} (\g^{mn})^{\a}_{\ \b} \lambda^{\b},\quad \g^{mn}=\hp (\g^m\g^n-\g^n\g^m).
\ee
In this subsection these components are given in terms of the $SU(5)$ components of $\lambda$ and $w$ in the gauge $w_a=0$. The $SU(5)$ components of \eqref{eq:nexpl} are given by
\bea
N_{kl}&=&\qu w_{\a} (\g_{[k})^{\a \b} (\g_{l]})_{\b \d} \lambda^{\d},\\
{N}^k_{\ l}&=&\qu\left( w_{\a} (\g^{k})^{\a \b} (\g_{l})_{\b \d} \lambda^{\d}-w_{\a} (\g_{l})^{\a \b} (\g^{k})_{\b \d} \lambda^{\d}\right),\\
N^{kl}&=&\qu w_{\a} ({\g}^{[k})^{\a \b} ({\g}^{l]})_{\b \d} \lambda^{\d}.
\eea
The explicit gauge invariant form is obtained by plugging in the expressions for $\g$
\bea
N_{kl}&=&-\hp w^-\lambda_{kl}- \qu w^{ab}\e_{abckl}\lambda^{c},\\
N^{kl}&=&\hp w^{kl}\lambda^{+}+\qu w_{a}\e^{abckl}\lambda_{bc},\\
{N}^k_{\ l}&=&-\qu\d^k_l\lambda^+w^--\qu\hp\d^k_lw^{ab}\lambda_{ab}+\hp w^{ak}\lambda_{al}+\qu w_a\lambda^a\d^k_l-\hp w_k\lambda^l,\\
N&=&{N}^k_{\ k}=-\fr{5}{4}w^-\lambda^+-\qu\hp w^{ab}\lambda_{ab}+\fr{3}{4}w_a\lambda^a,\\
(N_S)^k_{\ l}&=&{N}^k_{\ l}-\fr{1}{5}\d^k_lN
\\ \nn
&=&-\fr{1}{10}\d^k_l w^{ab}\lambda_{ab}+\hp w^{ak}\lambda_{al}+\fr{1}{10}w_a\lambda^a\d^k_l-\hp w_l\lambda^k.
\eea
After using the pure spinor solution and setting $w_a$ to zero
\bea \label{eq:N1}
N_{kl}&=&-\hp w^-\lambda_{kl}-\qu \fr{w^{ab}\lambda_{kl}\lambda_{ab}}{\lambda^+}+\hp \fr{w^{ab}\lambda_{ka}\lambda_{lb}}{\lambda^+},\\
N^{kl}&=&\hp w^{kl}\lambda^{+},\\
N&=&-\fr{5}{4}w^-\lambda^+-\qu\hp w^{ab}\lambda_{ab},\\
(N_S)^k_{\ l}&=&\hp(-\fr{1}{5}\d^k_lw^{ab}\lambda_{ab}+w^{ak}\lambda_{al}). \label{eq:N4}
\eea
$J$ in terms of the free variables is given by
\be
J=w_{\a}\lambda^{\a}=w^-\lambda^++\hp w^{ab}\lambda_{ab}+w_a\lambda^a.
\ee
In the gauge $w_a=0$ this becomes
\be \label{eq:Jw=0}
J=w^-\lambda^++\hp w^{ab}\lambda_{ab}.
\ee

\subsubsection{Lorentz currents with unconstrained spinors}
As mentioned in section \ref{sec:rev} the action for the eleven independent components of $\lambda$ and their conjugate variables can be used to prove the pure spinor OPE's from \eqref{eq:opewl}. This will be demonstarted below. First we consider two unconstrained bosonic Weyl spinors. The OPE for such fields is given by
\be
y_{\a}(z)\x^{\b}(w) \sim \d_{\a}^{\b}\frac{1}{z-w}.
\ee
The OPE of the pure spinor Lorentz current with itself is given by
\be
M^{m_1m_2}(z)M^{n_1n_2}(w) \sim \frac{1}{4}\frac{1}{z-w}(-y_{\a}(z){\g^{m_1m_2}}^{\a}_{\ \b}{\g^{n_1n_2}}^{\b}_{\ \g}\x^{\g}(w)+
\ee
\[
y_{\a}(w){\g^{n_1n_2}}^{\a}_{\ \b}{\g^{m_1m_2}}^{\b}_{\ \g}\x^{\g}(z))-\frac{1}{4}\frac{{\rm{Tr}}(\g^{m_1m_2}\g^{n_1n_2})}{(z-w)^2}.
\]
Using identities,
\bea
[\hp\g^{m_1m_2},\hp\g^{n_1n_2}]&=&\hp (\h^{n_1[m_2}\g^{m_1]n_2}-\h^{n_2[m_2}\g^{m_1]n_1}),\\
{\rm{Tr}}(\g^{m_1m_2}\g^{n_1n_2})&=&-16\h^{m_1[n_1}\h^{n_2]m_2},
\eea
the $MM$ OPE reduces to
\be \label{eq:nnope}
M^{m_1m_2}(z)M^{n_1n_2}(w) \sim\frac{-(\h^{n_1[m_2}M^{m_1]n_2}-\h^{n_2[m_2}M^{m_1]n_1})}{z-w}-4\frac{\h^{m_1n_2}\h^{m_2n_1}-\h^{m_1n_1}\h^{m_2n_2}}{(z-w)^2}.
\ee
We can read off the algebra of the Lorentz charges from the single pole in the OPE
\be \label{eq:comreln}
[M^{m_1m_2},M^{n_1n_2}]=-(\h^{n_1[m_2}M^{m_1]n_2}-\h^{n_2[m_2}M^{m_1]n_1}).
\ee
In case the worldsheet fields are fermionic, the OPE remains the same:
\be
p_{\a}(z)\q^{\b}(w) \sim \d^{\b}_{\a}\fr{1}{z-w}.
\ee
The Lorentz generator for the fermionic variables has a minus sign:
\be
M^{mn}=-p\g^{mn}\q.
\ee
This sign is necessary to reproduce the commutation relation \eqref{eq:comreln}. As a consequence the sign in the double pole in the OPE changes from -4 to +4. This coefficient is called the level. We would like the Lorentz current of the combined $p,\q$ and $\lambda,w$ sector to have level one, since this is the level of the $\y$ sector in the RNS formalism. This implies the $N_{(\lambda w)}$ generators must have level $-3$. In the next subsection we explain how such currents can be obtained from the pure spinor action after gauge fixing.

\subsubsection{Currents containing pure spinors}

In \cite{Hoogeveen:2007tu} we discuss how to gauge fix the gauge invariance for $w$ by setting
$w_a=0$. We ended up with a free action for the eleven independent components of $\lambda$ and their conjugate variables. One may anticipate that
one can set $w_a=0$ in all (gauge invariant) operators that depend on $w_{\a}$ without
Lorentz invariance being lost, and we will see that this is indeed the case.
We first study the OPE's of $N$ and $J$ with $\lambda$ and find no problems. Secondly we look at the $NN$ OPE. Here we find that the single pole is the same as in \eqref{eq:nnope}, but the level of the OPE depends on which $SU(5)$ components one chooses. This spoils Lorentz invariance, but it can be cured as demonstrated below.

The OPE of $J$ and $N^{mn}$ with $\lambda$ are given by
\be\label{OPE}
J(z)\lambda^{\a}(w) \sim \fr{1}{z-w}\lambda^{\a}(w), \quad N^{mn}(z)\lambda^{\a}(w) \sim \fr{1}{z-w}\hp{\g^{mn}}^{\a}_{\ \b}\lambda^{\b}(w).
\ee
In order to check these OPE's we set $w_a=0$ and use the free field OPE's
\be
w^-(z)\lambda^+(w)\sim \fr{1}{z-w},\quad w^{ab}(z)\lambda_{cd}(w)\sim \fr{1}{z-w}\d^{[a}_{c}\d^{b]}_d.
\ee
Let us start with $J$:
\be
J(z)\lambda^+(w)=w^-\lambda^+(z)\lambda^+(w) \sim \fr{1}{z-w}\lambda^+(w)
\ee
and similarly for $\lambda_{ab}$. $\lambda^a$ is more involved. By using
\be
w^-(z)\fr{1}{\lambda^+}(w)\sim\fr{1}{z-w}\fr{-1}{(\lambda^+)^2(w)},
\ee
we can reproduce the Lorentz invariant answer:
\be
J(z)\lambda^a(w)=( w^-\lambda^++\hp w^{ab}\lambda_{ab})(z)\fr{\e^{abcde}\lambda_{bc}\lambda_{de}}{8\lambda^+}(w) \sim \fr{1}{z-w}\fr{1}{8\lambda^+}\e^{abcde}\lambda_{bc}\lambda_{de}(w).
\ee
Let us continue with the trace of $N^{mn}$. In terms of unconstrained spinors it is given by
\be
N=-\fr{5}{2}\lambda^+w^--\hp\hp w_{ab}\lambda^{ab}+\fr{3}{2}w^a\lambda_{a}.
\ee
From here we can see that the expected charge of $\lambda^a$ is $\fr{3}{2}$. The OPE of $N$ with $\lambda^+$ or $\lambda_{ab}$ trivially reproduces the Lorentz invariant result, the OPE of $N$ with $\lambda^a$ is
\be
N(z)\lambda^a(w)=(-\fr{5}{2}\lambda^+w^--\hp\hp w_{ab}\lambda^{ab})(z) \fr{\e^{abcde}\lambda_{bc}\lambda_{de}}{8\lambda^+}(w)\sim
\ee
\[
\fr{1}{z-w}\fr{(\fr{5}{2}-\hp-\hp)\e^{abcde}\lambda_{bc}\lambda_{de}}{8\lambda^+}(w).
\]
All other components of the $N \lambda$ OPE can be checked along the same lines. The $N^{mn}N^{pq}$ OPE is a different story. The single pole always leads to the correct Lorentz algebra, but the coefficient of the double pole depends on which $SU(5)$ components we choose to take. For instance
\be
N(z)N(w) \sim -\fr{35}{16}\fr{1}{(z-w)^2}=-\fr{7}{4}\h^k_{\ l}\h^l_{\ k}\fr{1}{(z-w)^2}
\ee
\be
N^{12}(z)N_{12}(w)\sim \qu \fr{1}{(z-w)^2}+\fr{1}{z-w}\hp(N^1_{\ 1}(w)+N^2_{\ 2}(w))=
\ee
\[
-1\fr{1}{(z-w)^2}(-\h^{1}_{\ 1}\h^2_{\ 2})+\fr{1}{z-w}\hp(N^1_{\ 1}(w)+N^2_{\ 2}(w)).
\]
The first OPE would imply a Lorentz current level of $-\fr{7}{4}$ and the second one $-1$. It will be shown below that it is possible to deform the currents in equations \eqref{eq:N1}-\eqref{eq:N4} by conserved quantities such that the level of the $N^{mn}N^{pq}$ OPE is minus three \cite{Berkovits:2002zk}. This fixes the total derivatives one has to add to \eqref{eq:Jw=0} in order for the $JN$ OPE to be regular. Demanding $N^{mn}$ is a primary field of weight one determines the total derivatives in the stress energy tensor. If one now computes the $JT$ OPE, a ghost number anomaly value of minus eight follows. This cannot be adjusted.

The deformations are most easily given after bosonization of $\lambda$ and $w$, which is given by
\be
\lambda^+ \cong e^{\c-\f},\quad w^- \cong e^{-\c+\f}\partial \c,\quad \lambda^+w^- \cong \partial \f,
\ee
where $\f,\c$ are chiral bosons satisfying
\be
\f(z)\f(0) \sim -{\rm{ln}}z,\quad \c(z)\c(0) \sim {\rm{ln}}z.
\ee
Now define
\be
s=\c-\f,\quad 2t=\f+\c \leftrightarrow \f=\hp(2t-s),\quad \c=\hp(s+2t)
\ee
The OPE's for these new variables are
\be
s(z)s(0) \sim {\rm{regular}},\quad t(z)t(0) \sim {\rm{regular}}\ \ \ t(z)s(0) \sim {\rm{ln}}z.
\ee
The original worldsheet fields $\lambda$ and $w$ can be expressed in terms of $s,t$ as
\be
\lambda^+ \cong e^{s},\quad w^-\cong \hp e^{-s}(\partial s+2 \partial t),\quad \lambda^+w^- \cong \hp(2\del t - \del s).
\ee
The Lorentz currents of \eqref{eq:N1}-\eqref{eq:N4} in bosonized form are given by\footnote{In \cite{Berkovits:2002zk} the Lorentz currents which we call $(N^B)_{mn}$ have a different normalization. The relation with ours is given by
\be
N=-\fr{\sqrt{5}}{2}N^B,\quad N^{ab}=\hp (N^B)^{ab},\quad (N_S)^a_{\ b}=\hp (N^B_S)^a_{\ b},\quad N_{ab}=\hp (N^B)_{ab}.
\ee}
\bea
N&=&-\frac{5}{8}(2\partial t- \partial s) -\frac{1}{8} w^{ab}\lambda_{ab},\\
N^{ab}&=&\hp e^s w^{ab},\\
(N_S)^a_{\ b}&=&\hp(w^{ac}\lambda_{bc}-\frac{1}{5}\d^a_b w^{cd}\lambda_{cd}),\\
N_{ab}&=&e^{-s}[-\hp(\hp\partial s \lambda_{ab}+\partial t \lambda_{ab})-\frac{1}{4}w^{cd}\lambda_{ab}\lambda_{cd}+\hp w^{cd}\lambda_{ac}\lambda_{bd}].
\eea
The deformations one should add to \eqref{eq:N1}-\eqref{eq:N4} to make the $NN$ OPE Lorentz invariant are given by:
\bea
\D N&=&-\fr{5}{8} \partial s, \\
\D N^{ab}&=& 0,\\
\D (N_S)^a_{\ b}&=&0,\\
\D N_{ab}&=& e^{-s}(-\fr{3}{4}\partial s \lambda_{ab}+\del \lambda_{ab})= \del (e^{-s}\lambda_{ab})-\qu (\del e^{-s})\lambda_{ab}.
\eea
Note that the field equations imply the $\bar{\del}$ operator annihilates these deformations. Hence the deformed charges are still conserved. Furthermore the deformations do not modify the $N \lambda$ OPE, which is manifest in the $s,t$ variables.

\subsection{Lorentz invariant measures} \label{sec:lim}
The Lorentz invariant measures for both the weight zero field, $\lambda^{\a}$, and the weight one field, $N^{mn}$, are discussed below. Both these measure were first introduced in \cite{Berkovits:2004px} and the $\lambda$ zero mode measure is also discussed in \cite{Berkovits:2004bw}.
\subsubsection{Measure for the zero modes of $\lambda$}
From the ghost number anomaly in the $JT$ OPE \eqref{eq:opewl} we know a tree-level correlator can only be non-zero is the $J$ charge of the insertions is -8. Since there are no $w$ (or $N^{mn}$) zero modes at tree level, the measure for the $\lambda$ zero modes must have ghost number +8. In addition it must be Lorentz invariant. This results in
\be \label{eq:measl}
[d\lambda]\lambda^{\a}\lambda^{\b}\lambda^{\g}=X^{\a\b\g}_{\b_1\cdots \b_{11}}d\lambda^{\b_1}\wedge \cdots \wedge d\lambda^{\b_{11}}
\ee
for some invariant tensor $X$. The number of invariant $(3,11)$ tensors with spinor indices that are symmetric in the upper indices and antisymmetric in lower ones is one \cite{vanLeeuwen}. In other words there is only one possibility for $X$ which is given in \eqref{eq:defet}. Because the LHS of \eqref{eq:measl} is zero when contracted with $\g^m_{\a\b}$, the RHS should vanish too. It does because there are no scalars in ${\bf 10} \otimes {\bf 16} \otimes {\rm Asym}^{11} {\bf 16'}$. Thus
\be
\g^m_{\a\b}X^{\a\b\g}_{\b_1\cdots \b_{11}}=0.
\ee

In equation \eqref{eq:measl} one is free to choose $\a\b\g$. Different choices lead to different guises of the measure. In \cite{Nekrasov:2005wg} it was shown all these are related to each other by a coordinate transformation in pure spinor space. On the patch defined by $\lambda^+ \neq 0$ there is only one choice for $\a\b\g$ that results in a well defined measure on the whole patch which is $\a\b\g=+++$. This gives $[d\lambda]$ as
\be \label{eq:measlc}
[d\lambda]=\fr{d\lambda^+\wedge d\lambda_{12}\wedge \cdots \wedge d\lambda_{45}}{{\lambda^+}^3},
\ee
where we used $(\e T)^{+++}_{\b_1\cdots \b_{11}}$ is only non-zero if $\b_1, \ldots, \b_{11}=+,b_1b_2,b_3b_4,\cdots,b_{19}b_{20}$. This is a consequence of the $M$ charge conservation property of invariant tensors.

\subsubsection{Measure for the zero modes of $N^{mn}$} \label{measN}
The ghost number anomaly and Lorentz invariance imply the measure for the zero modes of $N$ must be of the form
\be
[dN]\lambda^{\a_1}\cdots \lambda^{\a_8}=X^{\a_1\cdots \a_8}_{m_1n_1\cdots m_{10}n_{10}}dN^{m_1n_1}\wedge \cdots \wedge dN^{m_{10}n_{10}}\wedge dJ.
\ee
There exists only one independent invariant tensor of this kind (cf.~\ref{sec:dyl}) and since \eqref{eq:defR}
provides an example of such tensor we obtain:
\be
[dN]\lambda^{\a_1}\cdots \lambda^{\a_8}=R^{\a_1\cdots \a_8}_{m_1n_1\cdots m_{10}n_{10}}dN^{m_1n_1}\wedge \cdots \wedge dN^{m_{10}n_{10}}\wedge dJ.
\ee
A more explicit form of $[dN]$ is obtained by choosing all $\a$'s equal to $+$. The relevant gamma matrix components are
\be
\g_{a_1\cdots a_5}^{++}=\e_{a_1\cdots a_5},\quad \g^{a_1\cdots a_5}_{++}=\e^{a_1\cdots a_5},
\ee
all other components of $\g^{++}_{mnpqr}$ vanish. Using these one sees $[dN]$ can be expressed as
\[
[dN]{\lambda^+}^8=\e_{a_1b_1a_2a_3a_4}\e_{a_5b_5b_2a_6a_7}\e_{a_8b_8b_3b_6a_9} \e_{a_{10}b_{10}b_4b_7b_9}dN^{a_1b_1}\wedge \cdots \wedge dN^{a_{10}b_{10}}\wedge dJ=
\]
\be  \label{eq:[dN]}
dN^{12}\wedge \cdots \wedge dN^{45}\wedge dJ={\lambda^+}^{11}d^{10}w^{ab}dw_+ \Rightarrow [dN]=(\lambda^+)^3dw_+d^{10}w^{ab},
\ee
where the gauge condition $w_a=0$ is imposed in the first equality of the second line.

\subsection{Gamma matrix traceless projectors} \label{sec:detrel}
The operator $\Lambda_{\a\b\g}$ is introduced in \eqref{eq:Lam}. This equation has a special form and
in this subsection we explain it. First note that $I^{\a\b\g}_{\a'\b'\g'}\equiv \int [d\lambda] \lambda^{\a}\lambda^{\b}\lambda^{\g}\Lambda_{\a'\b'\g'}$ must be a Lorentz invariant tensor. An invariant tensor forms invariant combinations with covariant objects if and only if all indices are contracted, otherwise the total object transforms according to the free indices. So if all indices on $\int [d\lambda] \lambda^{\a}\lambda^{\b}\lambda^{\g}\Lambda_{\a'\b'\g'}$ are contracted with covariant objects the total object is Lorentz invariant. After performing the integral the object is of course still Lorentz invariant and therefore $I$ must be an invariant tensor. Furthermore $I^{\a\b\g}_{\a'\b'\g'}$ must be symmetric in both its upstairs and downstairs indices and since $\lambda$ is a pure spinor $I$ must satisfy $\g^m_{\a\b}I^{\a\b\g}_{\a'\b'\g'}=0$. The $SO(10)$ invariant tensors of the form $T^{(\a\b\g)}_{(\a'\b'\g')}$ form a vector space which is two dimensional as can be computed by counting the number of scalars in ${\rm{Sym}}^3 {\bf{16}}\otimes {\rm{Sym}}^3 {\bf{16'}}$ \cite{vanLeeuwen}. A basis of this vector space is given by
\be \label{eq:ans}
\left\{ \d^{(\a}_{\a'}\d^{\b}_{\b'}\d^{\g)}_{\g'}, \g_m^{(\a\b}\g^m_{(\a'\b'}\d^{\g)}_{\g')} \right\}.
\ee
Hence
\be
\int [d\lambda] \lambda^{\a}\lambda^{\b}\lambda^{\g}\Lambda_{\a'\b'\g'}= c_1 \d^{(\a}_{\a'}\d^{\b}_{\b'}\d^{\g)}_{\g'}+ c_2\g_m^{(\a\b}\g^m_{(\a'\b'}\d^{\g)}_{\g')}.
\ee
Since $\lambda$ is a pure spinor
\be \label{eq:c=40}
0=\int [d\lambda] \lambda^{\a}\g^m_{\a\b}\lambda^{\b}\lambda^{\g}\Lambda_{\a'\b'\g'} =c_1\g^m_{\a\b}\d^{(\a}_{\a'}\d^{\b}_{\b'}\d^{\g)}_{\g'} +c_2\g^m_{\a\b}\g^{(\a\b}_n\g^n_{(\a'\b'}\d^{\g)}_{\g')}
\ee
\[
=(c_1+40c_2)\d^{\g}_{(\a'}\g^m_{\b'\g')},
\]
where we used (cf. \eqref{eq:r3td})
\be
\g_n^{\g\a}\g^m_{\a(\a'}\g^n_{\b'\g')}=2\d^{\g}_{(\a'}\g^m_{\b'\g')}.
\ee
We could have anticipated ending up with one equation for $c_1,c_2$ because ${\bf{10}}\otimes {\bf{16}}\otimes {\rm{Sym}}^3 {\bf{16'}}$ contains one scalar.

In summary the number of scalars in ${\rm{Sym}}^3 {\bf{16}}\otimes {\rm{Sym}}^3 {\bf{16'}}$ determined the number of degrees of freedom ($c_i$) and the number of scalars in ${\bf{10}}\otimes {\bf{16}}\otimes {\rm{Sym}}^3 {\bf{16'}}$ determined the number of relations between them.

\subsubsection{Arbitrary rank}
The tensor in equation \eqref{eq:c=40} can be denoted as
\be
\d^{((\a}_{\a'}\d^{\b}_{\b'}\d^{\g))}_{\g'}.
\ee
There is a unique such tensor because the number of scalars in ${\rm Gam}^3{\bf 16}\otimes ({\bf 16'})^3$ is one (cf.~\eqref{eq:gamn} for the meaning of Gam). In fact there is one scalar in ${\rm Gam}^n{\bf 16}\otimes ({\bf 16'})^n$ for any $n$. In order to write an explicit expression for $\d^{((\a_1}_{\b_1}\cdots \d^{\a_n))}_{\b_n}$ for any $n$ we look for a basis of rank $(n,n)$ invariant tensors that are symmetric in both their upper and lower indices. For even $n$ the number of scalars in ${\rm Sym}^n {\bf 16}\otimes {\rm Sym}^n {\bf 16'}$ is $\fr{n}{2}+1$. For odd $n$ the number of scalars in ${\rm Sym}^n {\bf 16}\otimes {\rm Sym}^n {\bf 16'}$ is $\fr{n-1}{2}+1$. Since odd $n$ is of more relevance to this work we explicitly give the basis for odd $n$. The $\fr{n-1}{2}+1$ basis elements are given by
\be
T_1=\d^{(\a_1}_{\b_1}\cdots \d^{\a_n)}_{\b_n},\quad T_2=\g^{(\a_1\a_2}_m\g_{(\b_1\b_2}^m\d^{\a_3}_{\b_3}\cdots\d^{\a_n)}_{\b_n)}
\ee
up to
\be
T_{k+1}=\g^{(\a_1\a_2}_{m_1}\g^{m_1}_{(\b_1\b_2}\cdots \g^{\a_{n-2}\a_{n-1}}_{m_k} \g_{\b_{n-2} \b_{n-1}}^{m_k}\d^{\a_n)}_{\b_n)}
\ee
where $k=\fr{n-1}{2}$. In order to see these tensors are independent compute the following components:
\be
T^{+\cdots+}_{+\cdots+}, T^{a_1+\cdots+}_{b_1+\cdots+}, \cdots, T^{a_1\cdots a_k+\cdots+}_{b_1 \cdots b_k+\cdots+}.
\ee
We can conclude
\be \label{gamdel}
\d^{((\a_1}_{\b_1}\cdots \d^{\a_{n}))}_{\b_{n}}=c_1T_1+\cdots+c_kT_k,
\ee
for some coefficients $c_i$, which can be explicitly computed as we did for the $n=3$ case. Note the above is for odd $n$. Even $n$ works very much in the same way, the only difference is the last $\d$ in all the $T$'s. If one removes this, the $T$'s form a basis for the even case.

\subsection{Chain of operators for $b$ ghost} \label{sec:appco}
The following chain of operators plays an important role in the $b$ ghost:
\begin{eqnarray}
QG^{\a}&=&\lambda^{\a}T,\\
QH^{\a\b}&=&\lambda^{\a}G^{\b}+g^{((\a\b))},\\
QK^{\a\b\g}&=&\lambda^{\a}H^{\b\g}+h_1^{((\a\b))\g}+h_2^{\a((\b\g))},\\
QL^{\a\b\g\d}&=&\lambda^{\a}K^{\b\g\d}+k_1^{((\a\b))\g\d}+k_2^{\a((\b\g))\d}+k_3^{\a\b((\g\d))},\\
0&=&\lambda^{\a}L^{\b\g\d\r}+l_1^{((\a\b))\g\d\r}+l_2^{\a((\b\g))\d\r}+l_3^{\a\b((\g\d))\r}+l_4^{\a\b\g((\d\r))}. \label{eq:L=0}
\end{eqnarray}
the last equation implies there exists an $S^{\a\b\g}$ such that
\be \label{eq:LS}
L^{\a\b\g\d}=\lambda^{\a}S^{\b\g\d}+s_1^{((\a\b))\g\d}+s_2^{\a((\b\g))\d}+s_3^{\a\b((\g\d))}.
\ee
The text below is essentially a summary of section 3 of \cite{Oda:2007ak}. The primary fields of weight two that solve the above equations are given by
\bea \label{eq:G}
G^{\a}&=&\hp \Pi^m(\g_md)^{\a}-\qu N_{mn}(\g^{mn}\del \q)^{\a}-\qu J \del \q^{\a}+\fr{7}{2}\del^2 \q,\\
H^{\a\b}&=&\fr{1}{16}\g_m^{\a\b}(N^{mn}\Pi_n-\hp J\Pi^m+2\del\Pi^m)
\\ \nn
&&+\fr{1}{96}\g^{\a\b\g}_{mnp}(\qu d\g^{mnp}d+6N^{mn}\Pi^c),\\
K^{\a\b\g}&=&-\fr{1}{48}\g^{\a\b}_m(\g_nd)^{\g}N^{mn}-\fr{1}{192}\g^{\a\b}_{mnp}(\g^md)^{\g}N^{np}\\ \nn
&&+\fr{1}{192}\g^{\b\g}_m\left[ (\g_nd)^{\a}N^{mn}+\fr{3}{2}(\g^md)^{\a}J-6(\g^m\del d)^{\a}\right]-\fr{1}{192}\g^{\b\g}_{mnp}(\g^md)^{\a}N^{np},\\
L^{[\a\b\g\d]}&=& -\fr{1}{3072}(\g_{mnp})^{[\a\b}(\g^{mqr})^{\g\d]}N^{np}N_{qr}. \label{eq:L}
\eea
NB1: Only the antisymmetric part of $L^{\a\b\g\d}$ is given because in \cite{Oda:2007ak} the full $L^{\a\b\g\d}$ is not given in terms of gauge invariant objects. An explicit expression is known within the $Y$ formalism \cite{ Oda:2007ak, Oda:2005wu, Oda:2005sd} and it is also proven all $Y$ dependence from $L^{\a\b\g\d}$ disappears when contracted with $Z_{\a\b\g\d}$. In \cite{Berkovits:2004px} $L^{\a\b\g\d}$ is given as
\be
L^{\a\b\g\d}={c_4}^{\a\b\g\d}_{mnpq}N^{mn}N^{pq}+{c_5}^{\a\b\g\d}_{mn}JN^{mn}+{c_6}^{\a\b\g\d}JJ+{c_7}^{\a\b\g\d}_{mn}N^{mn}+{c_8}^{\a\b\g\d}J,
\ee
with unknown coefficients.\\
NB2: the coefficients of the total derivative terms depend on the normal ordering prescription and the ones above are only consistent with the prescription of \cite{Oda:2007ak}.

\section{Detailed computations of $I_k$} \label{app:ik}
This appendix contains the details of the $\lambda$ integrals that appear
at one loop. We are especially interested in those that appear in computations involving a $Q$ exact state.

A typical integral one encounters in an amplitude in subsection \ref{sec:cfwus} is given by
\be
(I_k)_{a_1\cdots a_{2k}\b_2\cdots \b_{11}} =\int [d\lambda] \fr{1}{(\lambda^+)^{k-2}}\lambda^{\b_1}\lambda_{a_1a_2}\cdots \lambda_{a_{2k-1}a_{2k}}\Lambda_{\d_1\d_2\d_3}(\e T)^{\d_1\d_2\d_3}_{\b_1\cdots \b_{11}}.
\ee
By charge conservation we can conclude at most two choices for $\b_2, \ldots, \b_{11}$ lead to a non vanishing $I'_k$ for any $k$. This follows from
\be \label{eq:nik}
0=N(I_k)_{a_1\cdots a_{2k}\b_2\cdots \b_{11}}=[(k-3)\fr{5}{4}+k(-\qu)+N(\b_2\cdots \b_{11})](I_k)_{a_1\cdots a_{2k}\b_2\cdots \b_{11}}.
\ee
There are only two choices we can make. For example for $k=3$ equation \eqref{eq:nik} implies only the components with $N(\b_2\cdots \b_{11})=-\hp$ are non vanishing. Thus $\b_2\cdots \b_{11}$ must consist of either seven ${\bf{10}}$ indices and three ${\bar{\bf{5}}}$ or a $+$, five ${\bf{10}}$'s and four ${\bar{\bf{5}}}$'s.

In section \ref{app:coeflint} we first compute all integrals of the form
\be
(I'_k)^{\b_1}_{a_1\cdots a_{2k}\d_1\d_2\d_3}=\int [d\lambda] \fr{1}{(\lambda^+)^{k-2}}\lambda^{\b_1}\lambda_{a_1a_2}\cdots \lambda_{a_{2k-1}a_{2k}}\Lambda_{\d_1\d_2\d_3}.
\ee
Since $I_k$ vanishes for $k<3$ (cf. \eqref{eq:I0}-\eqref{eq:I2}), we are only interested in $I'_k$ for $k\geq 3$. By a similar argument the $I'_k$'s are also only non vanishing for at most two choices of $\d_1\d_2\d_3$. In the last subsection half of the non vanishing components of $I_3$ and all components of $I_5$ are computed.

\subsection{Coefficients in $\lambda$ integrals} \label{app:coeflint}
For a given $k$ at most two components of $\Lambda$ give non vanishing results. We can make three choices for $\b_1$ in $I'_k$, all three choices lead to an integral of the form (not necessarily for the same $k$):
\be
(I''_k)_{a_1\cdots a_{2k}\d_1\d_2\d_3}=\int [d\lambda]\fr{1}{(\lambda^+)^{k-3}} \lambda_{a_1a_2}\cdots \lambda_{a_{2k-1}a_{2k}}\Lambda_{\d_1\d_2\d_3}.
\ee
After some algebra one finds the only non vanishing components of the $I''_k$'s are:
\bea
(I''_4)_{a_1\cdots a_8+d_1d_2}&=&\fr{1}{20}\e_{a_1a_2a_3a_4(d_1}\e_{d_2)a_5a_6a_7a_8}+{\rm 2\ perms}, \label{eq:lint2}\\
(I''_4)_{a_1\cdots a_8\ \ \ \ \ \ \ \ d_5}^{\ \ \ \ \ \ d_1d_2d_3d_4}&=&\fr{1}{5}\e_{a_1a_2a_3a_4d_5}\d^{[d_1}_{a_5}\d^{d_2]}_{a_6}\d^{[d_3}_{a_7}\d^{d_4]}_{a_8}+{\rm 11\ perms},\label{eq:lint3}\\ &&-\fr{1}{20}\e_{a_1a_2a_3a_4d_5}\d^{[d_1}_{a_5}\d^{d_2}_{a_6}\d^{d_3}_{a_7}\d^{d_4]}_{a_8}+{\rm 5\ perms} \nn \\
(I''_5)_{a_1\cdots a_{10}d_1d_2}^{\ \ \ \ \ \ \ \ \ \ \ d_3d_4}&=&\fr{1}{20}\e_{(d_1|a_1a_2a_3a_4|}\e_{d_2)a_5a_6a_7a_8}\d^{[d_3}_{a_9}\d^{d_4]}_{a_{10}}+{\rm 14\ perms} ,\label{eq:lint4}\\
(I''_6)_{a_1\cdots a_{12}d_1d_2d_3}&=&\fr{1}{60} \e_{(d_1|a_1a_2a_3a_4|}\e_{d_2|a_5a_6a_7a_8|}\e_{d_3)a_9a_{10}a_{11}a_{12}}+{\rm 14\ perms}.\label{eq:lint5}
\eea
The first step to obtain these results is finding the number of invariant tensors with the appropriate symmetries, this is one in all cases but the second. Finding the coefficients requires more work, this is done in subsection \ref{app:coef3}. All these coefficients are fixed by \eqref{eq:Lam}, including the overall factor. Two corollaries are
\be
(I'_3)_{\ a_1\cdots a_6d_1d_2}^{b\ \ \ \ \ \ \ \ \ \ d_3d_4}= (5\d^b_{(d_1}\e_{d_2)a_1a_2a_3a_4} \d^{[d_3}_{a_5}\d^{d_4]}_{a_6} +\d^b_{[a_5} \d^{[d_3}_{a_6]} \d^{d_4]}_{(d_1}\e_{d_2)a_1a_2a_3a_4})+{\rm 2\ perms},
\ee
\be
(I'_4)^{b}_{\ a_1\cdots a_8d_1d_2d_3}=\fr{1}{12}\d^b_{(d_1}\e_{d_2|a_1a_2a_3a_4|}\e_{d_3)a_5a_6a_7a_8}+{\rm 2\ perms}  \label{eq:k=5abc}.
\ee

\paragraph{Proof of equations \eqref{eq:lint2} and \eqref{eq:lint3}} \label{app:coef3}
By Lorentz invariance we can write
\be \label{eq:wsde3}
\int [d\lambda]\fr{1}{\lambda^+} \lambda_{a_1a_2}\cdots \lambda_{a_7a_8}\Lambda_{+d_1d_2}=c_3\e_{a_1a_2a_3a_4(d_1}\e_{d_2)a_5a_6a_7a_8}+{\rm 2\ perms}
\ee
and
\be \label{eq:wsde45}
\int [d\lambda]\fr{1}{\lambda^+} \lambda_{a_1a_2}\cdots \lambda_{a_7a_8}\Lambda^{d_1d_2d_3d_4}_{\ \ \ \ \ \ \ \ d_5}=c_4(\e_{a_1a_2a_3a_4d_5}\d^{[d_1}_{a_5}\d^{d_2]}_{a_6}\d^{[d_3}_{a_7}\d^{d_4]}_{a_8}+{\rm 11\ perms})+
\ee
\[
c_5(\e_{a_1a_2a_3a_4d_5}\d^{[d_1}_{a_5}\d^{d_2}_{a_6}\d^{d_3}_{a_7}\d^{d_4]}_{a_8}+{\rm 5\ perms}).
\]
for some coefficients $c_3,c_4,c_5$. They can be determined from the defining equation of $\Lambda_{\a\b\g},$ \eqref{eq:Lam}. After evaluating the RHS of that equation for the relevant components we find
\bea
\int [d\lambda] \lambda^a\lambda^b\lambda^+\Lambda_{+d_1d_2}&=&\d^{(a}_{d_1}\d^{b)}_{d_2}-\fr{2}{5}\d^{(a}_{d_1}\d^{b)}_{d_2}=\fr{3}{5}\d^{(a}_{d_1}\d^{b)}_{d_2},\label{eq:juyh3}\\
\int [d\lambda] \lambda^a\lambda^b\lambda^+\Lambda^{d_1d_2d_3d_4}_{\ \ \ \ \ \ \ \ d_5}&=&\fr{1}{5}\e^{d_1d_2d_3d_4(a}\d^{b)}_{d_5} ,\label{eq:juyh4}\\
\int [d\lambda] \lambda_{a_1a_2}\lambda_{a_3a_4}\lambda^a\Lambda^{d_1d_2d_3d_4}_{\ \ \ \ \ \ \ d_5}&=&(\d^{d_1}_{[a_1}\d^{d_2}_{a_2]}\d^{d_3}_{[a_3}\d^{d_4}_{a_4]}\d^a_{d_5}+{\rm 1\ perm})-\fr{1}{5}\d^{d_1}_{[a_1}\d^{d_2}_{a_2}\d^{d_3}_{a_3}\d^{d_4}_{a_4]}\d^a_{d_5}+\nn \\
&&(\fr{1}{5}\d^{a}_{[a_1}\d^{[d_1}_{a_2]}\d^{d_2]}_{d_5}\d^{[d_3}_{a_3}\d^{d_4]}_{a_4}+{\rm 3\ perms}). \label{eq:juyh5}
\eea
If we now use equations \eqref{eq:wsde3} and \eqref{eq:wsde45} to evaluate the LHS of the above integrals we completely determine the values of $c_3,c_4,c_5$. In fact we find more than three equations, but they include only three independent conditions as they should. To obtain $c_3$ one has to write out $\lambda^a$ and $\lambda^b$ in \eqref{eq:juyh3} and then perform all the contractions of the two $\e$'s with the RHS of \eqref{eq:wsde3}:
\be
\fr{3}{5}\d^{(a}_{d_1}\d^{b)}_{d_2}=\int [d\lambda] \lambda^a\lambda^b\lambda^+\Lambda_{+d_1d_2}=12 c_3 \d^{(a}_{d_1}\d^{b)}_{d_2} \Rightarrow c_3=\fr{1}{20}.
\ee
Finding $c_4$ and $c_5$ is more involved. The LHS of \eqref{eq:juyh4} can be evaluated as
\be
\fr{1}{5}\e^{d_1d_2d_3d_4(a}\d^{b)}_{d_5}=\int [d\lambda] \lambda^a\lambda^b\lambda^+\Lambda^{d_1d_2d_3d_4}_{\ \ \ \ \ \ \ \ d_5}=(4c_4+12c_5) \d^{(a}_{d_5}\e^{b)d_1d_2d_3d_4}.
\ee
This gives us the first equation for $c_4,c_5$. In order to completely determine them, we have to work out the LHS of \eqref{eq:juyh5}:
\be \label{eq:koij}
\fr{1}{8}\e^{aa_5a_6a_7a_8}\int [d\lambda] \fr{1}{\lambda^+}\lambda_{a_1a_2}\lambda_{a_3a_4}\lambda_{a_5a_6}\lambda_{a_7a_8}\Lambda^{d_1d_2d_3d_4}_{\ \ \ \ \ \ \ d_5}=
\ee
\[
\fr{c_4}{8}((24 \d^a_{d_5}\d^{[d_1}_{a_1}\d^{d_2]}_{a_2}\d^{[d_3}_{a_3}\d^{d_4]}_{a_4}+{\rm 1\ perm})+8 \e^{ad_1d_2d_3d_4}\e_{a_1a_2a_3a_4d_5}+16 (\d^a_{d_5}\d^{[d_1}_{a_1}\d^{d_2]}_{a_2} \d^{[d_3}_{a_3}\d^{d_4]}_{a_4}+{\rm 1\ perm})+
\]
\[
(8 \d^a_{[a_1}\d^{[d_3}_{a_2]}\d^{d_4]}_{d_5}\d^{[d_1}_{a_3}\d^{d_2]}_{a_4}+{\rm 3\ perm}))+
\]
\[
\fr{c_5}{8}(24\d^a_{d_5}\d^{[d_1}_{a_1}\d^{d_2}_{a_2}\d^{d_3}_{a_3}\d^{d_4]}_{a_4}+24 \e^{ad_1d_2d_3d_4} \e_{a_1a_2a_3a_4d_5} +16\d^a_{d_5}\d^{[d_1}_{a_1}\d^{d_2}_{a_2}\d^{d_3}_{a_3}\d^{d_4]}_{a_4}+
\]
\[
(8\d^a_{[a_1}\d^{[d_3}_{a_2]}\d^{d_4}_{d_5}\d^{d_1}_{a_3}\d^{d_2]}_{a_4}+{\rm 1\ perm})).
\]
We want to be able to read off equations for the $c$'s when we compare to \eqref{eq:juyh5}. It turns out the space of invariant tensors with the indices and symmetries of \eqref{eq:juyh5} is four dimensional. We now write out our tensors on a basis that contains the three invariant tensors that are present in \eqref{eq:juyh5}. We are free to choose the fourth one as long as it does not lie in the span of the first three. After using
\be \e^{ad_1d_2d_3d_4}\e_{a_1a_2a_3a_4d_5}=\d^a_{d_5}\d^{[d_1}_{a_1}\d^{d_2}_{a_2}\d^{d_3}_{a_3}\d^{d_4]}_{a_4}+(\d^a_{[a_1}\d^{[d_1}_{a_2]}\d^{d_2}_{d_5}\d^{d_3}_{a_3}\d^{d_4]}_{a_4}+{\rm 1\ perm}),
\ee
\eqref{eq:koij} becomes
\be
(5c_4 \d^a_{d_5}\d^{[d_1}_{a_1}\d^{d_2]}_{a_2}\d^{[d_3}_{a_3}\d^{d_4]}_{a_4}+{\rm 1\ perm})+
(c_4 \d^a_{[a_1}\d^{[d_3}_{a_2]}\d^{d_4]}_{d_5}\d^{[d_1}_{a_3}\d^{d_2]}_{a_4}+{\rm 3\ perm})+
\ee
\[
(8c_5+c_4)\d^a_{d_5}\d^{[d_1}_{a_1}\d^{d_2}_{a_2}\d^{d_3}_{a_3}\d^{d_4]}_{a_4}+((c_4+4c_5)\d^a_{[a_1}\d^{[d_3}_{a_2]}\d^{d_4}_{d_5}\d^{d_1}_{a_3}\d^{d_2]}_{a_4}+{\rm 1\ perm}).
\]
Now we can read off four equations for $c_4,c_5$ by comparing to \eqref{eq:juyh5}. Combined with the equation we already found:
\be
5c_4=1,\quad c_4+8c_5=-\fr{1}{5},\quad c_4=\fr{1}{5},\quad c_4+4c_5=0,\quad 4c_4+12c_5=\fr{1}{5}.
\ee
These equations are solved by
\be
c_4=\fr{1}{5},\quad c_5=-\fr{1}{20}.
\ee
The coefficients in equations \eqref{eq:lint4} and \eqref{eq:lint5} follow in the same way.

\subsection{Computing the $I_k$'s} \label{sec:I3}
The idea of this section is simple, use the explicit form of the gamma matrices and the $\lambda$ integrals \eqref{eq:lint2}-\eqref{eq:k=5abc} to evaluate $I_k$. In practice this involves a lot of computation. We already know $I_0,I_1,I_2$ and $I_6$ all vanish. By the charge conservation property there is only one choice of $\b_2\cdots \b_{11}$ for which $I_5$ does not vanish. For $I_3$ and $I_4$ we can make two choices. We explicitly compute $I_3$ for
\be
\b_2,\cdots ,\b_{11}=+,c_1,c_2,c_3,c_4,b_1b_2,\cdots ,b_9b_{10}.
\ee
$I_3$ consists of three terms, two for $\b_1=b_1b_2$ and one for $\b_1=b_1$. The relevant components of $\e T$ are\footnote{To evaluate $\e T$ the following convention for $\e_{\b_1\cdots \b_{16}}$ is used, $(\e_{16})_{+a_1\cdots a_5}^{\ \ \ \ \ \ \ \ b_1b_2\cdots b_{19}b_{20}}=(\e_5)_{a_1\cdots a_5}(\e_{10})^{b_1\cdots b_{20}}$}
\bea
&&(\e T)^{+d_1d_2\ b_{11}b_{12}\ \ b_1b_2\cdots b_{9}b_{10}}_{\ \ \ \ \ \ \ \ \ \ \ \ \ +\ \ \ \ \ \ \ \ \ \ \ c_1c_2c_3c_4}=
\\ \nn
&&\fr{1}{16}8(\e_{10})^{b_1\cdots b_{20}}\e_{c_1\cdots c_5}\g^{k_1d_1}_{\ \ \ \ b_{13}b_{14}}\g^{k_2d_2}_{\ \ \ \ b_{15}b_{16}}\g^{\ \ +c_5}_{k_3}(\g_{k_1k_2}^{\ \ \ \ k_3})_{b_{17}b_{18}b_{19}b_{20}}=
\\ \nn
&&-\hp 8(\e_{10})^{b_1\cdots b_{20}}\e_{c_1c_2c_3c_4b_{17}}\d^{d_1}_{b_{14}}\d^{d_2}_{b_{16}}\e_{b_{13}b_{15}b_{18}b_{19}b_{20}},
\eea
\bea
&&(\e T)^{\ \ \ \ \ \ \ \ \ d_5\ b_{11}b_{12}\ \ b_1b_2\cdots b_{9}b_{10}}_{d_1d_2d_3d_4\ \ \ \ \ \ \ \ \ \ +\ \ \ \ \ \ \ \ \ \ c_1c_2c_3c_4}=
\\ \nn
&&8 \fr{1}{16}2(\e_{10})^{b_1\cdots b_{20}}\e_{c_1\cdots c_5}\g_{k_1d_1d_2b_{13}b_{14}}\g_{k_2d_3d_4b_{15}b_{16}}\g^{k_3d_5}_{b_{17}b_{18}}(\g^{k_1k_2}_{\ \ \ k_3})_{b_{19}b_{20}}^{\ \ \ \ c_5}
\\ \nn
&&+8\fr{1}{16} (\e_{10})^{b_1\cdots b_{20}}\e_{c_1\cdots c_5}\g^{k_1c_5}_{d_1d_2}\g_{k_2d_3d_4b_{13}b_{14}}\g^{k_3d_5}_{b_{15}b_{16}}(\g_{k_1\ \ k_3}^{\ \ k_2})_{b_{17}b_{18}b_{19}b_{20}}+(d_1d_2 \lera d_3d_4)=
\\ \nn
&&8\qu 2 (\e_{10})^{b_1\cdots b_{20}}\e_{c_1c_2c_3c_4b_{17}}\e_{b_{19}d_1d_2b_{13}b_{14}}\e_{b_{20}d_3d_4b_{15}b_{16}}\d^{d_5}_{b_{18}}+
\\ \nn
&&8\qu (\e_{10})^{b_1\cdots b_{20}}\e_{c_1c_2c_3c_4b_{19}}\e_{d_1d_2b_{13}b_{14}[b_{20}}\e_{b_{17}]d_3d_4b_{15}b_{16}}\d^{d_5}_{b_{18}}+
\\ \nn
&&8 \qu \hp (\e_{10})^{b_1\cdots b_{20}}\e_{c_1c_2c_3c_4[d_2}\e_{d_1]b_{15}b_{19}b_{20}b_{18}}\e_{b_{17}d_3d_4b_{13}b_{14}}\d^{d_5}_{b_{16}}+(d_1d_2 \lera d_3d_4),
\eea
\bea
&&(\e T)^{d_1d_2\ \ \ \ \ \ \ \ b_1\cdots b_{10}}_{\ \ \ \ d_3d_4\ b+\ \ \ \ \ \ \ c_1c_2c_3c_4}=
\\ \nn
&&-8 \fr{1}{32}(\e_{10})^{b_1\cdots b_{20}}\e_{c_1c_2c_3c_4b}\g^{k_1d_1}_{b_{11}b_{12}}\g^{k_2d_2}_{b_{13}b_{14}}\g_{k_3d_3d_4b_{15}b_{16}}(\g_{k_1k_2}^{\ \ \ \ k_3})_{b_{17}b_{18}b_{19}b_{20}}=
\\ \nn
&&8\qu \e_{c_1c_2c_3c_4b}\e_{d_3d_4b_{15}b_{16}b_{17}} \e_{b_{11}b_{13}b_{18}b_{19}b_{20}}\d^{d_1}_{b_{12}} \d^{d_2}_{b_{14}},
\eea
where we extracted the factor of eight coming from the $SU(5)$ decomposition (cf. \eqref{eq:u5ind}) and a power of $\hp$, which compensates for double counting in expressions like $x_{ab}y^{ab}$, in each line. Using the explicit form of the components of $(\e T)$ and the $\lambda$ integrals, $I_3$ can be written out as
\bea
I_3&=&\hp 3\int [d\lambda] \fr{1}{\lambda^+}\lambda_{b_{11}b_{12}}\lambda_{a_1a_2}\lambda_{a_3a_4}\lambda_{a_5a_6}\Lambda_{+d_1d_2}(\e T)^{+d_1d_2\ b_{11}b_{12}\ \ b_1\cdots b_{10}}_{\ \ \ \ \ \ \ \ \ \ \ \ \ +\ \ \ \ \ \ \ c_1c_2c_3c_4}+
\\ \nn
&&\fr{1}{8}3 \int [d\lambda] \fr{1}{\lambda^+}\lambda_{b_{11}b_{12}}\lambda_{a_1a_2}\lambda_{a_3a_4}\lambda_{a_5a_6}\Lambda^{d_1d_2d_3d_4}_{\ \ \ \ \ \ \ \ d_5}(\e T)^{\ \ \ \ \ \ \ \ d_5\ b_{11}b_{12}\ \ b_1\cdots b_{10}}_{d_1d_2d_3d_4\ \ \ \ \ \ \ \ \ +\ \ \ \ \ \ c_1c_2c_3c_4}+
\\ \nn
&&3\hp \int [d\lambda] \fr{1}{\lambda^+}\lambda^b\lambda_{a_1a_2}\lambda_{a_3a_4}\lambda_{a_5a_6}\Lambda^{\ \ \ d_3d_4}_{d_1d_2}(\e T)^{d_1d_2\ \ \ \ \ \ \ \ b_1\cdots b_{10}}_{\ \ \ \ d_3d_4\ b+\ \ \ \ \ \ \ c_1c_2c_3c_4}=
\\ \nn
&&\fr{3}{40}(\e_{a_1a_2a_3a_4(d_1}\e_{d_2)a_5a_6b_{11}b_{12}}+{\rm 2\ perms})
\\ \nn
&&[-\qu(\e_{10})^{b_1\cdots b_{20}}\e_{c_1c_2c_3c_4b_{17}}\d^{d_1}_{b_{14}}\d^{d_2}_{b_{16}} \e_{b_{13}b_{15}b_{18}b_{19}b_{20}}]+
\\ \nn
&&\fr{3}{40}((\e_{a_1a_2a_3a_4d_5}\d^{[d_1}_{a_5}\d^{d_2]}_{a_6}\d^{[d_3}_{b_{11}}\d^{d_4]}_{b_{12}}+{\rm 11\ perms}) )
\\ \nn
&&[4 (\e_{10})^{b_1\cdots b_{20}}\e_{c_1c_2c_3c_4b_{17}}\e_{b_{19}d_1d_2b_{13}b_{14}}\e_{b_{20}d_3d_4b_{15}b_{16}}\d^{d_5}_{b_{18}}+
\\ \nn
&&2 (\e_{10})^{b_1\cdots b_{20}}\e_{c_1c_2c_3c_4b_{19}}\e_{d_1d_2b_{13}b_{14}[b_{20}}\e_{b_{17}]d_3d_4b_{15}b_{16}}\d^{d_5}_{b_{18}}+
\\ \nn
&& (\e_{10})^{b_1\cdots b_{20}}\e_{c_1c_2c_3c_4[d_2}\e_{d_1]b_{15}b_{19}b_{20}b_{18}}\e_{b_{17}d_3d_4b_{13}b_{14}}\d^{d_5}_{b_{18}}+(d_1d_2 \lera d_3d_4)]+
\\ \nn
&&\fr{3}{2}(5 \d^b_{(d_1}\e_{d_2)a_1a_2a_3a_4}\d^{[d_3}_{a_5}\d^{d_4]}_{a_6}+ \d^b_{[a_5}\d^{[d_3}_{a_6]} \d^{d_4]}_{(d_1}\e_{d_2)a_1a_2a_3a_4} +{\rm 2\ perms})
\\ \nn
&&[2 \e_{c_1c_2c_3c_4b}\e_{d_3d_4b_{15}b_{16}b_{17}}\e_{b_{11}b_{13}b_{18}b_{19}b_{20}} \d^{d_1}_{b_{12}}\d^{d_2}_{b_{14}}]=
\\ \nn
&&-\fr{3}{5}\e_{a_1a_2a_3a_4b_{14}}\e_{b_{16}a_5a_6b_{11}b_{12}}(\e_{10})^{b_1\cdots b_{20}}\e_{c_1c_2c_3c_4b_{17}}\e_{b_{13}b_{15}b_{18}b_{19}b_{20}}+{\rm 2\ perms}+
\\ \nn
&&\fr{12}{5}(\e_{10})^{b_1\cdots b_{20}}\e_{b_{11}b_{12}a_3a_4b_{18}}\e_{c_1c_2c_3c_4b_{17}} \e_{b_{19}a_1a_2b_{13}b_{14}} \e_{b_{20}a_5a_6b_{15}b_{16}}+{\rm 2\ perms}+
\\ \nn
&&\fr{3}{5} (\e_{10})^{b_1\cdots b_{20}}\e_{b_{11}b_{12}a_3a_4b_{18}}\e_{c_1c_2c_3c_4b_{19}} \e_{a_1a_2b_{13}b_{14}[b_{17}}\e_{b_{20}]a_5a_6b_{15}b_{16}}+{\rm 2\ perms}+
\\ \nn
&&\fr{6}{5} \e_{a_1a_2a_3a_4b_{16}}(\e_{10})^{b_1\cdots b_{20}}\e_{c_1c_2c_3c_4b_{12}} \e_{b_{11}b_{15}b_{18}b_{19}b_{20}} \e_{b_{17}a_5a_6b_{13}b_{14}}+{\rm 2\ perms}+
\\ \nn
&&\fr{6}{5}\e_{b_{11}b_{12}a_1a_2b_{16}}(\e_{10})^{b_1\cdots b_{20}}\e_{c_1c_2c_3c_4[a_4} \e_{a_3]b_{15}b_{18}b_{19}b_{20}} \e_{a_5a_6b_{13}b_{14}b_{17}}+{\rm 2\ perms}+
\\ \nn
&&60(\e_{10})^{b_1\cdots b_{20}}\e_{b_{14}a_1a_2a_3a_4}\e_{c_1c_2c_3c_4b_{12}} \e_{a_5a_6b_{15}b_{16}b_{17}}\e_{b_{11}b_{13}b_{18}b_{19}b_{20}}+{\rm2\ perms}+
\\ \nn
&&12\e_{b_{14}a_1a_2a_3a_4}(\e_{10})^{b_1\cdots b_{20}}\e_{c_1c_2c_3c_4[a_5} \e_{a_6]b_{12}b_{15}b_{16}b_{17}}\e_{b_{11}b_{13}b_{18}b_{19}b_{20}}+{\rm2\ perms}=
\\ \nn
&&(-\fr{3}{5}(1)+\fr{12}{5}(-\hp)+\fr{3}{5}(-1)+\fr{6}{5}(-1)+\fr{6}{5}(\fr{27}{2})+60(1)+12(0))
\\ \nn
&&\e_{a_1a_2a_3a_4b_{14}}\e_{b_{16}a_5a_6b_{11}b_{12}}(\e_{10})^{b_1\cdots b_{20}}\e_{c_1c_2c_3c_4b_{17}}\e_{b_{13}b_{15}b_{18}b_{19}b_{20}}+{\rm 2\ perms}=
\\ \nn
&&\fr{129}{2}\e_{a_1a_2a_3a_4b_{14}}\e_{b_{16}a_5a_6b_{11}b_{12}}(\e_{10})^{b_1\cdots b_{20}}\e_{c_1c_2c_3c_4b_{17}}\e_{b_{13}b_{15}b_{18}b_{19}b_{20}}+{\rm 2\ perms}.
\eea
Since ${\rm Asym}^5 {\bf 10} \otimes {\rm Sym}^3 {\bf\bar{10}}\otimes {\rm Asym}^4 {\bf \bar{5}}$ contains one scalar all seven tensors in the penultimate step are proportional to each other. The constants of proportionality are obtained by computing components.

$I_5$ is only non vanishing if we choose
\be
\b_2,\cdots ,\b_{11}=b_3b_4,\cdots ,b_{11}b_{12},1,2,3,4,5.
\ee
This component of $I_5$ consists of two terms, one for $\b_1=b_1b_2$ and one for $\b_1=+$:
\be
(I_5)_{a_1\cdots a_{10}\ \ \ \ \ \ \ 12345}^{\ \ \ \ \ \ \ b_3\cdots b_{12}}=\int [d\lambda] \fr{1}{(\lambda^+)^2}\lambda_{a_1a_2}\cdots \lambda_{a_9a_{10}}\Lambda_{\d_1\d_2\d_3}(\e T)^{\d_1\d_2\d_3\ \ b_3\cdots b_{12}}_{\ \ \ \ \ \ +\ \ \ \ \ \ \ 12345}+
\ee
\[
\hp \int [d\lambda] \fr{1}{(\lambda^+)^3}\lambda_{b_1b_2}\lambda_{a_1a_2}\cdots \lambda_{a_9a_{10}}\Lambda_{\d_1\d_2\d_3}(\e T)^{\d_1\d_2\d_3\ b_1b_2b_3\cdots b_{12}}_{\ \ \ \ \ \ \ \ \ \ \ \ \ \ \ \ \ 12345}.
\]
The relevant components of $\e T$ are given by
\bea
&&(\e T)^{d_1d_2d_3\ b_1\cdots b_{12}}_{\ \ \ \ \ \ \ \ \ \ \ \ \ \ 12345}=
\\ \nn
&&-8\fr{1}{16}2(\e_{10})^{b_1\cdots b_{20}}\g^{ad_1}_{b_{13}b_{14}}\g^{bd_2}_{b_{15}b_{16}}\g^{cd_3}_{b_{17}b_{18}}\g_{abcb_{19}b_{20}}^{\ \ \ \ \ \ \ +}+
\\ \nn
&&-\fr{1}{16}8(\e_{10})^{b_1\cdots b_{20}}\hp \g_{a\ \ +}^{\ (d_1}\g^{|b|d_2}_{b_{13}b_{14}}\g^{|c|d_3)}_{b_{15}b_{16}}\g^a_{\ bcb_{17}b_{18}b_{19}b_{20}}=
\\ \nn
&&-(\e_{10})^{b_1\cdots b_{20}} \d^a_{[b_{13}}\d^{d_1}_{b_{14}]}\d^b_{[b_{15}}\d^{d_2}_{b_{16}]}\d^c_{[b_{17}}\d^{d_3}_{b_{18}]}(-1)\e_{abcb_{19}b_{20}}+
\\ \nn
&&-\qu (\e_{10})^{b_1\cdots b_{20}}\d^{(d_1}_a\d^{|b|}_{[b_{13}}\d^{d_2}_{b_{14}]}\d^{|c|}_{[b_{15}}\d^{d_3)}_{b_{16}]}\d^a_{[b_{17}}\e_{b_{18}]bcb_{19}b_{20}},
\eea
\bea
&&(\e T)^{d_1d_2\ \ \ \ \ \ \ b_3b_4\cdots b_{11}b_{12}}_{\ \ \ \ d_3d_4+\ \ \ \ \ \ \ \ \ \ \ \ 12345}=
\\ \nn
&&-8\fr{1}{32}(\e_{10})^{b_3\cdots b_{22}}\g_{ad_3d_4b_{13}b_{14}}\g^{bd_1}_{b_{15}b_{16}}\g^{cd_2}_{b_{17}b_{18}}\g^a_{\ bcb_{19}b_{20}b_{21}b_{22}}=
\\ \nn
&&-\qu (\e_{10})^{b_3\cdots b_{22}}(-1)\e_{ad_3d_4b_{13}b_{14}}\d^b_{[b_{15}}\d^{d_1}_{b_{16}]}\d^c_{[b_{17}} \d^{d_2}_{b_{18}]}2\d^a_{[b_{19}}\e_{b_{20}]bcb_{21}b_{22}},
\eea
where we extracted the factor of eight and the powers of $\hp$ again. In summary the two relevant components of $(\e T)$ are given by
\be
(\e T)^{d_1d_2d_3\ b_1\cdots b_{12}}_{\ \ \ \ \ \ \ \ \ \ \ \ \ \ 12345}=-8 \hp 5 (\e_{10})\d^{d_1}_{b_{13}}\d^{d_2}_{b_{15}}\d^{d_3}_{b_{17}}\e_{b_{14}b_{16}b_{18}b_{19}b_{20}}
\ee
and
\be
(\e T)^{d_1d_2\ \ \ \ \ \ \ b_3b_4\cdots b_{11}b_{12}}_{\ \ \ \ d_3d_4+\ \ \ \ \ \ \ \ \ \ \ \ 12345}=8 \qu  \e_{10} \e_{b_{17}d_3d_4b_{15}b_{16}}\e_{b_{18}b_1b_{13}b_{19}b_{20}}\d^{d_1}_{b_2}\d^{d_2}_{b_{14}}.
\ee
$I_5$ becomes
\bea \label{eq:k=5}
I_5&=&\int [d\lambda]\fr{1}{(\lambda^+)^3}\lambda^{\b_1}\lambda_{a_1a_2}\cdots \lambda_{a_9a_{10}}\Lambda_{\d_1\d_2\d_3}(\e T)^{\d_1\d_2\d_3\ b_3\cdots b_{12}}_{\ \ \ \ \ \b_1\ \ \ \ \ \ \ 12345}=
\\ \nn
&&\hp 3 \int [d\lambda] \fr{1}{(\lambda^+)^2}\lambda_{a_1a_2}\cdots \lambda_{a_9a_{10}}\Lambda_{d_1d_2}^{\ \ \ \ d_3d_4}(\e T)^{d_1d_2\ \ \ \ \ \ \ b_3\cdots b_{12}}_{\ \ \ \ d_3d_4\ +\ \ \ \ \ \ \ 12345}+
\\ \nn
&&\hp \int [d\lambda] \fr{1}{(\lambda^+)^3}\lambda_{b_1b_2}\lambda_{a_1a_2}\cdots \lambda_{a_9a_{10}}\Lambda_{d_1d_2d_3}(\e T)^{d_1d_2d_3\ b_1\cdots b_{12}}_{\ \ \ \ \ \ \ \ \ \ \ \ \ \ \ 12345}=
\\ \nn
&&\fr{3}{40}( \e_{(d_1|a_1a_2a_3a_4|}\e_{d_2)a_5a_6a_7a_8}\d^{[d_3}_{a_9}\d^{d_4]}_{a_{10}}+{\rm 14\ perms})(\e T)^{d_1d_2\ \ \ \ \ \ \ b_3\cdots b_{12}}_{\ \ \ \ d_3d_4\ +\ \ \ \ \ \ \ 12345}+
\\ \nn
&&\fr{1}{120}( \e_{(d_1|a_1a_2a_3a_4|}\e_{d_2|a_5a_6a_7a_8|}\e_{d_3)a_9a_{10}b_1b_2}+{\rm 14\ perms})(\e T)^{d_1d_2d_3\ b_1\cdots b_{12}}_{\ \ \ \ \ \ \ \ \ \ \ \ \ \ \ 12345}=
\\ \nn
&&\fr{3}{20}( \e_{d_1a_1a_2a_3a_4}\e_{d_2a_5a_6a_7a_8}\d^{[d_3}_{a_9}\d^{d_4]}_{a_{10}}+{\rm 14\ perms})(\e T)^{d_1d_2\ \ \ \ \ \ \ b_3\cdots b_{12}}_{\ \ \ \ d_3d_4\ +\ \ \ \ \ \ \ 12345}+
\\ \nn
&&\fr{1}{20}( \e_{d_1a_1a_2a_3a_4}\e_{d_2a_5a_6a_7a_8}\e_{d_3a_9a_{10}b_1b_2}+{\rm 14\ perms})(\e T)^{d_1d_2d_3\ b_1\cdots b_{12}}_{\ \ \ \ \ \ \ \ \ \ \ \ \ \ \ 12345}=
\\ \nn
&&\fr{3}{10}( \e_{d_1a_1a_2a_3a_4}\e_{d_2a_5a_6a_7a_8}\d^{[d_3}_{a_9}\d^{d_4]}_{a_{10}}+{\rm 14\ perms})
\\ \nn
&&[(\e_{10})^{b_1\cdots b_{20}} \e_{b_{17}d_3d_4b_{15}b_{16}} \e_{b_{18}b_1b_{13}b_{19}b_{20}} \d^{d_1}_{b_2}\d^{d_2}_{b_{14}}]
\\ \nn
&&-( \e_{d_1a_1a_2a_3a_4}\e_{d_2a_5a_6a_7a_8}\e_{d_3a_9a_{10}b_1b_2}+{\rm 14\ perms})
\\ \nn
&&[(\e_{10})^{b_1\cdots b_{20}}\d^{d_1}_{b_{13}}\d^{d_2}_{b_{15}}\d^{d_3}_{b_{17}}\e_{b_{14}b_{16}b_{18}b_{19}b_{20}}]=
\\ \nn
&& \fr{3}{5} \e_{b_2a_1a_2a_3a_4}\e_{b_{14}a_5a_6a_7a_8}(\e_{10})^{b_1\cdots b_{20}} \e_{b_{17}a_9a_{10}b_{15}b_{16}}\e_{b_{18}b_1b_{13}b_{19}b_{20}}+{\rm 14\ perms}
\\ \nn
&&- \e_{b_{13}a_1a_2a_3a_4}\e_{b_{15}a_5a_6a_7a_8}\e_{b_{17}a_9a_{10}b_1b_2}(\e_{10})^{b_1\cdots b_{20}}\e_{b_{14}b_{16}b_{18}b_{19}b_{20}}+{\rm 14\ perms}=
\\ \nn
&&-\fr{2}{5} \e_{b_{13}a_1a_2a_3a_4}\e_{b_{15}a_5a_6a_7a_8}\e_{b_{17}a_9a_{10}b_1b_2}(\e_{10})^{b_1\cdots b_{20}}\e_{b_{14}b_{16}b_{18}b_{19}b_{20}}+{\rm 14\ perms}.
\eea

\providecommand{\href}[2]{#2}\begingroup\raggedright\endgroup


\end{document}